\RequirePackage{fix-cm}
\documentclass[natbib,smallextended]{svjour3}       % onecolumn (second format)
\smartqed  % flush right qed marks, e.g. at end of proof
\usepackage{graphicx}
 \journalname{my journal}
\def\lesssim{\mathrel{\hbox{\rlap{\hbox{\lower4pt\hbox{$\sim$}}}\hbox{$<$}}}}
\def\gtrsim{\mathrel{\hbox{\rlap{\hbox{\lower4pt\hbox{$\sim$}}}\hbox{$>$}}}}

\begin{document}

\title{Perspectives on Gamma-Ray Burst Physics and Cosmology 
with Next Generation Facilities
%Future experiments 
}

\titlerunning{Gamma-Ray Bursts: Next Generation Facilities}        % if too long for running head

\author{
           Weimin Yuan \and
           Lorenzo Amati  \and 
	  John K. Cannizzo \and
           Bertrand Cordier   \and
           Neil Gehrels \and 
           Giancarlo Ghirlanda \and          
           Diego G\"otz \and
           Nicolas Produit \and
           Yulei Qiu \and
           Jianchao Sun \and
           Nial R. Tanvir \and
           Jianyan Wei \and
%           Weimin Yuan  \and                 
           Chen Zhang
}

\authorrunning{W. Yuan, et al.} % if too long for running head

\institute{
W. Yuan \at 
              Key Laboratory of Space Astronomy and Technology, 
              National Astronomical Observatories\\ 
              Chinese Academy of Sciences, Datun Rd 20A, Beijing 100012 China \\       
              \email{wmy@nao.cas.cn}
\and
L. Amati \at
              INAF - IASF Bologna, via P. Gobetti 101\\ 
              40129 Bologna, Italy \\
              \email{amati@iasfbo.inaf.it} 
 \and
        %  S. Author \at
          J.K. Cannizzo  \at
            %   second address
     CRESST/Joint Center for Astrophysics \\
      Univ. of Maryland, Baltimore County \\
       Baltimore, MD 21250, USA           \\
              \email{john.k.cannizzo@nasa.gov}
\and
B. Cordier \at
CEA-Irfu, Service d'Astrophysique \\
F-91191 Gif-sur-Yvette, France\\
 \email{bertrand.cordier@cea.fr}           
\and
N. Gehrels  \at
           Astroparticle Physics Division   \\
           NASA/Goddard Space Flight Center \\
           Greenbelt, MD 20771, USA         \\
            \email{neil.gehrels@nasa.gov}        
\and
 G. Ghirlanda   \at
            Istituto Nazionale di Astrofisica INAF - Osservatorio Astronomico di Brera \at
              Via E. Bianchi 46 - I 23807 Merate (LC), Italy \\
%              Tel.: +39-039-72320495\\
              \email{giancarlo.ghirlanda@brera.inaf.it}             
\and
D. G\"otz \at
CEA-Irfu, Service d'Astrophysique \\
F-91191 Gif-sur-Yvette, France\\
\email{diego.gotz@cea.fr}
\and
	 J. Sun \at
           Institute of High-Energy Physics, Chinese Academy of Sciences, \\
           Yuquan Lu 19B, Shijingshan District, Beijing, China\\
           \email{sunjc@ihep.ac.cn}           %  \\
\and
 N. R. Tanvir \at
              University of Leicester, Dept. of Physics and Astronomy, \\
              University Road, Leicester, LE1 7RH,              United Kingdom \\
%              Tel.: +44-116-2231217\\
%              Fax: +123-45-678910\\
              \email{nrt3@le.ac.uk}                           
\and
           N. Produit \at
           ISDC, Chemin d' Ecogia 16, CH-1290 Versoix, Switzerland\\
           \email{Nicolas.Produit@unige.ch}                            
\and
Y. Qiu \and J. Wei  \and C. Zhang \at
%Y. Qiu \and J. Wei  \and W. Yuan \and C. Zhang \at
              Key Laboratory of Space Astronomy and Technology,
              National Astronomical Observatories\\  Chinese Academy of Sciences,
              Datun Rd 20A, Beijing 100012 China \\       
\email{qiuyl@nao.cas.cn; wjy@nao.cas.cn; chzhang@nao.cas.cn}
%              \email{qiuyl@nao.cas.cn; wjy@nao.cas.cn; wmy@nao.cas.cn; chzhang@nao.cas.cn}
}

\date{Received: date / Accepted: date}

\maketitle

\begin{abstract}
High-redshift Gamma-Ray Bursts (GRBs) beyond redshift $\sim6$ are potentially
powerful tools to probe the distant early Universe. Their detections in large
numbers and at truly high redshifts call for the next generation of high-energy
wide-field instruments with unprecedented sensitivity at least one order of
magnitude higher than the ones currently in orbit. On the other hand, follow-up
observations of the afterglows of high-redshift GRBs and identification of
their host galaxies, which would be difficult for the currently operating
telescopes, require new, extremely large facilities of at multi-wavelengths.
This chapter describes future experiments that are expected to
advance this exciting field, both being currently built and being proposed. 
The legacy of Swift will be continued by SVOM, 
which is equipped with a set of space-based multi-wavelength instruments as well as
and a ground segment including a wide angle camera and two follow-up telescopes.
The established Lobster-eye X-ray focusing
optics provides a promising technology for the detection of faint GRBs at very
large distances, based on which the {THESEUS}, {Einstein Probe} and
other mission concepts have been proposed. Follow-up observations and
exploration of the reionization era will be enabled by large facilities such as
{SKA} in the radio, the 30m class telescopes in the optical/near-IR, and
the space-borne {WFIRST} and {JWST} in the optical/near-IR/mid-IR. In
addition, the X-ray and $\gamma$-ray polarization experiment {POLAR} is
also introduced.  
\keywords{Gamma-ray bursts \and high-redshift \and Gamma-ray \and X-ray \and instrumentation}
\end{abstract}

\section{Introduction}
\label{sect_intro}
Understanding the structure formation and the first stars around the epoch of
reionization is a major driver for the next generation of space- and
ground-based astronomical facilities. Being the brightest objects in the
Universe, Gamma-Ray Bursts (GRBs) beyond redshift $\sim6$ shine through the
dark age as cosmic beacons and are thus potentially powerful probes of the
distant early Universe. So far, seven GRBs with redshifts $6.2-9.4$ have been
detected with the Swift satellite in its $10$-year operation. 
New discoveries will continue to be made by the GRB mission SVOM in the foreseeable
future, which is equipped with a set of space-based multi-wavelength instruments as well as
a ground segment including a wide angle camera and two follow-up telescopes.
Looking far into the future, a full exploration of the early Universe requires a much
larger sample ($\sim$100) and events at even higher redshifts, i.e. redshift
$\sim10$ and beyond. This calls for the next generation of high-energy wide-field
instruments with unprecedented sensitivity one order of magnitude or more
higher than Swift/BAT. The proposed THESEUS and Einstein Probe
missions (and others alike), which are based on novel technology of X-ray
focusing, are promising concepts to achieve these goals. On the other hand,
follow-up observations in multi-wavebands, particularly in the near- and mid-
infrared, of the afterglows of high-redshift GRBs are key to identify their
host galaxies, to measure the redshifts, and to enable detailed astrophysical
and cosmological studies. This can be facilitated by the advent of the ground-
and space-based large telescopes at multi-wavelengths currently being under
construction, such as the 30-m class optical/near-IR telescopes, Square
Kilometer Array (SKA) in the radio, and the space-borne WFIRST and
JWST missions in the optical/near-IR/mid-IR. 
Last, complementary to the
exploration of the early Universe, future extensive measurement of the X-ray
and $\gamma$-ray polarization of GRBs, expected to be achieved by POLAR will open up
a new dimension in understanding the radiation region geometry and radiation
mechanism of GRB. 
This chapter describes  future experiments that are
expected to advance this exciting field, 
both currently being built and being proposed, including 
SVOM (Section\,\ref{sect:svom}; contributed by B. Cordier, D. G\"otz, Y. Qiu \& J. Wei),
POLAR (Section\,\ref{sect:polar}; J.C. Sun \& N. Produit), 
WFIRST \& JWST (Section\,\ref{sect:jwst}; N. Gehrels \& J.K. Cannizzo),
TMT, GMT \& E-ELT (Section\,\ref{sect:30m}; N. T. Tanvir),
SKA (Section\,\ref{sect:ska}; G. Ghirlanda),
Einstein Probe (Section\,\ref{sect:ep}; W. Yuan \& C. Zhang), and 
THESEUS (Section\,\ref{sect:theseus}; L. Amati). 

\section{SVOM\label{sect:svom}}
\label{intro}
\begin{figure*}
  \includegraphics[width=1.0\textwidth]{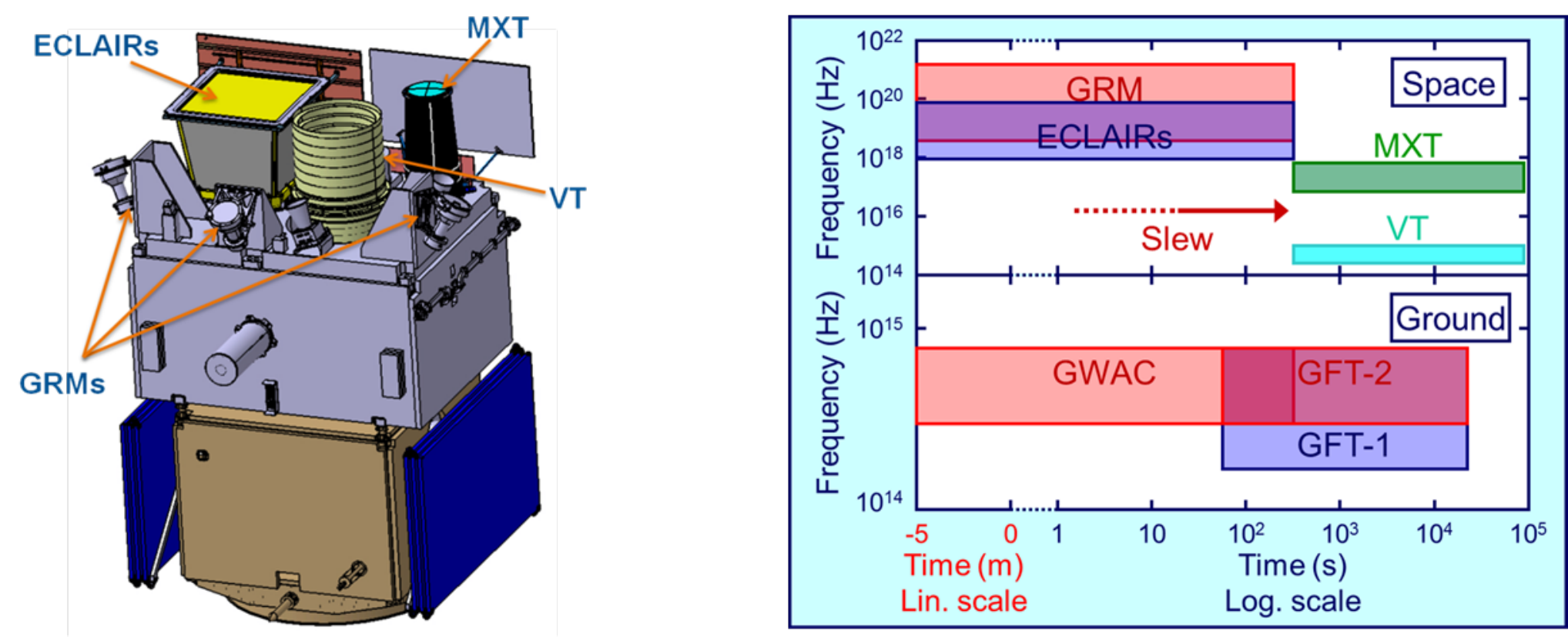}
\caption{Schematic showing the SVOM spacecraft with its multi-wavelength space payload. It consists of two wide-field instruments: ECLAIRs and the Gamma-Ray Monitor (GRM) for the observation of the GRB prompt emission and two narrow field instruments: the Micro channel X-ray Telescope (MXT) and the Visible Telescope (VT) for the observation of the GRB afterglow emission. Right: Space and ground instruments join to enable a unique coverage in time and wavelength.}
\label{fig:svom}       % Give a unique label
\end{figure*}

The success of the Swift mission for catching GRBs and other transients illustrates the benefit of its unique combination of space agility, fast communication with the ground, multi-wavelength observing capability, and long lifetime. These capabilities have permitted the detection of nearly 1000 GRBs and the measure of nearly 300 redshifts, have offered a new look at GRB progenitors and have led to several important discoveries, like the existence of GRBs at the epoch of the re-ionization of the universe.
The aim of SVOM (Space-based multi-band astronomical Variable Objects Monitor) is to continue the exploration of the transient universe with a set of space-based multi-wavelength instruments, following the way opened by Swift. SVOM is a space mission developed in cooperation between the Chinese National Space Agency (CNSA), the Chinese Academy of Science (CAS) and the French Space Agency (CNES). The mission features a medium size satellite, a set of space and ground instruments designed to detect, locate and follow-up GRBs of all kinds, a anti-Sun pointing strategy allowing the immediate follow-up of SVOM GRBs with ground based telescopes, and a fast data transmission to the ground. The satellite (see Figure \ref{fig:svom}) carries two wide field high energy instruments: a coded-mask gamma-ray imager called ECLAIRs, and a gamma-ray spectrometer called GRM, and two narrow field telescopes that can measure the evolution of the afterglow after a slew of the satellite: an X-ray telescope called MXT and an optical telescope called VT. The ground segment includes additional instrumentation: a wide angle optical camera (GWAC) monitoring the field of view of ECLAIRs in real time during part of the orbit, and two 1-meter class robotic follow-up telescopes (the GFTs). SVOM has some unique features: an energy threshold of ECLAIRs at 4 keV enabling the detection of faint soft GRBs (e.g. XRFs and high-redshift GRBs); a good match in sensitivity between the X-ray and optical space telescopes which permits the detection of most GRB afterglows with both telescopes; and a set of optical instruments on the ground dedicated to the mission. The mission has recently been confirmed by the Chinese and French space agencies for a launch in 2021, and it has entered in an active phase of development. We present an overview of the scientific objectives of SVOM in the next sub-section, and a brief description of the instrumental aspects of the mission in the following ones.

\subsection{Scientific objectives}
\label{sec:sc_obj}
The main science goal of SVOM is the study of cosmic transients detected in hard X-rays and in the optical domain. While the mission has been designed for the study of GRBs, it is also well suited for the study of other types of high-energy transients like Tidal Disruption Events (TDEs), Active Galactic Nuclei (AGNs), or galactic X-ray binaries and magnetars. For this type of sources, SVOM is both a ''discovery machine'', with wide field instruments that survey a significant fraction of the sky (ECLAIRs, GRM and GWAC), and a ''follow-up machine'', with fast pointing telescopes in space and on the ground (MXT, VT, and GFTs) that provide a multi-wavelength follow-up of different kinds of sources, with good sensitivity and a high duty cycle. The follow-up can be triggered by the satellite itself or from the ground, upon reception of a request for target of opportunity observations (ToO). %The science goals of SVOM are presented in the document "SVOM: a space mission to probe fundamental physics and to explore the young universe".
In this section we concentrate on the objectives of SVOM for GRBs. 
Through GRB population simulations we have evaluated a GRB detection rate of 70--80 GRBs/yr for ECLAIRs and ≥ 90 GRBs/yr for GRM. One essential goal of SVOM is to provide GRBs with a redshift measurement. The redshift is required to measure the energetics of the burst and the epoch at which the GRB occurred in the history of the universe. Four elements in the design of the mission concur to facilitate the measure of the redshift for SVOM GRBs: a near anti-solar pointing during most of the orbit ensuring that SVOM GRBs can be quickly observed with ground based telescopes, a good sensitivity of the on-board optical telescope permitting the rapid identification of high-z candidates (which are not detected at visible wavelengths), Near Infra-Red follow-up on the ground to look for the afterglows of dark/high redshift GRBs, and agreements with the community to promote the optical spectroscopy of SVOM GRBs using large telescopes. With this strategy we expect to measure the redshift of more than 50\% of SVOM GRBs, constructing a sample that could potentially be more representative of the intrinsic GRB population than the Swift sample. One drawback of this pointing strategy is that it avoids the galactic plane. Galactic sources can nevertheless be observed during target of opportunity pointings. In the following, we discuss a few selected topics about GRBs for which SVOM may bring decisive progress. 

\subsubsection{GRB Progenitors}
\label{sec:progenitors}
In order to get a better understanding of the GRB phenomenon SVOM was designed to detect all kinds of GRBs, and to provide extensive multi-wavelength observations of the prompt GRB and its afterglow. In addition to classical long GRBs that will be detected by the two wide field instruments (ECLAIRs and GRM), SVOM will routinely detect short bursts with its gamma-ray monitor and soft GRBs (XRFs or highly redshifted GRBs) with ECLAIRs. The ability to detect soft GRBs with a spectral energy distribution (SED) peaking below 20-30 keV will favour the detection of X-ray flashes. As shown by BeppoSAX and HETE-2 \citep{heise,sakamoto}, these faint GRBs can only be detected if they are close enough and if their spectral energy distribution peaks at low energies (because faint GRBs with a SED peaking at high energies radiate too few photons to be detected). The detection of X-Ray Flashes in the local universe (z $<$ 0.1) will permit detailed studies of their associated supernovae (as was the case for XRF 020903; \citealt{soderberg}), providing crucial clues to understand the broader context of the SN-GRB connection.

\subsubsection{GRB Physics}
The instrument suite of SVOM will provide good multi-wavelength coverage of GRBs. For those occurring in the field of view of GWAC, the prompt emission will be measured from 1 eV to 5 MeV, with the GWACs, ECLAIRs, and the GRM. The prompt optical emission may also be observed by one GFT for GRBs lasting longer than 40 seconds. GRB afterglows will be observed with the two narrow field instruments on-board the satellite and with the ground follow-up telescopes on Earth. For some GRBs, SVOM will provide a very complete view of the phenomenon and its evolution, hopefully bringing new insight into the complex physics at work in these events. A lesson learned from Swift is that a few well observed GRBs may crucially improve our understanding of GRB physics, as was the case for GRB 130427, a bright nearby burst detected by Swift, Fermi and various optical and radio telescopes on the ground \citep{Maselli:2014db,Perley:2014hl}.
The physical processes at work within the jet remain not well understood even after the observation of hundreds of GRBs. Comprehensive discussions of the theoretical challenges connected with the understanding of the prompt GRB emission can be found in \citet{kumar} and \citet{zhang}. The authors show that several crucial questions connected to the physics of the ultra-relativistic jet and its interaction with the surrounding medium remain unanswered, like the nature and content of the jet (is the energy stored in the baryons or in magnetic fields?), the mechanisms of particle acceleration, the micro physics and the dominant radiation processes, the importance of the reverse shock, the role of pairs, etc. Performing multi-wavelength observations during the prompt emission and the early afterglow, SVOM will provide key observations to understand the physics of relativistic jets. GRBs detected with SVOM will also benefit from contemporaneous or follow-up observations with a novel generation of powerful instruments, like CTA (the Cerenkov Telescope Array) for very high energy photons, LSST (Large Synoptic Survey Telescope) for optical transients associated with on-axis and off-axis GRBs, and the precursors of SKA (the Square Kilometer Array) in radio.

\subsubsection{Cosmology}
%GRBs are like “fireworks” in the distant universe. 
Their extreme luminosity of GRBs permits their detection in hard X-rays up to very high redshifts, possibly beyond z = 10 and the spectroscopy of the optical afterglow provides the redshift of the burst and a tomographic vision of the line of sight to the burst. With Swift and the measure of about 300 redshifts, GRBs are providing new diagnostics of the distant Universe. When the signal to noise ratio (SNR) of the optical spectrum of the afterglow is sufficient, we get detailed information on the circumburst medium, on the gas and dust in the host galaxy, on the intergalactic medium and the intervening systems. With a smaller SNR, the measure of the redshift allows locating the time of the explosion in the history of the universe and reconstructing the history of the GRB formation rate, which traces the formation rate of massive stars.
There is of course a special interest in very distant GRBs (z $>$ 5), which provide a unique view on the young Universe, especially since they occur in galaxies which are undetectable with other methods of observation. One exciting challenge of GRB missions is the detection of GRBs resulting from the explosion of population III stars (the first generation of stars formed with pristine gas containing no metals). Such events are expected to be rare, to occur at high redshifts, to have no detectable hosts (except in absorption in the spectrum of their optical afterglow) and afterglows that are only detectable in the near infrared. They could be similar to some very long GRBs found at lower redshift \citep{meszaros,gendre}. We expect to detect about 5 GRBs/yr at redshift z > 5 with ECLAIRs, but they will be useful only if we can measure their redshift. One difficulty is that the optical afterglows of GRBs fade very quickly, and after a few hours they are often too faint to permit measuring the redshift of the burst. In order to quickly identify high-z candidates that deserve deep spectroscopy in the NIR, we rely on the sensitivity of VT and on fast NIR follow-up telescopes on the ground. Dark GRBs, whose afterglows are not detected in the VT, are good candidates for high-z bursts, but they can also be extinct by dust in the vicinity of the source. The nature of these events (distant or extinct GRB) will be confirmed quickly with fast visible/NIR photometry from the ground, allowing the most appropriate spectroscopic follow-up. In some cases, GRBs detected by SVOM could trigger follow-up observations by the JWST (James Webb Space Telescope) for the study of the hosts.

%\subsubsection{Gravitational Wave Sources}
%SVOM will also contribute to clarify the origin of short GRBs, especially with the possibility to search for GRBs in coincidence with the signals detected by advanced gravitational waves (GW) detectors. The favorite scenario for the production of short GRBs is the coalescence of two compact objects (two neutron stars or a neutron star plus a black hole), which predicts that short GRBs are accompanied by powerful bursts of gravitational waves. One complication of these searches is that GRBs are much more beamed than gravitational waves. Considering that we detect 1 short GRB out of $\sim$50 mergers [28; 29], and a moderate enhancement of the gravitational wave emission along the jet, we expect that about 10\% of the mergers detected by advanced detectors of gravitational waves could be associated with a GRB. Assuming a rate of binary mergers of $\sim$50/yr within the horizon of GW detectors ($\sim$400 Mpc), we expect to detect with ECLAIRs $\sim$3 events coincident with GW triggers, and $\sim$9 with GRM, in 5 years of operation. SVOM will also have the capability to point its narrow field instruments towards candidate sources of GWs. We evaluate that 15 events approximately can be followed quickly (6 hours) with SVOM narrow field instruments in 5 years of operation.

\subsection{Space Borne Instruments}
In this section and the following, we give a brief description of the instruments of SVOM. In order to be both a discovery machine and a follow-up machine, SVOM features two types of instruments: space based instruments and ground based instruments. We briefly describe these instruments below. %A more complete description of the French payload can be found in [18].

\subsubsection{ECLAIRs the hard X-ray coded mask imaging camera}
\begin{figure*}
  \includegraphics[width=1.0\textwidth]{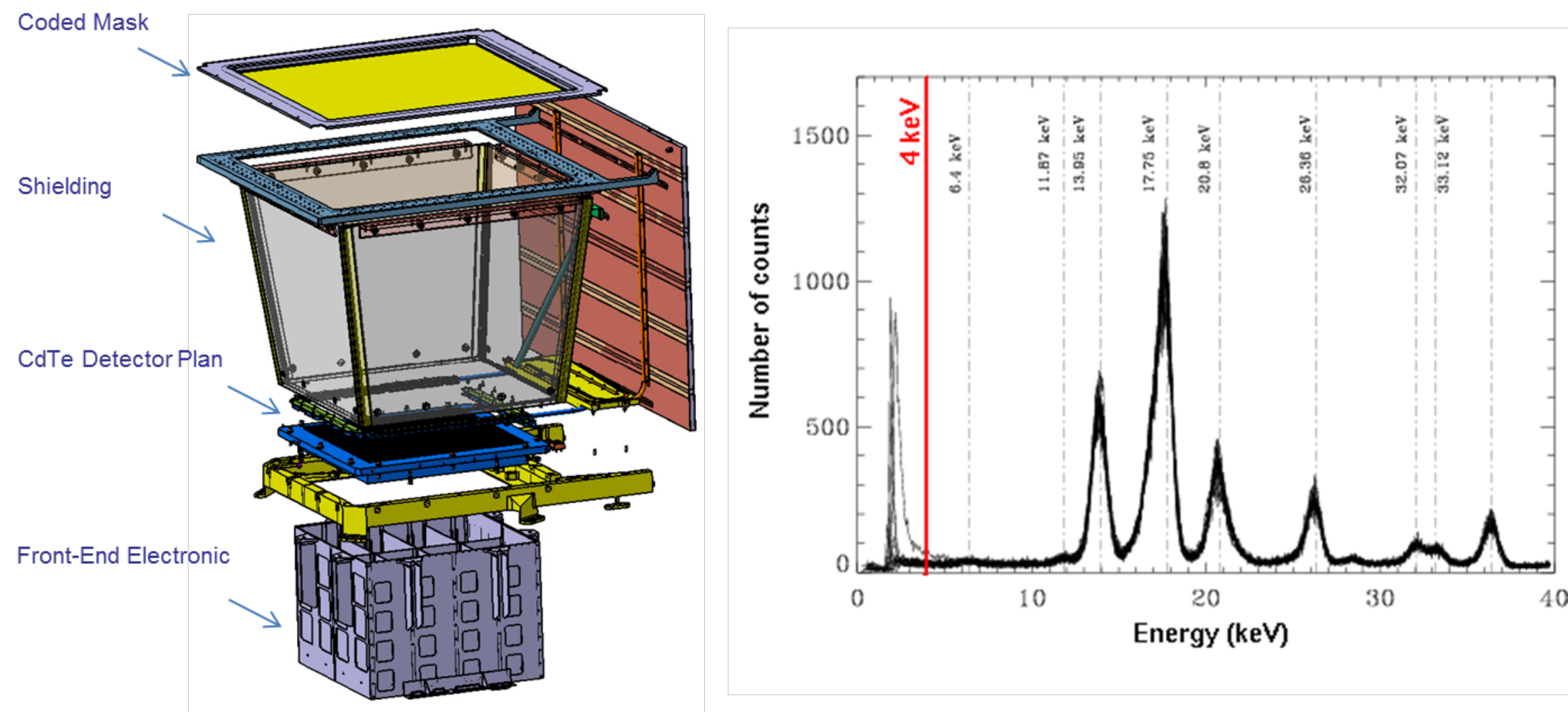}
\caption{Schematics showing the different sub-systems of ECLAIRs except the data processing unit in charge of the GRB detection and localisation. Right: Laboratory spectral measurements performed on a detector module prototype (a 32 CdTe pixel matrix) with radioactive sources (241Am). The red vertical line corresponds to the expected low energy threshold of the ECLAIRs camera [6].}
\label{fig:eclairs}       % Give a unique label
\end{figure*}

ECLAIRs (see Figure \ref{fig:eclairs}) is the instrument on-board the satellite that will detect and locate the GRBs. ECLAIRs is made of four parts: a pixellated detection plane (1024 cm$^{2}$) with its readout electronics, a coded mask, a shield defining a field of view of 2 steradians (89$^{\circ}$ $\times$ 89$^{\circ}$), and a processing unit in charge of detecting and locating transient sources. The detection plane is made of 200 modules of 32 CdTe detectors each, for a total of 6400 detectors of size 4$\times$4$\times$1 mm. Each module is read by a customized ASIC connected to an electronics that encodes the position, the time and the energy of each photon. One of the requirements of ECLAIRs is to reach an energy threshold of 4 keV, in order to study soft GRBs like X-Ray Flashes and highly redshifted GRBs. As shown in figure 2, the first modules satisfy the low energy threshold requirement \citep{godet}.
 The coded mask is a square of side 54 cm located at a distance of 46 cm from the detection plane; it has an opening fraction of 40\% and provides a localization accuracy of several arc minutes ($\sim$14$^{\prime}$ for a source at the limit of detection). The instrument features a count rate trigger and an image trigger, like Swift. These triggers are computed, from the photon data, in several energy bands and on time-scales ranging from 10 ms to several minutes. Our simulations show that ECLAIRs will detect 70-80 GRBs/yr. 
 %For more details on the trigger process, see Antier et al. (these proceedings) and for a complete description of ECLAIRs see Schanne et al. (these proceedings).

\subsubsection{GRM: the Gamma-ray Monitor}
GRM consists of a set of three detection modules. Each of them is made of a scintillating crystal (sodium iodide), a photomultiplier and its readout electronics. Each detector has a surface area of 200 cm$^{2}$ and a thickness of 1.5 cm. One piece of plastic scintillator in front of NaI(Tl) is used to distinguish low energy electrons from normal X-rays. The three modules are pointed at different directions to form a total field of view of 2 sr, within which rough (of the order of 10 degrees radius) localization of transient sources can be achieved on-board. The energy range of the GRM is 15-5000 keV, extending the energy range of ECLAIRs towards high energies to measure E$_{Peak}$ for a large fraction of SVOM GRBs. We expect that GRM will detect $>$ 90 GRBs/yr. GRM will have a good sensitivity to short/hard GRBs, like the GBM of Fermi. GRM can generate on-board GRM-only triggers, taking use of only GRM detectors. Such triggers with localization information will be transferred to ECLAIRs for trigger enhancement on the short GRBs, and to ground facilities (e.g. GWAC, GW experiments) for joint observations. 
A calibration detector containing one radioactive $^{241}$Am isotope is installed on the edge of each detection module, for the purpose of gain monitoring and energy calibration. In addition, a particle monitor auxiliary to GRM can generate South Atlantic Anomaly alerts and help protecting the detection modules.

\subsubsection{MXT: the Microchannel X-ray Telescope}
\begin{figure*}
  \includegraphics[width=1.0\textwidth]{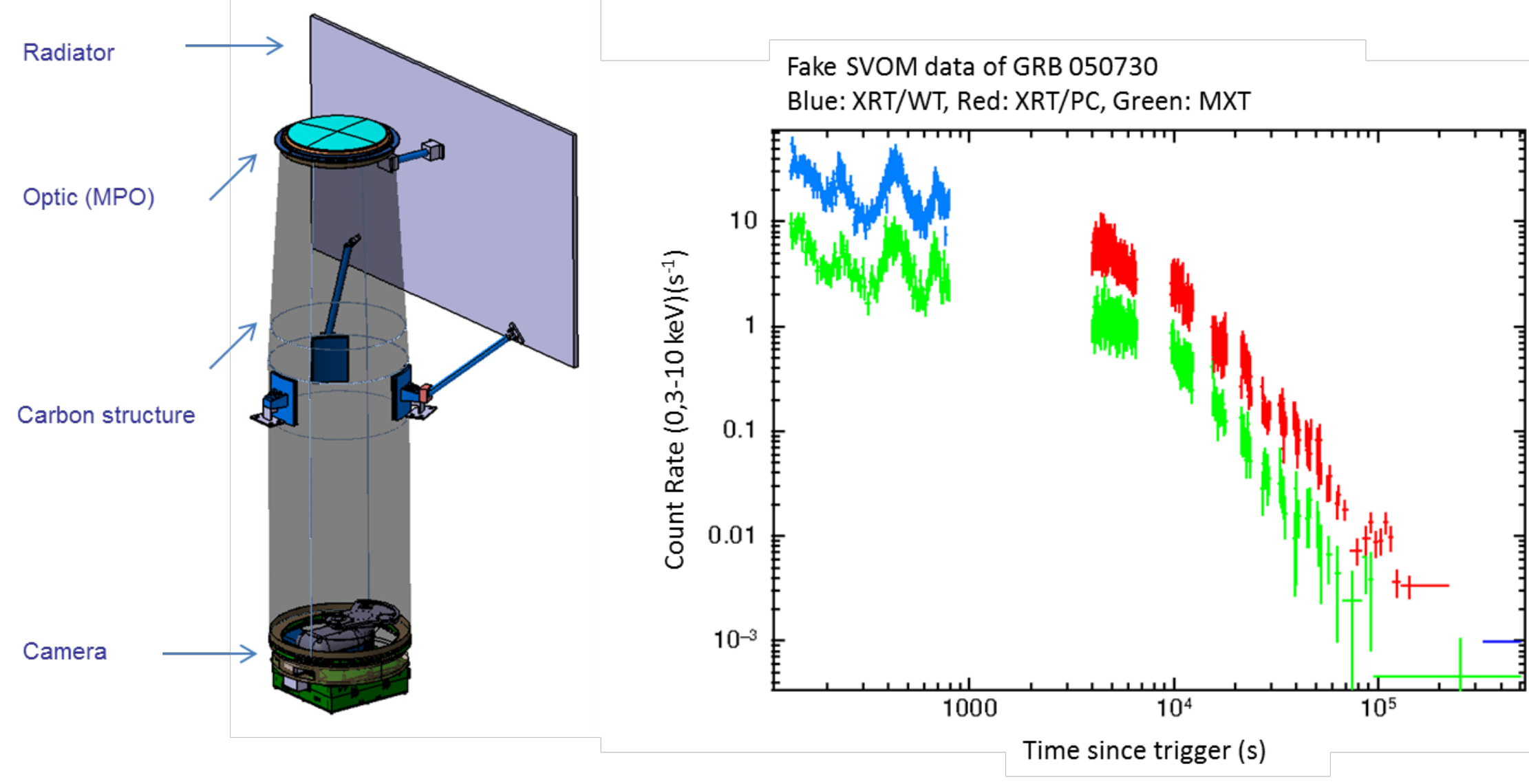}
\caption{Schematics showing the different sub-systems of the MXT.  Right: XRT measured (blue and red points) versus MXT simulated (green points) light curve of GRB 050730, a GRB with a median flux in the Swift/XRT afterglow database.}
\label{fig:mxt}       % Give a unique label
\end{figure*}

MXT (see Figure \ref{fig:mxt}) is a light and compact focusing X-ray telescope designed to observe and measure the properties of the GRB X-ray afterglow after a slew of the satellite. The telescope will implement a novel technique to focus X-rays, based on micropore optics arranged in a lobster eye geometry. The use of micropore optics instead of classical X-ray electro-formed mirrors permits a significant reduction of the size and weight of the telescope, fitting on a medium size satellite like SVOM. The optics has a diameter of 24 cm and a focal length of 1 meter \citep{mxt}. 
MXT will make use of the radiation hard pn-CCD detector developed for the DUO mission and adopted in a larger version for the eROSITA mission \citep{meidinger,predehl}. The MXT detector is made of 256 x 256 Si pixels of 75 $\mu$m side and has an expected energy resolution of 75 eV at 1 keV.
MXT will be operated in the energy range of 0.2-10 keV, will have an effective area of 45 cm$^{2}$ at 1 keV, and a field of view of 64$\times$64 arc minutes. Despite the smaller effective area with respect to XRT on board Swift at 1 keV (about 120 cm$^{2}$) MXT will be able to fully characterize the GRBs light curves (as shown in Figure \ref{fig:mxt}). In fact with a sensitivity of 7$\times$10$^{-13}$ erg cm$^{-2}$ s$^{-1}$ in 10$^{4}$ s, MXT will detect the afterglows of more than 90\% of SVOM GRBs. Indeed its sensitivity is well adapted to early GRB afterglow observations, and its PSF of about 4 arc min will allow to significantly reduce the ECLAIRs error boxes. Simulations based on the Swift-XRT database show that the localization accuracy for the MXT is $\sim$13'' for 50\% of the bursts, 10 minutes after the trigger (statistical uncertainty only). 

\subsubsection{VT: the visible Telescope}

\begin{figure*}
  \includegraphics[width=1.0\textwidth]{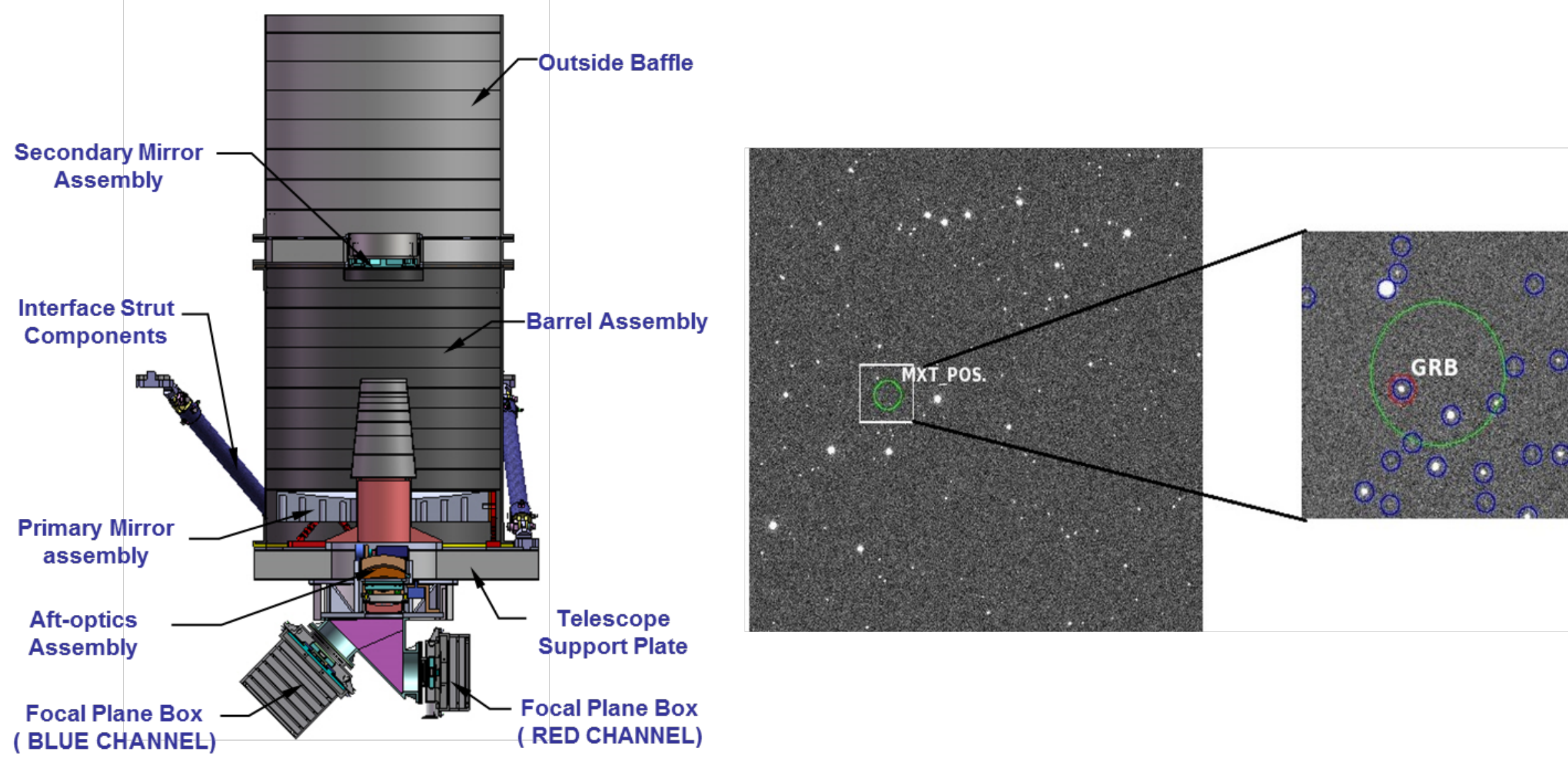}
\caption{Schematic showing the different sub-systems of the VT. Right: a simulated image of the VT and its associated finding chart.}
\label{fig:vt}       % Give a unique label
\end{figure*}

The Visible Telescope (VT) is a dedicated optical follow-up telescope on board the SVOM satellite. Its main purpose is to detect and observe the optical afterglows of gamma-ray bursts localized by ECLAIRs. It is a Ritchey-Chretien telescope (see Figure \ref{fig:vt}) with a diameter of 40 cm and an f-ratio of 9. Its limiting magnitude is about 22.5 (M$_{V}$) for an integration time of 300 seconds
VT is designed to maximize the detection efficiency of GRB's optical afterglows. Instead of a filter wheel, a dichroic beam splitter is used to divide the light into two channels, in which the GRB afterglow can be observed simultaneously. Their wavelength ranges are from 0.4$\mu$m to 0.65$\mu$m (blue channel) and from 0.65$\mu$m to 1$\mu$m (red channel). Each channel is equipped with a 2k $\times$ 2k CCD detector. While the CCD for the blue channel is a normal thinned back-illuminated one, a deep-depleted one is adopted for the red channel to obtain high sensitivity at long wavelength. The Quantum Efficiency (QE) of the red-channel CCD at 0.9$\mu$m is over 50\%, which enables VT to have the capability of detecting GRBs with the redshift larger than 6.5. The field of view of VT is about 26$^{\prime}$ $\times$ 26$^{\prime}$, which can cover the error box of ECLAIRs in most cases. Both CCDs have a pixel size of 13.5$\mu$m $\times$ 13.5$\mu$m, corresponding to spatial resolutions of 0.77 arc second. This ensures the GRB positioning accuracy to be greatly improved by VT from several arc minutes (ECLAIRs) and tens of arc-seconds (MXT) to a level of sub-arc second.

In order to promptly provide the GRB alerts with the sub-arc second accuracy, VT will do some data processing on board. After a GRB has been localized by the co-aligned MXT, lists of sources are extracted from the VT sub-images whose centres and sizes are determined by the GRB positions and the corresponding error boxes provided by MXT. The lists are immediately downlinked to the ground through the VHF network. Then, the ground software will make finding charts with these lists (see Figure \ref{fig:vt}) and search the optical counterparts of the GRB by comparing the lists with the existing catalogues. If a counterpart is identified, an alert will be then produced and distributed to world-wide astronomical community, which is useful for triggering large ground-based telescopes to measure the redshifts of the GRBs by spectroscopy.
VT is expected to do a good job on detecting high-redshift GRBs. The confirmed high-redshift GRBs are rare in the Swift era, in contrast to a theoretical prediction of a fraction of more than 10\% \citep{salvaterra15}. This is probably due to the fact that for most Swift GRBs the early-time optical imaging follow-up is not deep enough for a quick identification and some faint GRBs cannot be spectroscopically observed in time by the large ground-based telescopes. This passive situation will be significantly improved by SVOM, due to the high sensitivity of VT, in particular at long wavelength, and the prompt optical-counterpart alerts. Additionally, the anti-solar pointing strategy of SVOM allows GRBs to be spectroscopically observed by large ground-based telescopes at the early time of the bursts. Consequently, more high-redshift GRBs are expected to be identified in the SVOM era.
VT is also used to support the platform to achieve the required high pointing stability. A Fine Guidance Sensor (FGS) is mounted on the VT focal plane to measure relative image motions. Its images are processed in real time by a specialized data processing unit to get the centroid positions of several stars brighter than the magnitude of 15 (M$_{V}$). The results are sent at a frequency of 1Hz to the platform to improve the pointing stability, which enables VT to have a good performance in a long exposure time.

\subsection{Ground Based Instruments}
The ground follow-up instruments constitute an important part of the mission. Three instruments are developed for the follow-up of SVOM GRBs: a wide angle camera that surveys a significant fraction of the sky for transients, and two robotic telescopes. In addition to these dedicated instruments, the SVOM collaboration will seek agreements with various existing telescopes or networks willing to contribute to the follow-up of SVOM GRBs.

\subsubsection{GWAC: the Ground Wide Angle Camera}

\begin{figure*}
  \includegraphics[width=1.0\textwidth]{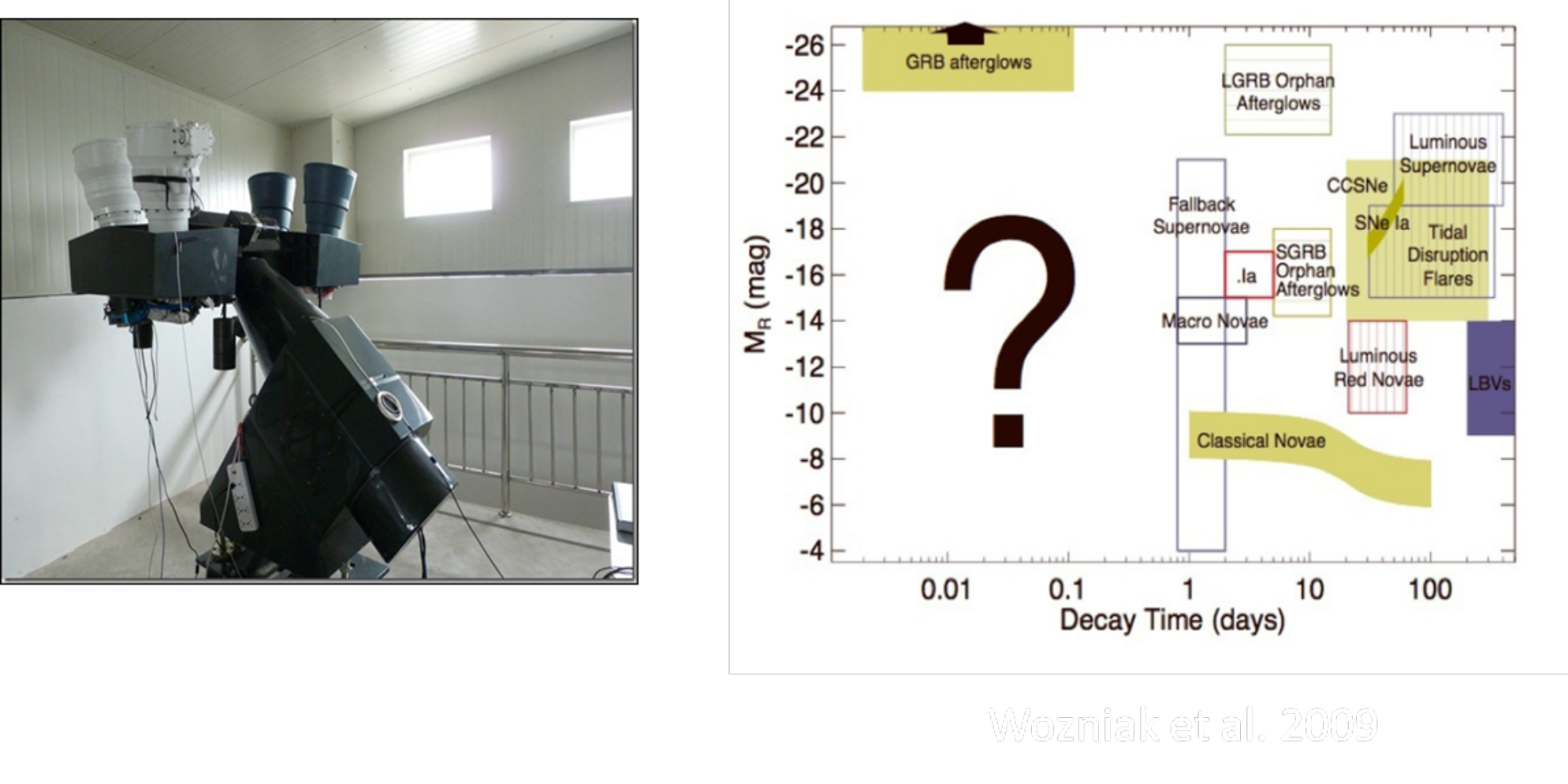}
\caption{First prototype of one GWAC module. The final system will be composed by 9 modules. Right: the figure shows the discovery space of an instrument dedicated to short time-scale optical transients \citep{wozniak}.}
\label{fig:gwacs}       % Give a unique label
\end{figure*}

GWAC (see Figure \ref{fig:gwacs}) provides a unique way to survey a large field of view for optical transients. The instrument will monitor 63\% of ECLAIRs field of view, looking for optical transients occurring before, during and after GRBs. GWAC will also have its own trigger system, providing alerts to the world. GWAC is a complex system: the heart of the system is a set of 36 wide angle cameras with a diameter of 18 cm and a focal length of 22 cm, together these cameras cover a field of view of 5000 sq. deg. They use 4k $\times$ 4k CCD detectors, sensitive in the range of wavelength 500-800 nm. These cameras reach a limiting magnitude V=16 (5 $\sigma$) in a typical 10 second exposure. This set of cameras is completed by two 60 cm robotic telescopes. equipped with EMCCD cameras. These telescopes will provide multicolour photometry of the transients discovered by GWAC with a temporal resolution $\leq$ 1 second.

\subsubsection{GFTs: the Ground Follow-up Telescopes}

The ground follow-up telescopes have two main goals. Firstly, they measure the photometric evolution of the optical afterglow in the first minutes after the trigger in a broad range of visible and NIR wavelengths, with a temporal resolution of few seconds. Secondly, when an afterglow is detected, they provide its position with arc second precision within 5 minutes of the trigger. Some essential features of the GFTs are their field of view, their size, and their sensitivity in the near infrared. The field of view ($\sim$30 arc minutes) enables observing quickly the entire error boxes of ECLAIRs. The size, typically 1 meter, allows the detection of all visible (i.e. non-dark) afterglows at the condition to arrive within few minutes after the trigger \citep{akerlof,klotz}. Finally, the near infrared sensitivity permits the detection of high-z GRBs and GRBs extinct by dust, whose afterglow are obscured in the visible domain \citep{greiner}. GFTs are especially useful for the study of the early afterglow during the slew of the satellite, and for the rapid identification of the optical afterglow in various cases: when SVOM cannot slew to the burst or when the slew is delayed due to pointing constraints, and when the optical afterglow is only visible in the NIR.
One telescope is located in China at Xinglong Observatory and the other one will be located in Mexico at San Pedro M\'artir.

\subsection{The System}

In order to facilitate measuring the redshifts of GRBs detected with ECLAIRs, the instruments of SVOM will be pointed close to the anti-solar direction. Most of the year the optical axis of the SVOM instruments will be pointed at about 45$^{\circ}$ from the anti-solar direction. This pointing is interleaved with avoidance periods during which the satellite passes away from the Sco X-1 source and the galactic plane. This strategy ensures that SVOM GRBs will be in the night hemisphere and quickly observable from the ground by large telescopes. More details on the SVOM pointing strategy, can be found in \citet{cordier}.

As soon as a GRB will be located, its coordinates and its main characteristics will be sent to the ground within seconds with a VHF antenna. The VHF signal will be received by one of the $\sim$ 40 ground stations distributed around the Earth below the orbit. The data will then be relayed to the Operation Center, which will send SVOM alerts to the internet via the GCN and VO Event networks (http://gcn.gsfc.nasa.gov/), and to the ground instruments GWAC and the GFTs. SVOM can also perform target of opportunity observations (with MXT and VT for instance), with a delay of few hours, which depends on the availability of uplink communication with the satellite.

SVOM will try to select high-z GRB candidates by analysing MXT and VT data, and multi-band photometry data from GFTs. If a GRB is detected by MXT in the soft X-rays, but not by VT in the optical band, it will be selected as a candidate of high-z or an optically dark GRB. Ground telescopes with NIR capabilities are encouraged to follow up these candidates as soon as possible to try to measure their redshift. GFTs will be able to measure photometric redshifts by observing GRB afterglows with multiple filters from the visible to the NIR bands.

\subsubsection{Conclusion}

SVOM will be a highly versatile astronomy satellite, with built-in multi-wavelength capabilities, autonomous re-pointing and dedicated ground follow-up. With its peculiar pointing strategy and the low energy detection threshold, SVOM is expected to improve the number of GRBs detected at high redshift, and hence to contribute to the use of GRBs as probes of the young Universe.

Beyond the GRB studies emphasized here, SVOM will bring new observations about all types of high energy transients, in particular those of extragalactic origin (TDEs, AGNs, etc.).

\section{POLAR - Space-Borne Gamma-Ray Bursts Compton Polarimeter}
\label{sect:polar} 

%\subsection{Introduction}
Until recent years, only a few polarization
measurement results on Gamma-Ray Burst (GRB) prompt emissions have
been published \citep{Ref1}, while in the same time more and more
hard X-ray/Gamma-ray polarization measurement instruments have been
put into operation or under development. All these efforts will
accumulate a larger observation database on GRB's polarization.\\
\newline
POLAR \citep{Ref2} is a novel space-borne Compton polarimeter
dedicated for the precise measurement of the polarization of GRB's
prompt emission, which is expected to be on-board the Chinese space laboratory ``Tiangong-2 (TG-2)'' with a scheduled launch date in late 2016. The future detailed measurement of the
polarization of GRBs will lead to a better understanding of its
radiation region geometry and emission mechanisms. POLAR will
contribute greatly to two aspects on the polarization observation of
GRBs. One is its high sensitivity which is minimum detectable
polarization (MDP). The MDP of POLAR can be as low as
$\sim$10\% \citep{Ref3}. The other one is its high observation
statistics. $\sim$50 GRBs/year are expected to be detected by POLAR.
This will provide significant results for restricting the radiation
mechanism of GRBs.

\subsection{Design of POLAR detector}
\label{sec:1} 
\begin{figure}[b]
  \begin{center}
  \includegraphics[width=0.85\textwidth]{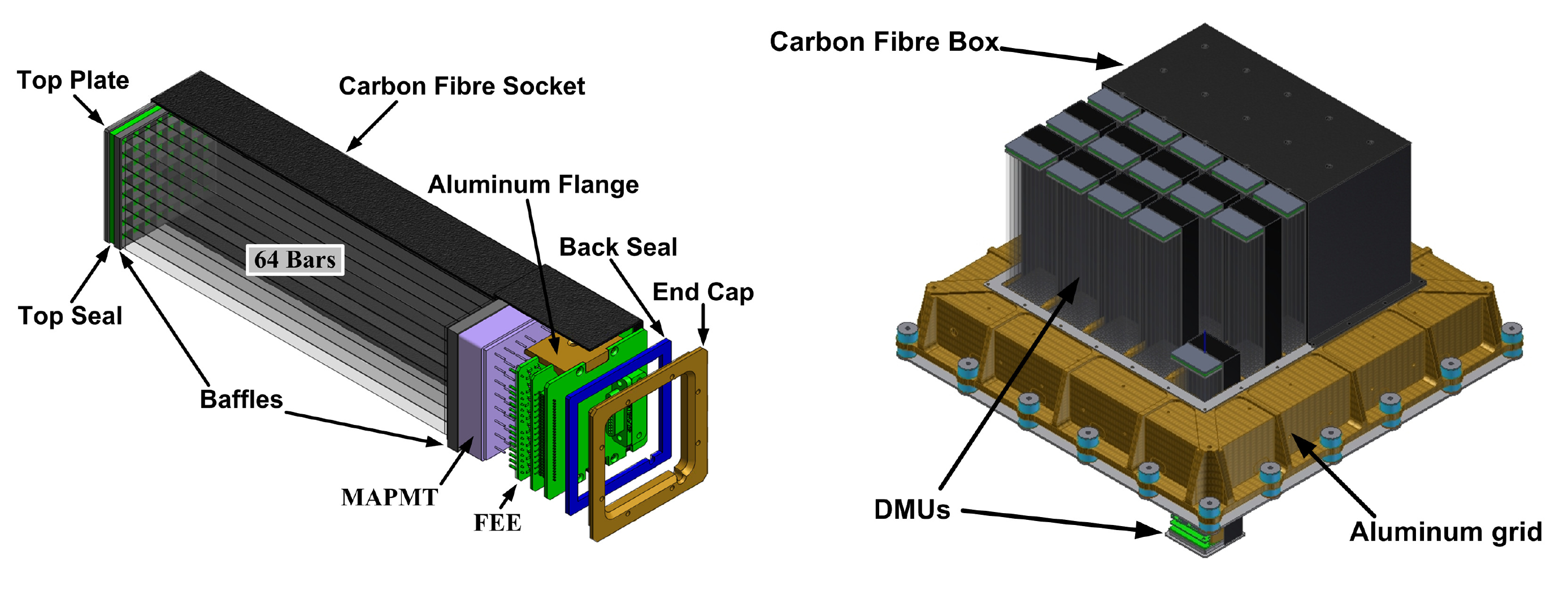}
  \caption{Detector design of POLAR. Left: Detector modular unit; Right: polarization detector OBOX}
  \label{fig:polar1}
  \end{center}
\end{figure}

POLAR is composed of the polarization detector (OBOX)
and electronic cabinet (IBOX). OBOX consists of 25 detector modular
units (DMU). Each DMU is composed of 64 low-Z material plastic
scintillator (EJ-248M) bars, read-out by a flat-panel multi-anode
photomultipliers H8500 and ASIC front-end electronics. The CAD view
of DMU and OBOX are shown in Figure\,\ref{fig:polar1}\\

\noindent The linear polarization degree and polarization direction
of the polarized Gamma-rays can be reconstructed by POLAR through
Compton scattering effect. In order to improve the Compton
scattering efficiency, the low-Z material plastic scintillator has
been selected as the detection material of POLAR as it is more
stable chemically and mechanically.

\subsection{Detection principle}
\label{sec:2} 
\begin{figure}[h!]
 \begin{center}
  \includegraphics[width=0.45\textwidth]{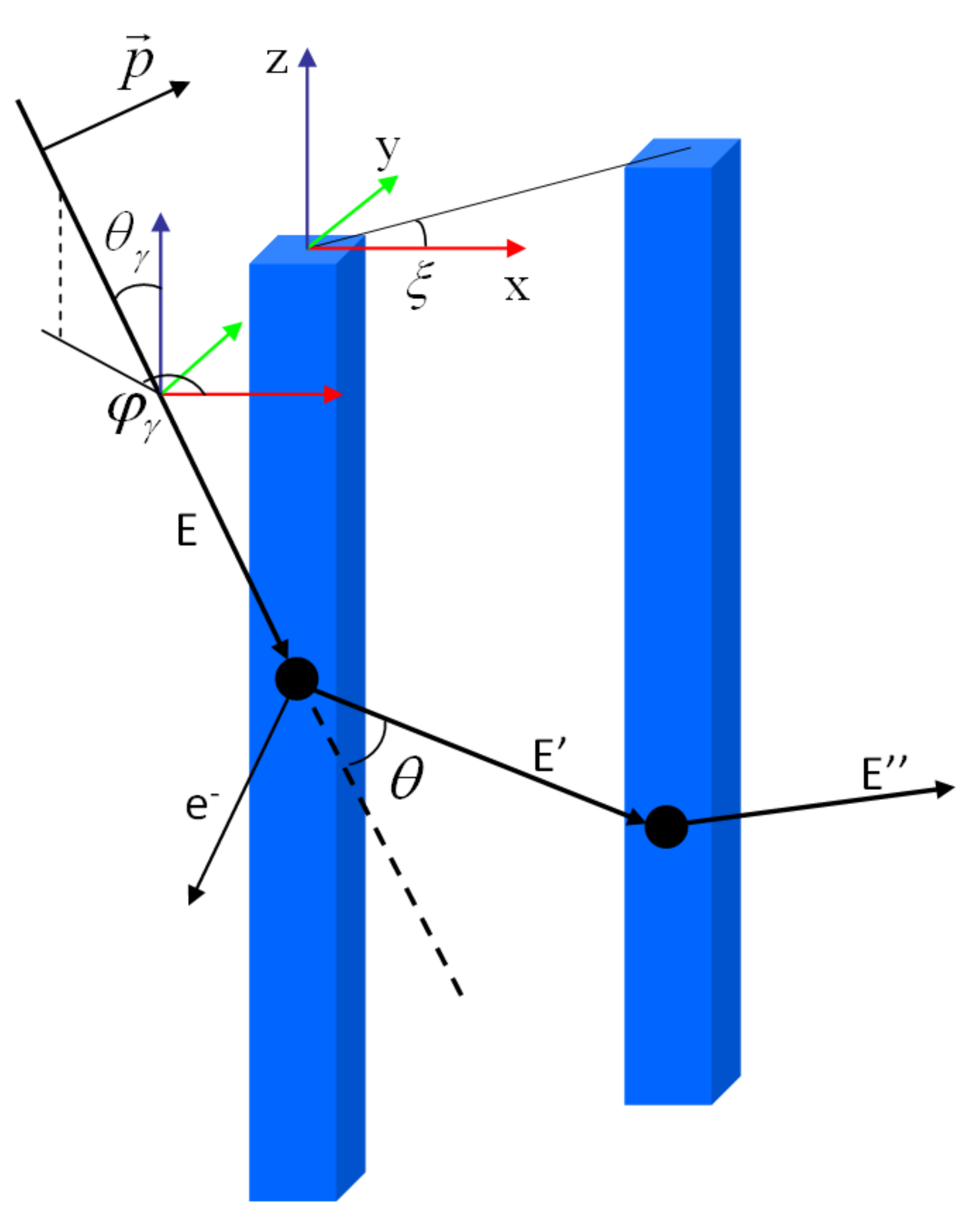}
  \caption{Detection principle of POLAR by Compton scattering effect}
  \label{fig:polar2}
  \end{center}
\end{figure}

The main interaction between the hard
X-rays/Gamma-rays within the energy range 50 keV$\sim$500 keV and
plastic scintillator material is Compton scattering. The
differential cross-section is given by the Klein-Nishina formula for
the polarized Compton scattering interaction process:

\begin{eqnarray}
\label{eq1.10} \frac{d{\sigma_p}}{d{\Omega}}&=&\frac{r_0^2}{2}{\varepsilon}^2[\varepsilon+{\varepsilon}^{-1}-2\sin^2\theta\cos^2\eta] %
\nonumber\\[1mm] %
&=&\frac{r_0^2}{2}{\varepsilon}^2[\varepsilon+{\varepsilon}^{-1}-\sin^2\theta+\sin^2\theta\cos(2(\eta+\frac{\pi}{2}))], %
\end{eqnarray}
\\
\newline
where $r_0$ is the classic radius of an electron and
$\varepsilon$=E'/E is the ratio of the energy of the scattered
photon and the energy of the incident photon, $\theta$ is the
Compton scattering angle and
$\eta$ is the scattering azimuth angle.\\
\newline
An incident photon interacts with one of the 1600 PS bars in POLAR
through Compton scattering effect; the recoil electron will be
absorbed by the PS bar and its deposited energy will be detected by
POLAR and readout by the following electronics. The scattered photon
will interact with next PS bars until being absorbed or escaping the
detector array. For the unpolarized Gamma-rays, the scattering
azimuth angle is isotropic, while for the polarized Gamma-rays, the
scattering azimuth angle distribution is related to the incident
photons' polarization degree and polarization angle. Therefore, the
polarization property of the incident Gamma-rays can be
reconstructed by analyzing the distribution of the Compton
scattering azimuth angles. Figure\,\ref{fig:polar2} shows the detection principle of
POLAR.\\

\noindent In Figure\,\ref{fig:polar2}, $\vec{P}$ is the polarization vector of the
incident Gamma-rays, $\theta_{\gamma}$ and $\varphi_{\gamma}$
describe the incident photons' direction, $e^{-}$ is the recoil
electron, $\xi$ is the projection of the Compton scattering azimuth
angle in the detector plane.

\subsection{Physics performance study with Monte-Carlo method}
\label{sec:3} 
\begin{figure}[h]
  \begin{center}
  \includegraphics[width=1.00\textwidth]{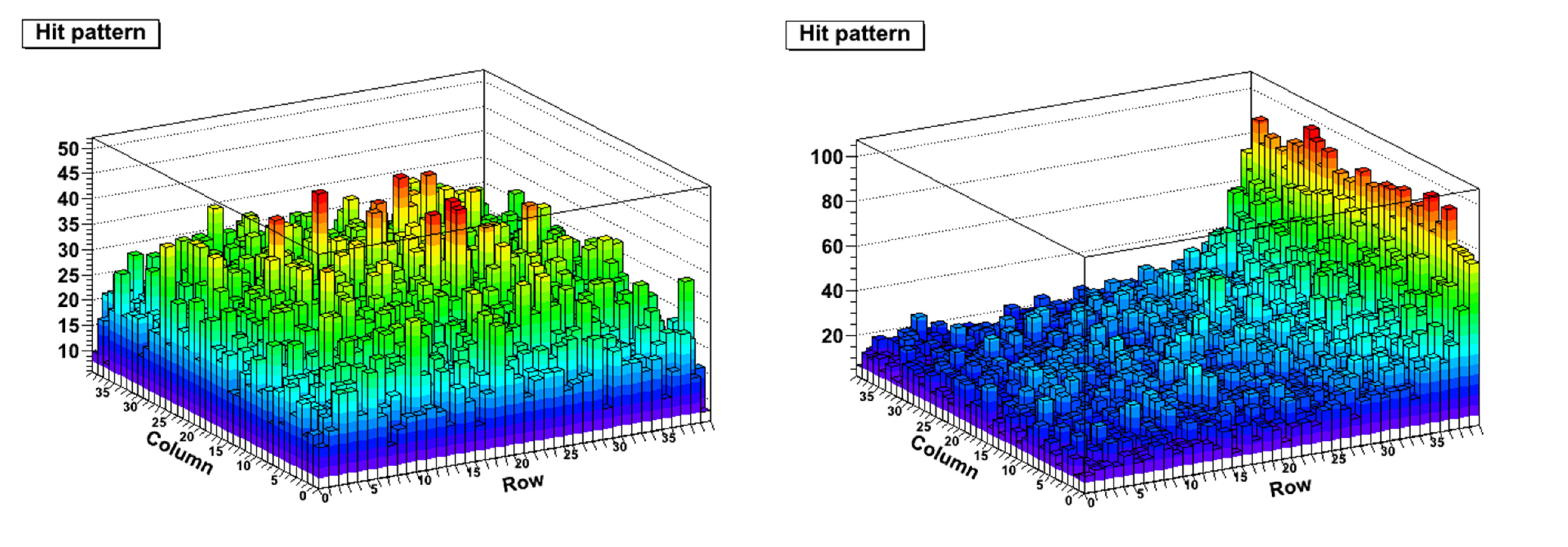}
  \caption{Triggering pattern of POLAR detector with standard incident GRB photons. Left: $\theta_{\gamma}$=$0^{\circ}$, $\varphi_{\gamma}$=$0^{\circ}$; Right: $\theta_{\gamma}$=$45^{\circ}$, $\varphi_{\gamma}$=$0^{\circ}$}
  \label{fig:polar3}
  \end{center}
\end{figure}

The performance of POLAR has been studied using the
Monte-Carlo simulation method. The GEANT4 toolkit is used for the
simulations. The Band \citep{Ref4} function is used to create the
spectrum of an GRB. Typically, the standard GRB spectrum
($\alpha$=-1.0, $\beta$=-2.5 and $E_{peak}$=200 keV) is used for the
study. Figure\,\ref{fig:polar3} shows the triggered events on the 1600 detection
channels of POLAR with different GRB incident angles.\\

\begin{figure}[b]
  \begin{center}
  \includegraphics[width=0.75\textwidth]{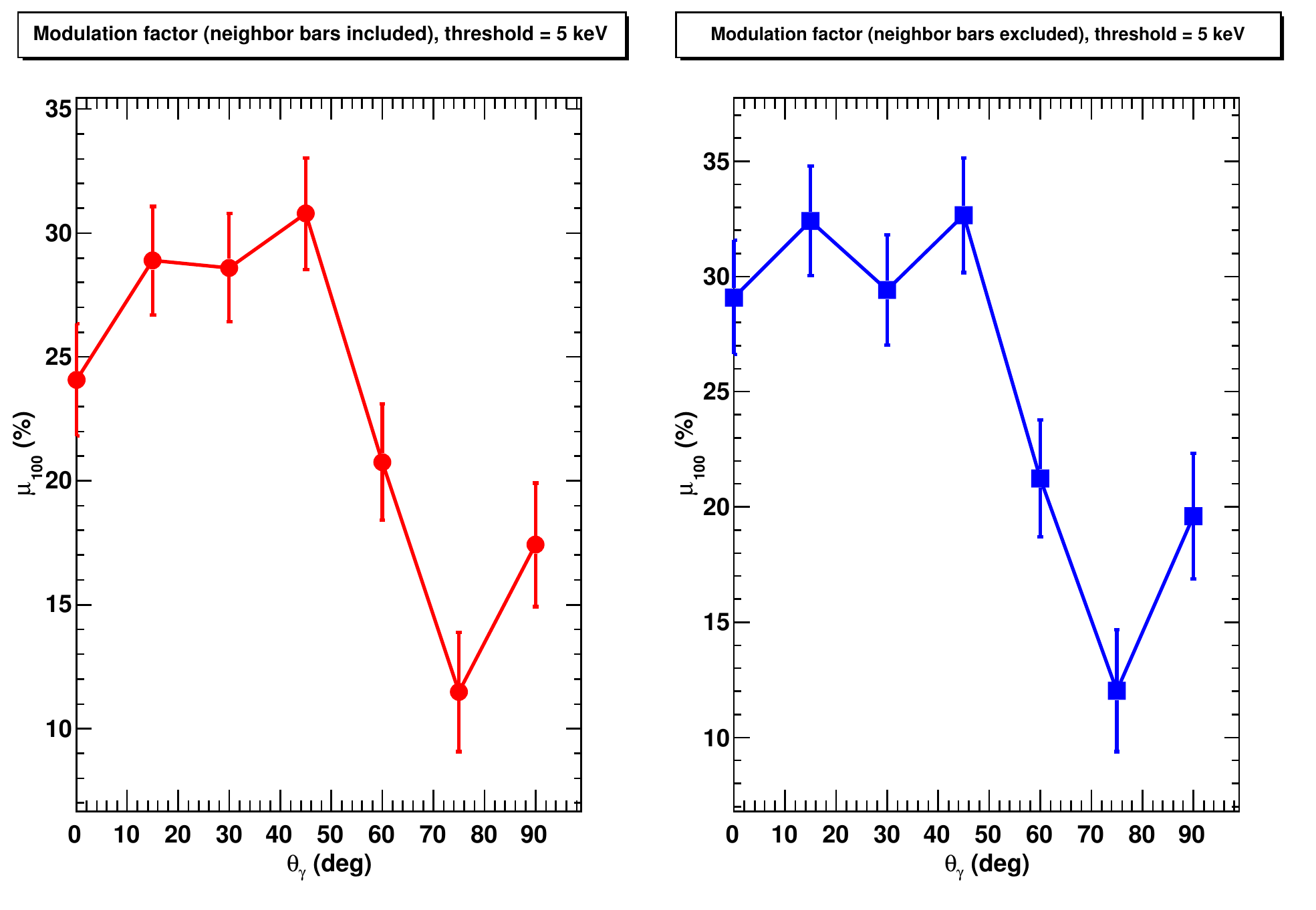}
  \caption{Simulated modulation factor (100$\%$ polarized) as a function of the incident angle $\theta_{\gamma}$ of GRB; $\varphi_{\gamma}$ is set to $0^{\circ}$ for simplicity. Left: neighboring bars included; Right: neighboring bars excluded }
  \label{fig:polar4}
  \end{center}
\end{figure}

\noindent For different incident angles of GRBs, the reconstructed
modulation factor by POLAR can be quite different. Figure\,\ref{fig:polar4} shows the
modulation factor measured by POLAR as a function of the incident
angle of GRB when the GRB photons are 100$\%$ polarized. More
simulation results are discussed else where
 \citep{Ref3, Ref5, Ref6}.

\subsection{Calibration tests}
\label{sec:4} 
\begin{figure}[b]
  \begin{center}
  \includegraphics[width=1.00\textwidth]{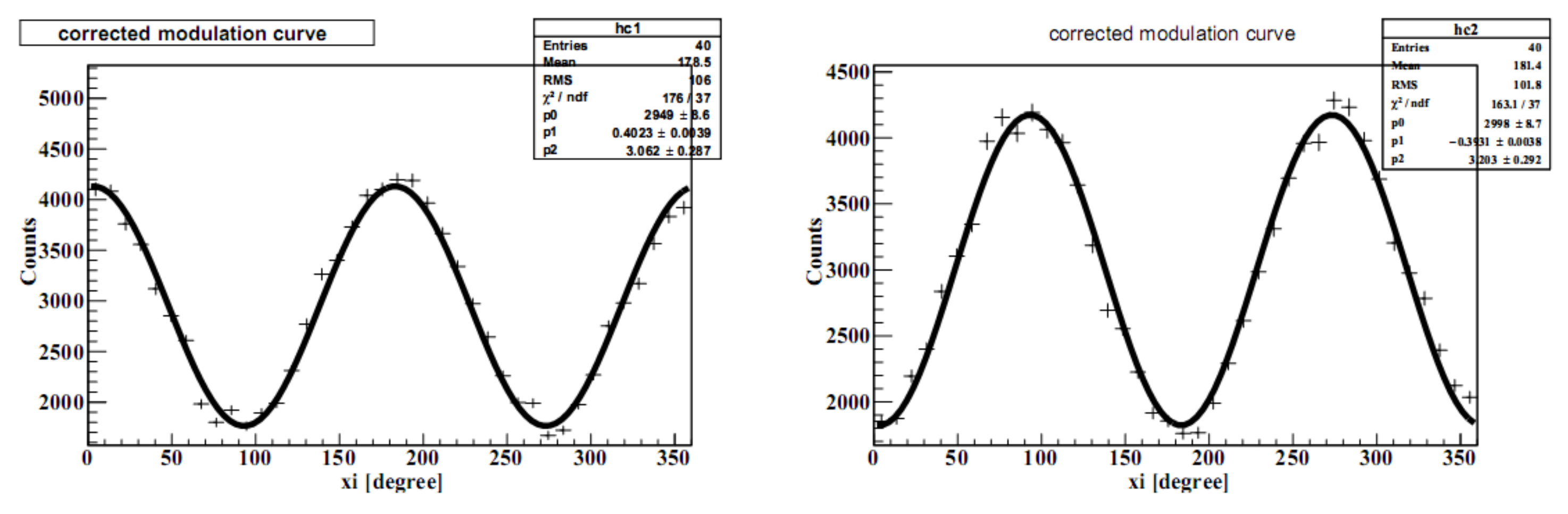}
  \caption{Modulation curve measured during the ESRF beam tests. Left: the polarization direction of the beam relative to POLAR detector is 90$^{\circ}$; Right: the polarization direction of the beam relative to POLAR detector is 0$^{\circ}$}
  \label{fig:polar5}
  \end{center}
\end{figure}

In 2015, the full flight model of POLAR has been
constructed. A series of calibration activities have been performed
to verify the basic working function of the instrument, test and
optimize its working parameters, study the polarization measurement
performance of the detector, verify and improve the data analysis
procedure and algorithm, as well as to optimize the Monte-Carlo
simulation model of POLAR which will be used in the in-orbit data
analysis. Among all of the calibration tests, the most important one
is the polarization measurement of POLAR performed with the European
Synchrotron Radiation Facility (ESRF) beam. The beamline used for
the test is ID11 which can provide us with 100$\%$ polarized hard
X-rays between 35 keV$\sim$140 keV. The beam of ID11 is horizontally
polarized. The size of the beam used for POLAR calibration is
0.5$\times$0.5 mm$^{2}$ with original beam intensity $\sim$10$^{7}$
phs/s, which was reduced to about 10$^{3}$$\sim$10$^{4}$ phs/s
with absorber to avoid the data pileup in the electronics.\\
\newline
During the ESRF beam test, four different beam energies were used,
i.e., 140 keV, 110 keV, 80 keV and 60 keV. For each beam energy,
three different beam incident angles (0$^{\circ}$, 30$^{\circ}$ and
60$^{\circ}$) were chosen, where 0$^{\circ}$ is on-axis,
30$^{\circ}$ and 60$^{\circ}$ are off-axis. Figure\,\ref{fig:polar5} shows the very
preliminary analysis results of the reconstructed Compton scattering
azimuth angle distributions (modulation curves) with 140 keV beam.
When fitting the modulation curves, the calculated modulation
factors for 90$^{\circ}$ and 0$^{\circ}$ polarized beam are
40.23$\%$$\pm$0.0039$\%$ and 39.31$\%$$\pm$0.0038$\%$, respectively.
The measurement results are quite similar to the Monte-Carlo
simulation result of 40.90$\%$, which verifies the accuracy of the
model for the Monte-Carlo simulations.

\subsection{Summary and discussions}
\label{sec:5} For GRB prompt emissions, the most interesting energy
range is about 100 keV$\sim$200 keV, within which POLAR has its best
performance. The Monte-Carlo simulation results show that for the
GRBs with total energy fluence $F_{\rm total}$ larger than
3$\times$10$^{-5}$ erg$\cdot$cm$^{-2}$, the minimum detectable
polarization (MDP) of POLAR can reach below $\sim$10$\%$. In
addition, for strong GRBs POLAR can localize their positions with
accuracy better than 5$^{\circ}$ ($F_{\rm total}$$\geq$10$^{-5}$
erg$\cdot$cm$^{-2}$). The main technical properties of POLAR are
summarized in Table\,\ref{tab:polar1}. The calibration test results show that POLAR
performs as expected on the polarization measurement. After
eliminating all kinds of system effects of the instrument during the
data analysis procedure and optimising the Monte-Carlo simulation
model, the final measurement results and simulation results are
quite similar. The calibration results as well as the optimised
instrument working parameters will be used in the further
Monte-Carlo simulations and real in-orbit
operations.\\
\newline
POLAR has also undergone and passed the space qualification tests,
such as thermal cycling, vibration, shock and thermal vacuum, etc.
These tests can verify the reliability of POLAR instrument and
guarantee its long-term operation and observation.

\begin{table}[h]
\begin{center}
\caption{Main technical properties of POLAR}
\label{tab:polar1}       % Give a unique label
\begin{tabular}{lll}
\hline\noalign{\smallskip}
No. & Property & Performance  \\
\noalign{\smallskip}\hline\noalign{\smallskip}
1 & Detector material & Plastic scintillator (EJ-248M) \\
2 & Yearly detectable GRBs & $\sim$50 \\
3 & GRB localization accuracy & $\leq$5$^{\circ}$ ($F_{\rm total}$$\geq$10$^{-5}$ erg$\cdot$cm$^{-2}$) \\
4 & Detection energy range & 50 keV$\sim$500 keV \\
5 & Field of view & $\pm$70$^{\circ}$$\times$$\pm$70$^{\circ}$ \\
6 & Minimum detectable polarization & $<$10$\%$ ($F_{\rm total}$$\geq$3$\times$10$^{-5}$ erg$\cdot$cm$^{-2}$) \\
\noalign{\smallskip}\hline
\end{tabular}
\end{center}
\end{table}

  % POLAR
\section{\emph{WFIRST} and \emph{JWST}}
\label{sect:jwst}

\subsection{\emph{WFIRST}}
%\subsection{\emph{WFIRST}}

%\subsubsection{Instrumentation}
\subsubsection{Instrumentation}

 \emph{WFIRST} consists of two instruments, a Wide-Field Instrument
 (WFI) and a coronagraph.
 
   The WFI will be capable of imaging and spectroscopy
   over thousands of square degrees
    and monitoring of SNe and microlensing fields.
     It has sensitivity in the $0.7-2\mu$m bandpass
 and a 0.28 deg$^2$ field-of-view (FOV), about a hundred times that
of \emph{JWST}.
   The WFI has 18 H4RG\footnote{H4RG is a hybrid CMOS (Complimentary Metal-Oxide-Semiconductor)
      4K$\times$4K optical imager made
  by 
    Teledyne Scientific \& Imaging.}
             detectors (288 Mpixels)
   and utilizes 6 filter imaging.
    The wide field instrument includes two channels, a wide field channel 
          and an integral field unit (IFU) spectrograph channel. 
   The wide field channel includes three mirrors
  and a filter/grism wheel to provide an imaging mode covering
   $0.76-2.0$ $\mu$m and a spectroscopy mode covering $1.35-1.95$ $\mu$m. 
  The IFU channel uses an image slicer and spectrograph to provide 
     individual spectra of each 
  0.15 arcsec wide strip covering the $0.6-2.0$ $\mu$m 
         over a $3.00 \times 3.15$ arcsec FOV.

The coronagraph can image ice and gas giant exoplanets
 as well as debris disks.
     It has a $400-1000$ nm bandpass, 
$\le10^{-9}$ contrast, and a 100 msec inner working
   angle (at 400 nm).
The coronagraph instrument includes an imaging mode and a 
spectroscopic mode to perform exoplanet direct imaging and 
  spectroscopic characterization of planets and debris disks 
 around nearby stars. 
\subsubsection{Science}

\begin{figure}[h!]
\begin{centering}
%includegraphics[width=4.20truein]{fig1.pdf}
%includegraphics[width=4.20truein]{fig1.jpg}
\includegraphics[width=3.50truein]{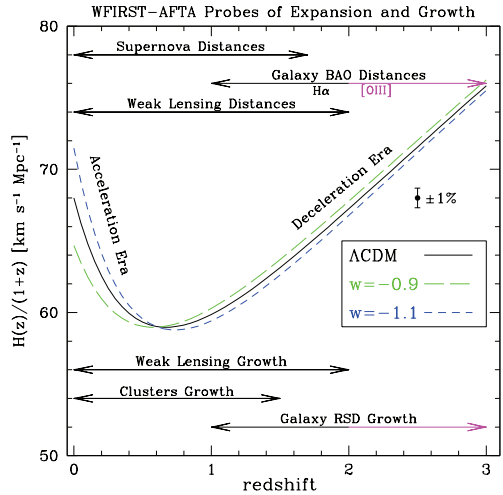}
%\vskip -1.0truein
\caption{
   Dark energy techniques used by \emph{WFIRST}
  (Spergel et al. 2015).
    }
\label{fig:wfirst1}   % Fig 
\end{centering}
 \end{figure}

\emph{WFIRST} has a 2.4 m wide-field IR telescope ($0.7-2$ $\mu$m)
       and an exoplanet imaging coronagraph instrument ($400-1000$ nm).
    It will 
        (i) carry out galaxy surveys over thousands of square degrees
  down to $J=27$ AB to study dark energy  (Figure\,\ref{fig:wfirst1})
         weak lensing and baryon acoustic oscillations,
      (ii) monitor a few square degrees for dark energy SN Ia studies,
    (iii) perform microlensing
 observations of the galactic bulge for an exoplanet census, 
       and
    (iv) undertake direct imaging observations of
 nearby exoplanets with a pathfinder coronagraph.
    The mission will have a robust and 
 well-funded guest observer program 
       for 25\% of the observing time. 
 %
  %   With its 2 hr TOO response time,
      With a $<1$ hr TOO response time (TBD, under study),
  \emph{WFIRST} will be a powerful
 tool for time domain astronomy and for 
  coordinated observations with gravitational wave
 experiments. Gravitational wave events produced 
      by mergers of nearby binary neutron stars
 (LIGO-Virgo, cf. Kanner et al. 2012) or extragalactic supermassive 
       black hole binaries (\emph{LISA}) will produce
 electromagnetic radiation 
       that \emph{WFIRST} can observe.

%Wide Field InFrared Space Telescope
%  
%NASA plus TBD partners
%
%Wide-field (~30 arcmin)
%
%0.7 – 2.0 microns + optical exoplanet coronagraph
%
%2.4m mirror
%
%2 hour (TBD) TOO response time
%
%Launch 2024

 

%\subsection{\emph{JWST}}
\subsection{\emph{JWST}}

%\subsubsection{Instrumentation}
\subsubsection{Instrumentation}

%Instruments

%The science instruments are the heart of the James Webb Space Telescope.
  % The f
%  Four science instruments will be contained within 
  The 
 Integrated Science Instrument Module (ISIM)  %. The ISIM 
  will house the four main instruments that will detect light 
  from distant stars and galaxies, and planets orbiting other stars.
%
%he ISIM includes the following instruments:
The ISIM will include:

(i) the 
    Near-Infrared Camera, or NIRCam (University of Arizona),

(ii) the 
    Near-Infrared Spectrograph, or NIRSpec (ESA, with components provided by NASA/GSFC),

(iii) the
    Mid-Infrared Instrument, or MIRI (European Consortium,   
                       European Space Agency (ESA), and NASA Jet Propulsion Laboratory), 

and (iv) the 
    Fine Guidance Sensor/ Near InfraRed Imager and Slitless Spectrograph,
    or FGS/NIRISS  (Canadian Space Agency). 

%Instruments integrated in ISIM Structure
 
% NASA, ESA, CSA partnership

% Narrow-field (~3 arcmin)

% 0.6 – 27. microns

% 6.5m mirror

% 2 day TOO response time

% Launch 2018

%The JWST originated in 1996 as the Next Generation Space 
%  Telescope (NGST). In 2002 it was renamed after NASA's second 
%  administrator (1961–1968) James E. Webb (1906–1992), noted for 
%  playing a key role in the Apollo program and establishing scientific 
%  research as a core NASA activity.[12] The JWST is a project of 
%  the National Aeronautics and Space Administration, the United 
%  States space agency, with international collaboration from 
%  the European Space Agency and the Canadian Space Agency.

The 6.5 m primary mirror,  a gold-coated beryllium reflector,  will have a 
%collecting area about five times that of \emph{Hubble}. 
 collecting area about five times that of \emph{HST}. 
   \emph{JWST} will be oriented towards near-IR astronomy, but 
  with sensitivity also in orange and red light, as well as the mid-IR. 
        Its coverage will span $0.6-27$ $\mu$m.
  The motivation for an emphasis on near-IR to mid-IR
  is three-fold: 
high$-z$ objects have their visible emissions shifted into the 
  IR, cold objects such as debris disks and planets
   emit most strongly in the IR, and this band has been difficult to study from the ground 
 or by previous space missions. 
%The telescope has an expected mass about half of Hubble's, but its primary mirror (a 6.5 meter diameter gold-coated beryllium reflector) will have a collecting area about five times larger (25 m2 vs. 4.5 m2). The JWST is oriented towards near-infrared astronomy, but can also see orange and red visible light, as well as the mid-infrared region, depending on the instrument. The telescope will focus on the near to mid-infrared for three main reasons: high-redshift objects have their visible emissions shifted into the infrared, cold objects such as debris disks and planets emit most strongly in the infrared, and this band is difficult to study from the ground or by existing space telescopes such as Hubble.

%The
  \emph{JWST} will operate near the Earth-Sun L2 Lagrange point, 
 %approximately 1,500,000 kilometres (930,000 mi) beyond the Earth. 
           $\sim$$1.5\times10^{11}$ m beyond the Earth. 
  Objects near this point can orbit the Sun 
  synchronously with the Earth, allowing the telescope to remain 
       at a roughly constant distance, using a single sunshield 
  to block heat and light from the Sun and Earth. This will keep the temperature 
  of the spacecraft below 50 K which is necessary for IR observations.

Launch is scheduled for 2018 on an Ariane 5 rocket. 
   Its nominal mission length is five years, with a goal of ten years.
  %    [16][17] The prime contractor is Northrop Grumman.

%\section{Section title}
%\label{sec:1}
%Text with citations \cite{RefB} and \cite{RefJ}.

%WFIRST  -- wide

%Swift  -- deep

%JWST
%\subsubsection{Science}
\subsubsection{Science}

    \emph{JWST}'s scientific mission has four main components:
     (i) 
   to search for light from the first stars and galaxies, 
  (ii)
    to study the formation and evolution of galaxies, 
   (iii) to understand the formation of stars and planetary systems,
   and 
  (iv)
    to study planetary systems and the origins of life.   
          This science wil be facilitated 
    by \emph{JWST}'s near-IR capabilities.
%The JWST's primary scientific mission has four main components: to search for light from the first stars and galaxies that formed in the Universe after the Big Bang, to study the formation and evolution of galaxies, to understand the formation of stars and planetary systems and to study planetary systems and the origins of life.[30] These goals can be accomplished more effectively by observation in near-infrared light rather than light in the visible part of the spectrum. For this reason the JWST's instruments will not measure visible or ultraviolet light like the Hubble Telescope, but will have a much greater capacity to perform infrared astronomy. The JWST will be sensitive to a range of wavelengths from 0.6 (orange light) to 28 micrometers (deep infrared radiation at about 100 K (−170 °C; −280 °F)).
%
 
\begin{figure}[h!]
\begin{centering}
%includegraphics[width=4.20truein]{fig_ng.pdf}
%includegraphics[width=4.20truein]{fig1.pdf}
%includegraphics[width=4.20truein]{fig1.jpg}
\includegraphics[width=3.50truein]{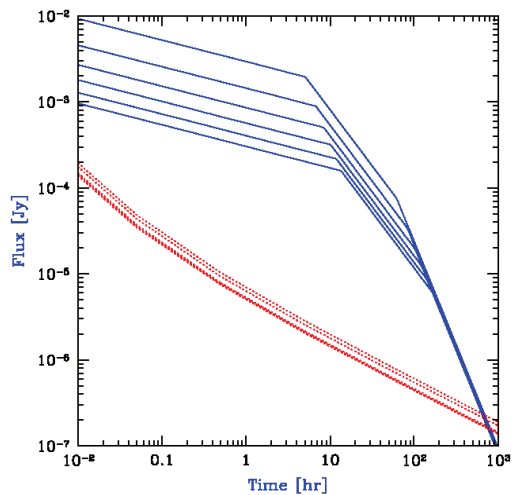} 
%\vskip -1.0truein
\caption{
% bromm loeb:
 % Detectability of high-redshift GRB afterglows as a ftjnction of time since the GRB explo-
   Detectability of high$-z$  GRB afterglows versus observer time $t-T0$ 
         (Bromm \& Loeb 2007; from Barkana \& Loeb 2004).
           GRB afterglow light curves are presented
 %  flux is shown (solid curves) 
    at the
redshifted Ly$\alpha$ wavelength
     ({\it solid blue curves}).
   \emph{JWST}/NIRSpec detection threshold 
            ({\it dotted red curves})
  are shown %(dotted curves) is the detection threshold iorJWST 
   adopting a
spectral resolution $\lambda/\Delta\lambda = 5000$, 
   % with the near infrared spectrometer,
    a signal to noise ratio of 5 per spectral
resolution element, and an exposure time of $0.2(t-T0)$.
   %20\% of the time since the GRB explosion. In 
  For each set
      of curves $z = 5$, 7, 9, 11, 13, and 15, respectively
      (top to bottom).
%  barkana loeb:
%
% Detectability of high-redshift GRB afterglows as a function of
% time since the GRB explosion as measured by the observer. The GRB after-
% glow ﬂux is shown (solid curves) at the redshifted Ly÷avelength. Also
% shown (dotted curves) is the detection threshold for JWST assuming a spectral
% resolution R ¼ 5000 with the near-infrared spectrometer, a signal-to-noise
% ratio of 5 per spectral resolution element, and an exposure time equal to 20%
% of the time since the GRB explosion. In each set of curves, a sequence of
% redshifts is used, z ¼ 5, 7, 9, 11, 13, and 15 (top to bottom). [See the electronic
% edition of the Journal for a color version of this ﬁgure.]
%
    }
\label{fig:wfirst2}   % Fig 
\end{centering}
 \end{figure}

 \emph{JWST} is expected to see 
      the very first galaxies,
             forming just a few 100 Myr after the Big Bang.
%The distant universe: The more distant an object is, the younger it appears: its light has taken longer to reach us. Because the universe is expanding, as the light travels it becomes red-shifted, and these objects are therefore easier to see if viewed in the infrared.[33] JWST's infrared capabilities are expected to let it see all the way to the very first galaxies forming just a few hundred million years after the Big Bang.[34]
%
%
IR observations allow the study of objects and regions of space 
  which would be obscured by gas and dust in the visible, 
   such as the molecular clouds where stars are born, 
 the circumstellar disks that give rise to planets, and the cores of active galaxies.
%Dust penetration: Infrared radiation is better able to pass freely through dusty regions of space that scatter radiation in the visible spectrum. These two images of the Carina Nebula (right margin) were taken with the HST. The top image was photographed utilizing the visible spectrum whereas the bottom image was taken in the infrared using the HST's WFC3 upgrade. Many more stars can be counted in the infrared image than in the visible light image. Observations in infrared allow the study of objects and regions of space which would be obscured by gas and dust in the visible spectrum,[33] such as the molecular clouds where stars are born, the circumstellar disks that give rise to planets, and the cores of active galaxies.[33]
%
 Relatively cool objects emit primarily in the IR.  Most objects cooler than 
  stars are better studied in the IR, including ISM clouds,  brown dwarfs, planets -- both in our own and other solar systems, 
   and comets and Kuiper belt objects.
%Cool objects: Relatively cool objects (temperatures less than several thousand degrees) emit their radiation primarily in the infrared, as described by Planck's law. As a result, most objects that are cooler than stars are better studied in the infrared. This includes the clouds of the interstellar medium, the "failed stars" called brown dwarfs, planets both in our own and other solar systems, and comets and Kuiper belt objects.
  In addition, \emph{JWST} will be a valuable tool in high$-z$ GRB follow-up (Figure\,\ref{fig:wfirst2}) 
    thanks to its ability to see back 
  to early time.

\subsection{Summary}

 \emph{JWST}  and \emph{WFIRST} will bring in a new era
of NIR and MIR astrophysics.
  \emph{JWST} will provide very deep ($J=29$)
observations with multiple instruments.
   \emph{WFIRST} will provide wide-field (1000's 
  of square degrees)
     deep ($J=27$) surveys.
For both missions
GRB follow-up science will be an integral
    part of science programs;
         \emph{JWST} will have a 2 d TOO response time,
 % while \emph{WFIRST} will have a    2 hr response time.
   while \emph{WFIRST} will have a $<1$ hr response time.
   TOO  follow-ups for both missions 
  will be limited to few per year.

    % WFIRST and JWST
\section{The 30\,m class ground-based telescopes}
\label{sect:30m}
\subsection{Introduction}
\label{intro}

As of 2015, construction has begun for three giant 30\,m class ground-based optical/infra-red
telescopes: the Giant Magellan Telescope (GMT) at Las Campanas, Chile,
the Thirty Meter Telescope (TMT) at Mauna Kea, Hawaii, and the European Extremely
Large Telescope (E-ELT) at Cerro Armazones, also in Chile.
Thus by 2025 it is expected astronomers will have access to facilities significantly 
more powerful than the existing state-of-the-art 8\,m generation.
Their advantage goes beyond simply enhanced light-gathering-power, since each telescope
is also planned to have sophisticated capabilities for high-order adaptive optics, thus providing
spatial resolutions even better than the space-based {\em HST} and {\em JWST}
for many observations \citep[e.g.,][]{gilmozzi2007}. A summary of the  basic vital statistics for each telescope is given in Table~\ref{tab:elt1}.

\begin{table}[!h]
\caption{Summary of planned 30\,m class telescope characteristics}
\label{tab:elt1}       % Give a unique label
\begin{tabular}{lllll}
\hline\noalign{\smallskip}
Telescope & Location & Altitude  & Diameter  & Mirror technology  \\
 & & (m) & (m) &   \\
\noalign{\smallskip}\hline\noalign{\smallskip}
GMT & Las Campanas & 2550  & 24.5 & 7 monolithic mirrors, common focus\\
TMT & Mauna Kea & 4050 & 30 & Segmented primary \\
E-ELT & Cerro Armazones & 3060 & 39.3 & Segmented primary \\
\noalign{\smallskip}\hline
\end{tabular}
\end{table}

A major driver for construction of these facilities is the exploration of early structure
formation, during and before the epoch of reionization.  This pushes towards optimisation
in the near infra-red, since the Gunn-Peterson trough renders the universe opaque
at rest-frame wavelengths below 1215\AA, which corresponds to $\lambda>1\,\mu$m
at $z\gtrsim7$.
In recognition of this, all three telescope consortia have prioritised building
of intermediate resolution, integral field spectrographs with nIR capability; namely HARMONI on the E-ELT, IRIS on the TMT and GMTIFS on the GMT.
HARMONI and IRIS, indeed, have been selected as  first light  instruments.
With spectral resolutions of $R\sim$4000--10000, the majority of the available spectral range
is between bright night-sky lines, and therefore provides low background and consequently high sensitivity.

\subsection{High redshift transient science}
\label{sec:transients}
% Paras on the potential

\subsubsection{Science drivers}
\label{sec:drivers}
Gamma-ray bursts are extremely bright, and hence visible in principle to very high redshifts.
This, combined with their very broad spectral energy distributions, and association with massive
star death, has long been recognised as making them potentially very powerful probes of
the early universe \citep{lamb2000,tanvir2007}.
They trace early star formation, pin-point the positions of primeval galaxies, and
provide luminous backlights for absorption spectroscopy \citep{chornock2013,sparre2014,hartoog2015}.

To-date, the highest spectroscopic redshift for a GRB is $z=8.2$ for GRB 090423 \citep{tanvir2009},
whilst a photometric redshift of $z\approx9.4$ was found for GRB 090429B \citep{cucchiara2011}.
However, despite their high intrinsic luminosities, at these distances follow-up with the
current generation of telescopes and instruments is hard for all but the most extreme events.  Thus it
is likely that some high-$z$ GRBs detected by {\em Swift} have not been recognised as such,
and those that have been recognised have usually lacked the high-S/N data to fully exploit their potential.

\begin{figure}
\begin{center}
\includegraphics[width=1.0\textwidth]{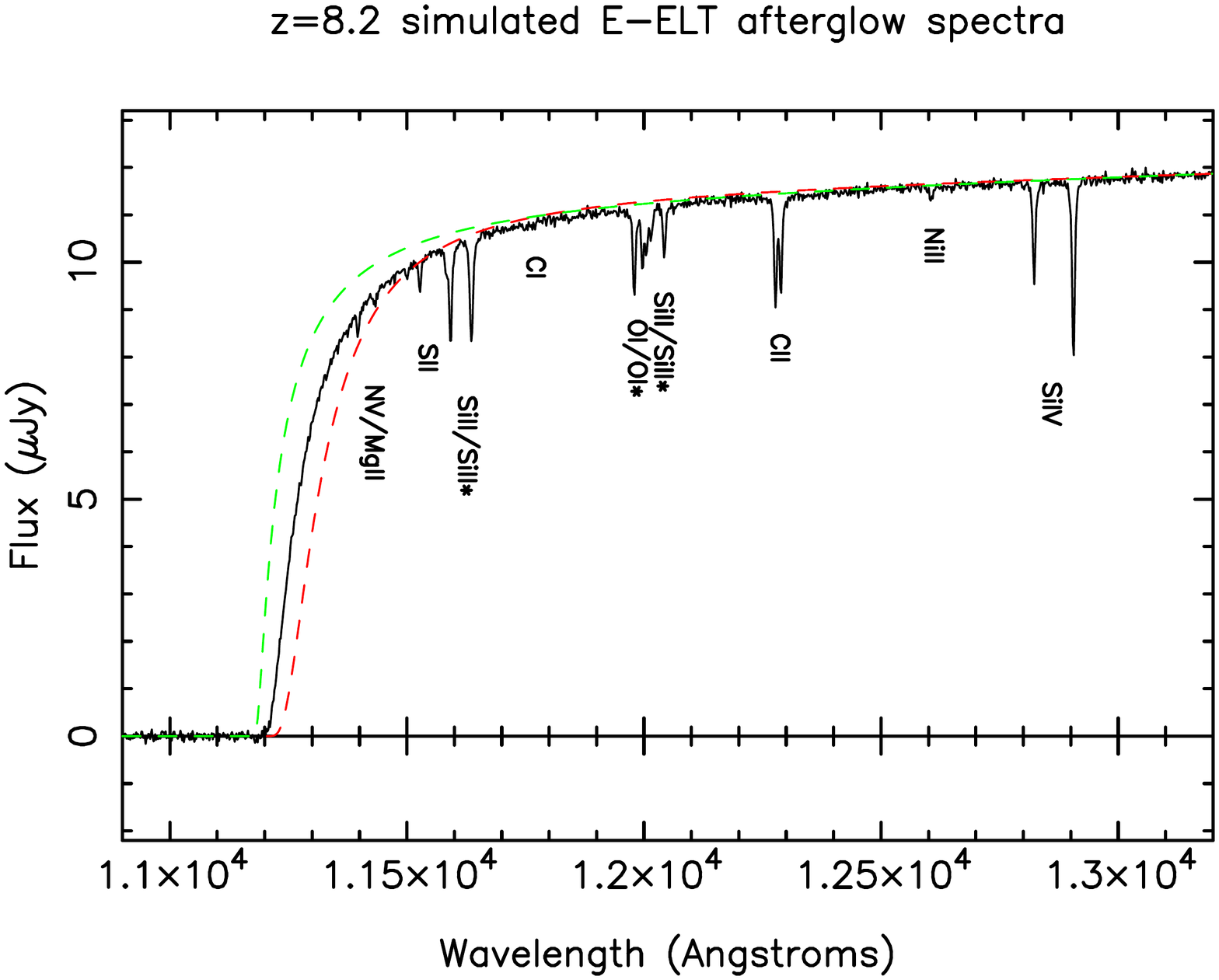}
\caption{Simulated spectrum (black solid curve) of typical GRB afterglow around the Ly-$\alpha$ line,
as it would be seen in a 30\,min exposure with E-ELT.  The high signal-to-noise allows
precise determination of neutral hydrogen column in both the host and intervening 
intergalactic medium, together with abundances for many metal species (the simulation is of
1\% of Solar metallicity). Attempts to
fit the damping wing with a host HI only model  is shown by the red dashed curve, and a 100\% neutral IGM only model 
as a green dashed
curve; both provide poor fits.  This shows that with such S/N it is possible to decompose the
host HI column density and IGM neutral fraction.}
\label{fig:eltsim}       % Give a unique label
\end{center}
\end{figure}

The next generation telescopes combine near infra-red optimisation with exquisite point-source
sensitivity, thanks to their large collecting area and AO-assisted high spatial resolution.
Thus afterglows for which only redshifts could be obtained previously, will in the future  
be used to  study abundances, neutral hydrogen, dust and molecular content in even very faint hosts.
All of these properties are extremely difficult to study by other means.
This potential is illustrated graphically in Figure~\ref{fig:eltsim}, which shows a simulated E-ELT spectrum of a typical
GRB afterglow at $z=8.2$ (in fact based on observed luminosity of GRB\,090423).
The measurement of HI column in both host and intergalactic medium (IGM) in a statistical sample of high-$z$
afterglows would be of considerable
importance in understanding the sources driving reionization.
The former provides the distribution of opacities, and hence escape fractions for ionizing radiation,
along the lines of sight to massive stars in the early universe \citep{chen2007,fynbo2009}, 
while the latter would allow us to map the progress of reionization and its variation from field to field
\citep{mcquinn2008}.

Other transients that may be studied by 30\,m class telescopes include supernovae of all types,
particularly superluminous supernovae that, although rare, are bright enough to be studied in detail
at $z>6$. This class of event may include pair-instability supernovae, the likely end-point of many
massive population III stars \citep{heger2003}.

\subsubsection{Challenges}
\label{sec:challenges}
% Paras on the challenges
Observing transients with 30\,m class telescopes offers great promise, but also
presents particular challenges.
Some challenges are for the wider astronomical community,
in particular it is obviously essential that we maintain a capability to discover
transient sources to feed into down-stream follow-up.
In the 2020s new sources of transient discovery should be
operating, notably the Large Synoptic Survey Telescope (LSST),
the advanced generation gravitational wave detectors (e.g., aLIGO, aVirgo),
and the SVOM satellite. However, of these only SVOM will be capable
of detecting high redshift sources, specifically gamma-ray bursts (GRBs)
and related high-energy transients.  Since SVOM's capabilities are
broadly similar to those of {\em Swift} we still only expect a modest rate
of high-$z$ bursts to be detected.
Another challenge for transient follow-up, particularly of GRBs, is that
their rapid time-evolution means that observations should ideally be done
on a time-scale of hours after trigger.  With the 30\,m telescopes being restricted
to just two geographical regions of the Earth, this limits the potential for
follow-up, and, especially if one also considers weather and technical down-time,
strongly motivates the continued operation of smaller telescopes as part of
a transient follow-up network.

Several other challenges are at least more within the control of the 
consortia building the 30\,m class telescopes.  First and foremost is the provision of
suitable imaging and spectroscopic capability.  Unlike some other areas of
astronomy, wide field is not key here, but good spatial resolution and (simultaneous)
wide wavelength coverage is important.
Since speed is of the essence, operational efficiency is also very important,
both to realise rapid responses to trigger requests (quick instrument/mode
changes, active optics reconfiguring, target acquisition, etc.), and also to expedite quick-look  reduction
and transmission of data to observers (or their software agents) allowing
further follow-up to be initiated with minimal delay.  These capabilities require
attention even in the design phase, for example to ensure that suitable
data calibration products will exist to support the desired quick-look pipeline processing.

\subsection{Host galaxies}
\label{sec:hosts}
In addition to the transients themselves, the 30\,m class telescopes will provide unprecedented
insights into their host galaxies.
In general, high-$z$ galaxies  found in Lyman-break surveys are compact \citep[e.g.,][]{curtislake2014},
typically a few tenths of an arcsec, and hence only crudely  resolved at {\em HST} resolution in the nIR.
GRB hosts seem to be even fainter \citep{tanvir2012}, and probably smaller, but with with spatial
resolutions as good as $\sim0.04$\,arcsec for E-ELT, significant detail will be resolvable, allowing
us to see  probe the physical conditions leading to early star formation.

%\paragraph{Paragraph headings} Use paragraph headings as needed.

%\begin{equation}
%\end{equation}

%\begin{figure*}
%  \includegraphics[width=0.75\textwidth]{example.eps}
%\end{figure*}
%

\subsection{Conclusions}
\label{sec:conclusions}
Within ten years, if these ambitious plans come to fruition, the astronomical community will have
access to three 30\,m class telescopes. 
Thanks to their huge grasp, adaptive optics capability, and near-infrared optimisation, these facilities will 
revolutionise our view of early structure formation in and before the era of reionization.
Observations of transients and their host galaxies has a particularly important role to play,
as they provide a route to studying in remarkable detail the physical nature of
the very small star-forming galaxies that are thought to have dominated star formation
at that time.
        % ELT
\section{Radio observations of GRBs and orphan afterglows}
\label{sect:ska}
\subsection{Radio observations of GRBs}
\label{intro}

Radio observations of GRBs dates back to the end of the nineties when the afterglow discovery \citep{Costa:1997la,van-Paradijs:1997it} opened a new era in the field of GRBs with the follow up of their long wavelength long lasting afterglow emissions. The study of X--ray and optical emission from GRBs was revolutionised, starting in 2005, by the Swift satellite \citep{Gehrels:2009fk} thanks to its fast (within one minute) pinpointing (with subarcmin precision) of the location of the $\gamma$--ray event. Currently, $\sim$95\% of the bursts detected by the $\gamma$--ray detector (BAT -- Burst Alert Telescope -- 15--150 keV) are detected on board by the X--Ray Telescope (XRT; 0.1--3 keV) and $\sim$ 75\% of these are also detected at optical wavelengths. 

\begin{figure*}
\hskip -2.0 truecm
\includegraphics[scale=0.6]{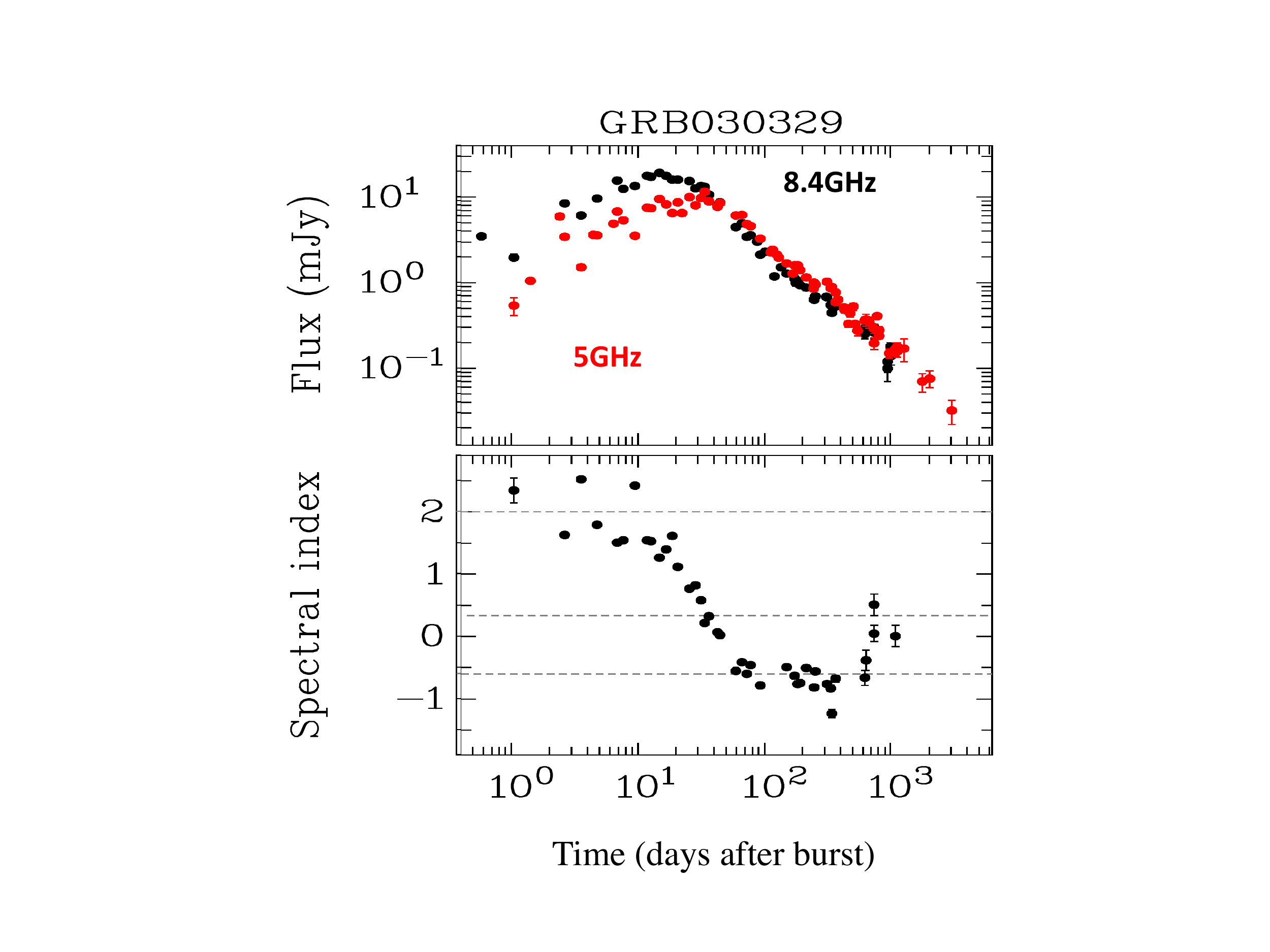}
\caption{Radio light curve (top panel) at 8.5GHz (black symbols) and 5GHz (red symbols) of GRB 030329. Evolution of the spectral index (computed between 5 and 8.5 GHz) as a function of time (bottom panel). Dashed lines in the bottom panel show reference slopes of -2/3, 1/3 and 2 (from bottom to top). Figure adapted from \cite{Granot:2014fk}. }
\label{fg1}       % Give a unique label
\end{figure*}

Until the last two years, the detection rate of GRBs in the radio was around $\sim$ 30\%. This small percentage (if compared to the optical and X--ray) remained unchanged before and after the launch of Swift \citep{Chandra:2012qf}. The ameliorated sensitivity of the upgraded Very Large Array (i.e. JVLA) increased the detection rate to  $60$\% in the last two years. 
This seems to support the idea that the small detection rate was mainly due to the sensitivity limits of past radio follow up programs \citep{Ghirlanda:2013bq}. Most often, these detections were at the level of sub-$\rm mJy$ flux densities in the 0.6--660 GHz (53\% at 8.5 GHz) frequency range and were obtained within dedicated observing programmes with the largest radio telescopes (VLA, WSRT, ATCA, GMRT). The median peak luminosity of the detected long GRBs is 10$^{31}$ erg s${-1}$ Hz$^{-1}$ at 8.4GHz while the few detected short GRBs and SN/GRBs are a factor 100  and 10 less luminous, respectively.  Regarding the distance scale of radio detected GRBs there seems to be no preference: both low redshift (e.g. GRB 980425 - \cite{Kulkarni:1998jk}) and high redshift (GRB 090423 - \cite{Chandra:2010xy}) events have been detected in the radio band.

In those events with multi epoch detections (i.e. radio light curve) a typical peak luminosity of $\sim 2\times 10^{31}$  erg s$^{-1}$ Hz$^{-1}$ is reached after about 10--20 days (3-6 days in the source rest frame) since the burst trigger. Typical post peak decay index is of order unity. Figure\,\ref{fg1} shows the radio light curve (top panel) of GRB 030329, one of the GRBs with the densest/longest monitoring in the radio band so far. The light curve shows a peak at $\sim$15 days at 8.5 GHz and slightly later ($\sim$45 days) at 5 GHz, a typical behaviour of the radio light curve. Within the standard afterglow model where the radiation is produced by synchrotron emission at the external shock (produced by the deceleration of the jet by the interstellar medium), the broad band spectral energy distribution (SED) can be described by a characteristic peak and at least three break frequencies: the self--asborption frequency $\nu_{\rm sa}$ below which the spectrum is steep (with typical slopes $\nu^{2-5/2}$), the electron minimum frequency $\nu_{\rm m}$  corresponding to the minimum energy ($\gamma_{\rm m}$) of the relativistically shock accelerated electrons and the cooling frequency $\nu_{\rm c}$ which corresponds to the minimum energy of  the electrons that cool on dynamical timescales. The typical ordering is $\nu_{\rm sa}<\nu_{\rm m}<\nu_{\rm c}$ at relatively early times (days--weeks) and  $\nu_{\rm m}<\nu_{\rm sa}<\nu_{\rm c}$ at late times.  The evolution of these frequencies and of the peak flux with time depends on the jet dynamics. The standard afterglow model (e.g.  \cite{van-Eerten:2012oj}) links the spectrum (dependent on the the micro physical parameters describing the shock - i.e. electron energy index, fraction of energy in relativistic electrons and in magnetic field) and the jet dynamics (dependent on the macro physical parameters, i.e. the kinetic energy of the jet and the density of the external medium).  

The radio band, therefore, is affected by $\nu_{\rm sa}$ and $\nu_{\rm m}$ and their relative position. The appearance of a peak in the radio light curve is produced by the passage across the observing band of these frequencies. In the example of GRB 030329 shown in Figure\,\ref{fg1} the bottom panel shows that the spectrum is self absorbed at early times (with index -2 typical of the SED below $\nu_{\rm sa}$) and becomes softer at later times reaching a slope -0.6 which can be used to infer the electron energy power law slope, $p\sim2.2$ in this case.  The early time spectral variation is due to interstellar scintillation. In general, in the radio band, the time of the peak increases and its brightness decreases with decreasing radio frequency. 

Radio emission, when detected, can be followed on very long timescales. While the X--ray (optical) afterglow fades below the sensitivity limits (below the host galaxy flux level) within weeks since the burst, radio emission has been observed years after (e.g. GRB 970508 - \cite{Frail:2000qv}; GRB 980703 - \cite{Mesler:2013nr}) with GRB 030329 (Figure\,\ref{fg1}) still detected at 5 GHz almost a decade after the $\gamma$--ray event. In this respect, radio observations are unique in probing the transition of the blast wave to the non--relativistic regime (i.e. trans--relativistic phase). In particular, during this phase estimates of the outflow parameters such as the kinetic energy \citep{Frail:2001sf,Berger:2004ty,Frail:2005sf,Shivvers:2011rz} do not depend on relativistic effects and jet aperture (if spherical symmetry is reached) differently from the same parameters\' estimates performed during the early relativistic phase (although radio calorimetry is not assumptions free - e.g. \cite{Eichler:2005rm}). 

Another unique feature of the radio band is the possibility to study the broad band SED of dark bursts (see e.g. \cite{Melandri:2012zl} for a recent compilation of dark bursts properties): by combining multifrequency radio observations with X--ray data it is possible to model the time evolution of the SED and infer, for bursts without optical detection, an estimate of the optical extinction. One of the best cases, GRB 051022 \citep{Rol:2007yg}, turned out to be highly extinguished ($\sim$2.3 mag in the infrared  and at least 5.4 mag in the optical) possibly due to some dust clumps located along the line of sight within the host galaxy.   

Polarisation is a powerful diagnostic of the jet composition. In particular, the reverse shock emission (produced by the deceleration shock propagating backwards into the jet outflow) and its polarisation level can constrain the magnetic  field structure of the jet. Reverse shock emission in the optical band, characterzed by a flare like light curve (i.e. with rise and decay steeper than the typical slopes predicted by the afterglow emission), has been observed in a handful of bursts (e.g. 080319B - \cite{Racusin:2008fv}; see \cite{Japelj:2014eu}). Radio flares from the reverse shock has been observed only in a couple of bursts (990123 - \cite{Kulkarni:1999lq}) with the most recent case of 130427A \citep{Laskar:2013fp}. This is a relatively nearby GRB (at $z=0.34$) with an extraordinary energetic budget (exceeding 10$^{54}$ erg). Its emission has been detected up to the GeV bad by Fermi and broadband SED modelling has produced different interpretations (e.g. \cite{Maselli:2014db}). The radio band offered the opportunity to interpret the early time radio flare as due to the reverse shock (also observed in the optical - Verstrand et al. 2013) suggesting a low density environment \citep{Laskar:2013fp,Perley:2014hl}. However, additional radio data \citep{van-der-Horst:2014qd} suggested also a possible alternative interpretation of a double jet component. This bursts is one of the best examples where the combination of multi wavelength data allow us to break the degeneracy among the free parameters of the standard model while it still  represents a challenge for the standard model itself \citep{Maselli:2014db}.

Radio afterglow observations offer in principle the unique opportunity to measure the jet expansion and constrain the GRB size. This can be done (i) directly through imaging as for (the only case) GRB 030329 which was observed to expand over a couple of months thus indicating an apparent velocity 3--5 the speed of light \citep{Taylor:2004qr,Pihlstrom:2007rc,Mesler:2012rw} or (ii) indirectly through radio scintillation (i.e. from the measurement of the time it is quenched when the projected source size is larger than the galactic ISM inhomogenities causing it).

At very late times, when the afterglow emission has faded below the host galaxy, radio observations allows us to derive an independent estimate of the star formation in  GRB hosts thus providing an estimate of the dust obscuration. Systematic search of radio host emission provided several upper limits at the level of 10--100 $\mu$Jy and only few detections at redshift z$<$1 \citep{Michaowski:2012dn,Perley:2013xe,Perley:2015uo}. Despite the small detection rates, these studies indicate that the host star formation rates derived from the radio, which have the advantage of probing the unobscured SFR are $\ll$100 M$_{\odot}$ yr$^{-1}$. One major source of uncertainty in these estimates is the unknown host radio spectrum, which requires multifrequency observations.Radio hosts observations will shed light     on the class of dark GRBs ($\sim$20--30) i.e. bursts that are not detected in the Optical band \citep{Perley:2015uo}. 

\subsection{The future of radio observations of GRBs}

\begin{figure*}
\hskip -2.0 truecm
\includegraphics[scale=0.6]{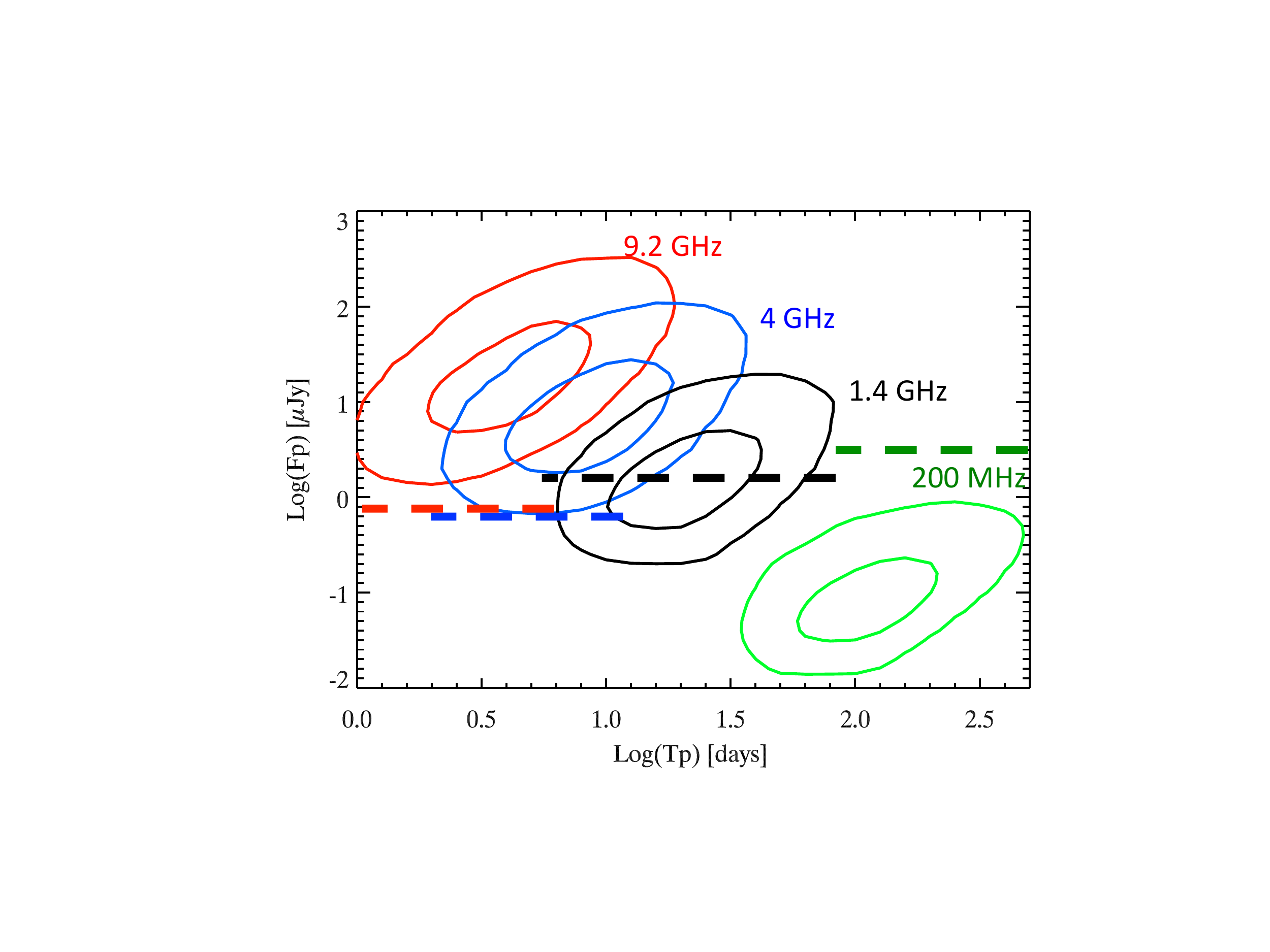}
\vskip -2 truecm
\caption{Contour levels (68\% and 95\%) of the distribution of the population of GRBs corresponding to the time when the peak of the afterglow is reached and the corresponding flux. Four characteristic frequencies corresponding to the SKA1-MID (REF) are shown with different colors. The dashed lines show the expected continuum sensitivity limits at 5$\sigma$ for a 12h exposure.  }
\label{fg2}       % Give a unique label
\end{figure*}

We are now entering the era of  advanced radio facilities. The updated JVLA, the largest radio telescopes working in interferometric mode (VLBI, VLBA, EVN) and the advent of the SKA precursors are in the reach of detecting and monitoring GRBs. In particular, the Square Kilometer Array (SKA - \cite{Carilli:2003ef}) will be the most powerful radio facility for continuum, spectroscopy and survey studies of the radio sky. The observatory will be composed of two sites (Australia and South Africa) where thousands of different antennas (from aperture arrays to single dishes) will be operated over a wide frequency range extending to 9 GHz at the highest frequencies. 

Through a population synthesis code it is possible to predict the radio flux density at the time when the radio light curve peaks for a synthetic population of GRBs \citep{Ghirlanda:2013bq}. The assumptions of this simulations are discussed in \citep{Ghirlanda:2013bq,Ghirlanda:2014wt,Ghirlanda:2015la}. Figure\,\ref{fg2} shows the contour levels corresponding to the distribution of the simulated population. The axes represent the flux of the radio emission at the time when the peak of the  emission is reached. The different contour levels refer to four characteristic frequencies of the SKA-MID configuration\footnote{https://www.skatelescope.org/news/ska-science-book/}. The dashed lines are the 5$\sigma$ sensitivity limits obtained for a 12h exposure. Figure\,\ref{fg2} shows that the characteristic timescale of the peak of the afterglow light curve moves to longer times at lower frequencies.  Current estimates with the SKA1-MID sensitivity limits (Burlon et al. 2015) show that it will be possible to detect most of the population of GRBs at relatively large frequencies (from 4 to 9.2 GHz, blue and red curves in Figure\,\ref{fg2}). Still the self absorption dominating the lowest frequencies will make accessible with the SKA-1 roughly half of the bursts at 1.4GHz (black curves) and almost make it tough to detect any GRB at 200 MHz (green curves). These estimates only describe the power of the radio observations to study the radio afterglow emission of GRBs. In general the follow up campaigns of GRBs will strongly benefit from the increased sensitivities of present and forthcoming radio facilities provided that the detection rate of GRBs in the $\gamma$--ray band (i.e. the triggers) is maintained/increased with current/future GRB detectors (from Swift and Fermi now to SVOM in the future). 

The high sensitivity of the SKA will make it possible to perform late time radio observations of the bursts when they have transitioned to the non--relativistic regime. We have computed \citep{Ghirlanda:2013bq,Burlon:2015it} the expected flux density at the non--relativistic transition and compared to the SKA1-MID expected limits. Only a handful of objects reach the sensitivity limit of current facilities (a few $\mu$Jy at best). We foresee that the full SKA will routinely observe a significant fraction (15--25\%) of the whole GRB afterglow population at late times. 

One challenging aspect of GRBs in the forthcoming era of extended radio facilities and in the final SKA era will the possibility to access their parent population. We know that GRBs are jetted sources with a typical opening angle of a few degree. We detect only those bursts whose jet is closely aligned with our line of sight. This means that for  burst that we detect in $\gamma$--rays there are hundreds of events that point their jet elsewhere. Depending on the jet structure these events can still be detected as slow transients in wide field surveys. Orphan afterglow properties have been studied in the literature \cite{Rossi:2008om,Ghirlanda:2014wt,Ghirlanda:2015la}. Despite their large number no convincing evidence of a radio orphan GRB has been reported so far. This is most likely due to the sensitivity limits of past radio surveys. Forthcoming radio surveys like the VAST/ASKAP (operating at 1.4 GHz) or the MeerKAT or EVLA (operating at 8.4 GHz) could detect 3$\times10^{-3}$ and 3$\times10^{-1}$ OA deg$^{-2}$ yr$^{-1}$, respectively. The deeper SKA survey, reaching the $\mu$Jy flux limit, could detect up to 0.2-1.5 OA deg$^{-2}$ yr$^{-1}$ \citep{Ghirlanda:2014wt,Metzger:2015eq}.

The SKA era will be transformational for the study of GRBs. The SKA will access close to 100\% of the GRBs detected in $\gamma$--rays and will substantially contribute to the multi wavelength follow up of these sources. Radio observations will extend the timescale of follow up to very late times when the afterglow emission at other frequencies will be already undetectable. We will get an unprecedented insight into the true energy budget of GRBs and it will become possible to probe both the macrophysics of the ambient medium (e.g. the density profile) and the microphysics of the shocks
Finally, it will become possible to detect the so far elusive class of Òorphan afterglowsÓ that should appear as slowly evolving transients detectable with the SKA on weekly basis in its wide field survey.

\section{Lobster-eye micro-pore optics for X-ray focusing and the Einstein Probe mission}
\label{sect:ep}

%\subsection{Introduction}
%\label{intro}

Time-domain astronomy will see its golden era towards the end of this decade with the advent of major wide-field facilities across the electromagnetic spectrum and in the multi-messenger realms
including gravitational wave and neutrinos.
In the X-ray regime, the driving science calls for new generation instruments with high  sensitivity, good angular resolution (a few arc-minutes or less) and a large sky coverage
(field of view of order of thousands square degrees).
These requirements can be fulfilled by wide-field X-ray focusing optics---the emerging lobster-eye
Micro-Pore Optics (MPO),
as focusing imaging results in enormously enhanced gain in signal to noise,
and thus high detecting sensitivity.
The Einstein Probe (EP), which is a candidate mission of priority
of the Chinese Academy of Sciences with an intended launch date around 2020,
is based on this lobster-eye MPO technology.
Its aim is to monitor a large fraction of the whole sky
at high cadences with sensitivity in X-ray at least one order of magnitude deeper than
the most sensitive all-sky monitoring type instruments ever built.

Among various types of faint transients that EP is expected to discover,
high-redshift gamma-ray bursts
are of particular interest since, as the most luminous objects in the Universe at their peaks,
they provide a unique tool to study the early Universe beyond redshifts 6 and up to $\sim$20.
During this epoch the first luminous objects as Pop-III stars (as well as Pop-II stars)
are expected to form and start
to re-ionise the Universe out of the dark ages \citep{cf05}.
Whilst synthesizing metals they quickly evolved and exploded, and thereby polluted their environment.
The detection of these stars individually would be extremely difficult or almost impossible,
however, even with the next generation of space observatories like JWST.
The only way of observing Pop-III stars in action is to detect their explosive deaths.
They have been predicted to produce a GRB-like event \citep{bl06,wang12}.
By using high-redshift GRBs as beacons we could identify and probe the regions where the first stars and their remnants (the first black holes)  were formed.

\subsection{Micro-pore optics X-ray focusing technology}
\label{sec:MPO}

%Text with citations \cite{RefB} and \cite{RefJ}.
%\subsection{Subsection title}
%\label{sec:2}

\begin{figure}
\begin{center}
  \includegraphics[width=0.7\textwidth]{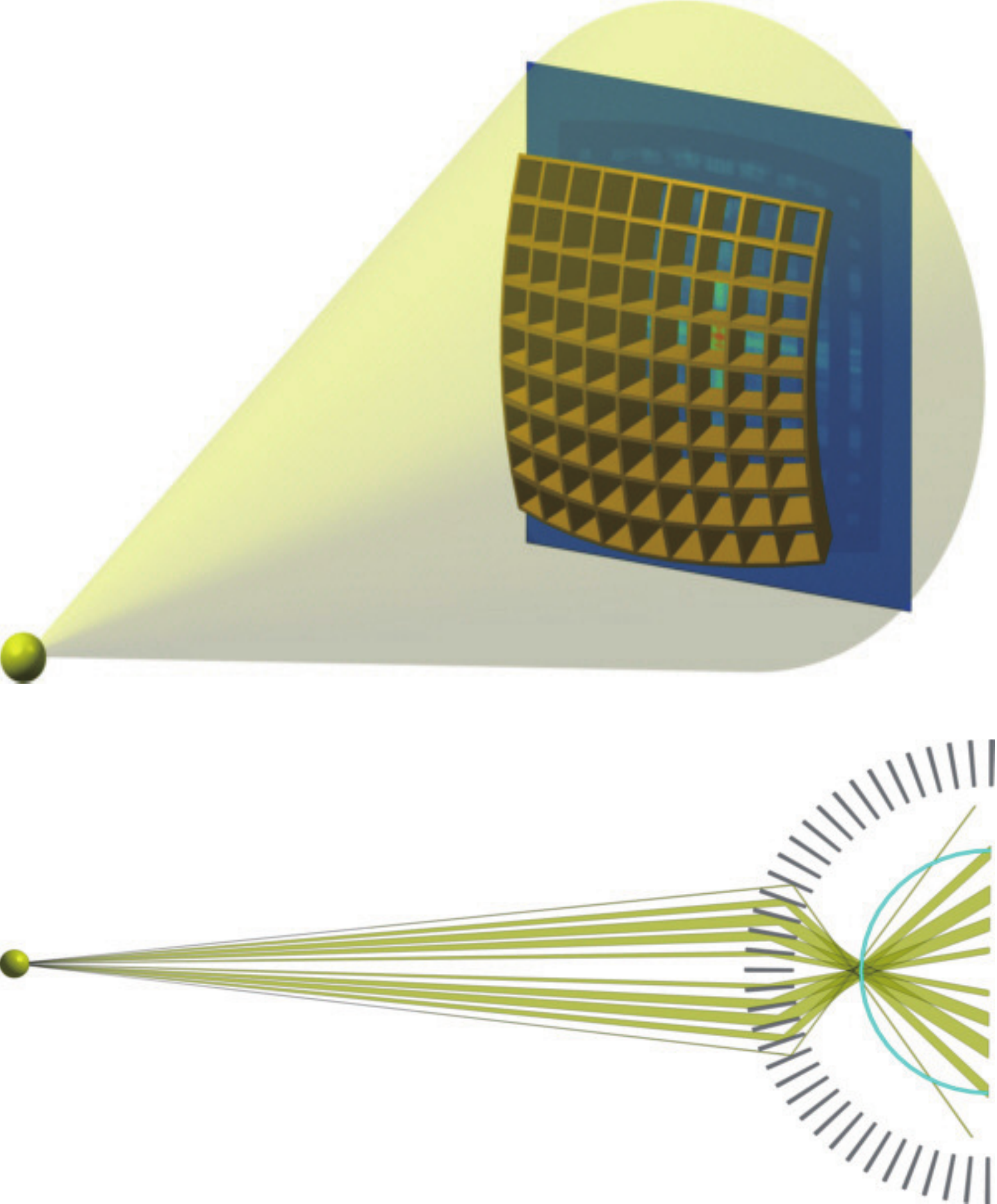}
\caption{Upper panel: Lobster eye optics: X-rays from source illuminate micro channel grid (gold colour) and are focused on detector plane with cross-like Pointing Spreading Function (right, blue). Lower panel: Optical path in a lobster eye optics. (Credit: adopted from http://www.x-ray-optics.de/.}
\label{fig:lobster_eye}       % Give a unique label
\end{center}
\end{figure}

X-ray focusing instruments rely on multiple smooth refection surfaces almost parallel to incident X-rays, which can be arranged in several different configurations.
One is Lobster-eye optics \citep{angel1979},
which mimics the imaging principle of the eyes of lobsters as shown in Figure \ref{fig:lobster_eye}.
Incoming light is reflected off the walls of many tiny square pores arranged on a sphere and
pointed towards the cocentric spherical center.
The reflection surfaces are configured orthogonal to each other without a specific optical axis, and thus the
FOV can in principle subtend the entire solid angle of $4\pi$.
In contrast,  the conventional Wolter-I optics \citep{wolter1952}
(used for Chandra, XMM-Newton, Swift/XTR, etc.) has an tubular, rotationally symmetric elliptical or parabolic reflection surface followed by a hyperbolic surface,
which  is impossible to achieve a FOV larger than  one degree.

\begin{figure}
 \begin{center}
  \includegraphics[width=1.0\textwidth]{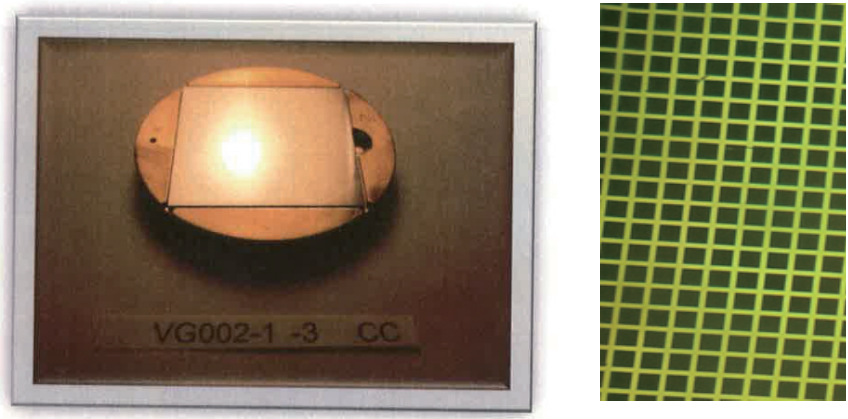}
\caption{MPO device (left) and a close-up under microscope (right). The pore size is of 20 micrometers. }
\label{fig:MPO}       % Give a unique label
\end{center}
\end{figure}

Lobster-eye telescopes can be constructed using novel X-ray focusing device, the so-called Micro-Pore Optics (MPO) \citep{fraser1993}.
This technology integrates millions of square pores as shown in Figure \ref{fig:MPO} with sizes of about tens of micrometers on a very thin ($\sim$1\,mm) glass piece of several square centimeters.
The walls of pores are smooth enough to reflect X-ray photons up to several Kilo-eV. MPO  is one of the most compact X-ray optics with the largest effective area to weight ratio.
MPO is employed for the MIXS X-ray telescope onboard EAS's Mercury mission BepiColombo \citep{Fraser201079}, where a payload mass is strictly constrained, as well as for the MXT X-ray telescope onboard SVOM (Section \ref{sect:svom}).
A lobster-eye telescope with MPO devices can achieve a huge instantaneous FOV
with very light weight, which is unique for X-ray imaging telescopes.

\begin{figure}
 \begin{center}
  \includegraphics[width=0.9\textwidth]{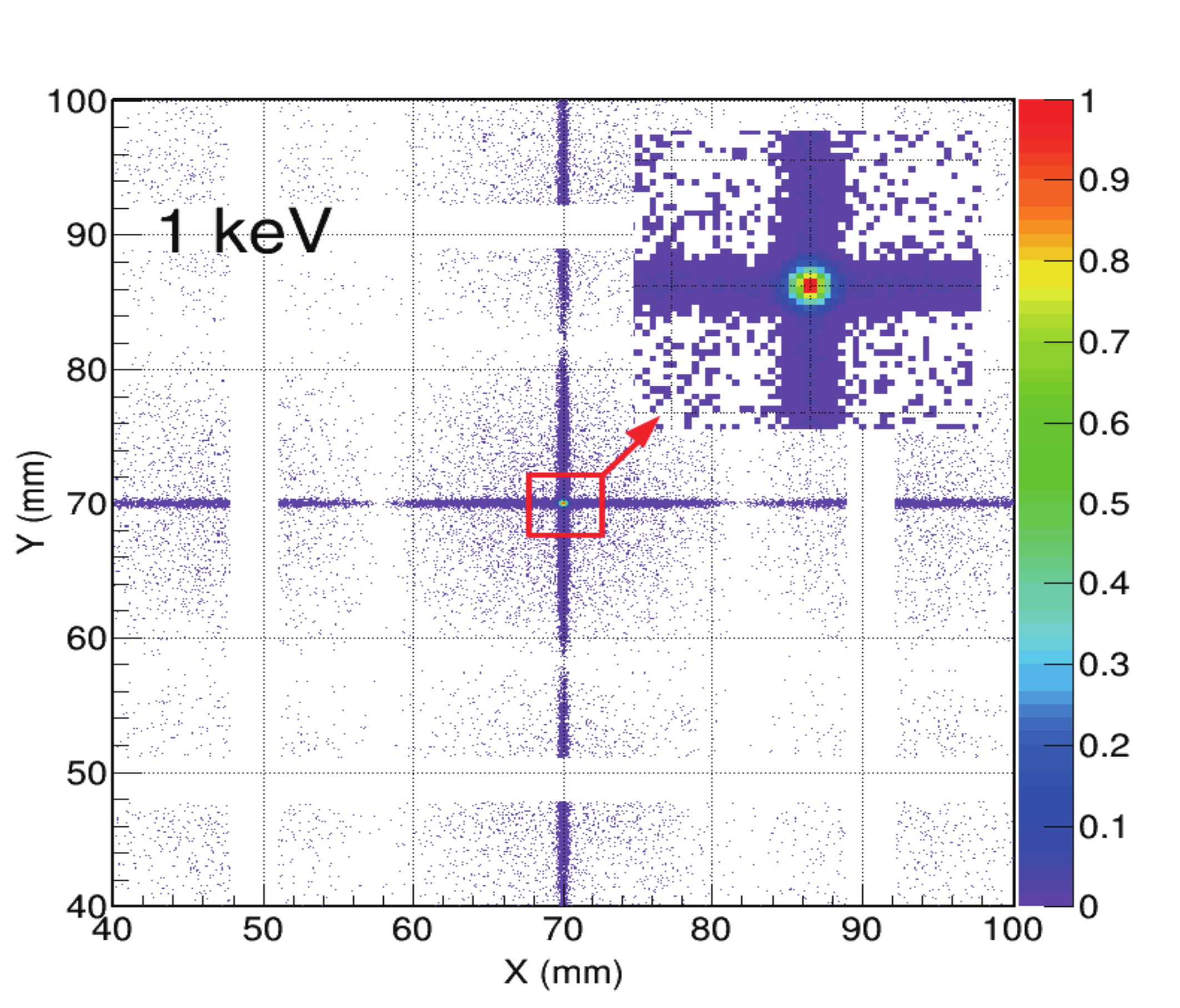}
\caption{Simulated PSF at 1\,keV of a Lobster-eye telescope with a focal length of 375\,mm. }
\label{fig:psf}       % Give a unique label
 \end{center}
\end{figure}

A simulated point spread function (PSF) of a Lobster-eye telescope is shown in Figure \ref{fig:psf}.
The cross-like feature is caused by complex reflection processes of X-ray photons
across the square pore.
The incoming X-rays are split into four beams corresponding respectively to no reflection (straight through, forming background), single reflection (cross arms) or double reflections off
adjacent walls of the pore (focal spot).  The FoV of the optical arrangement indicated in Figure \ref{fig:lobster_eye} is only limited by the
size of the optics (the number of MPO pieces) or the size of the detector.
The PSF remains almost unchanged over the entire FOV without vignetting of the effective area.
Such a wide-field lobster-eye telescope provides the technological basis
of the next generation wide-field X-ray monitors to detect faint and short-lived phenomena
like high-redshift Gamma-Ray Bursts, distant X-ray novae and tidal disruption events.

\subsection{Einstein Probe}
\label{sec:ep}

%\subsection{Instruments}
The Einstein Probe is a candidate mission of priority in the Advanced Study Phase
of the Space Science Pilot Programme of the Chinese Academy of Sciences (CAS), with an intended launch date around 2020.
It will  discover and characterise
high-energy transients and monitor variable objects in the soft X-ray band
with unprecedented sensitivity.
Its primary scientific goals are to: (1) reveal quiescent black holes at almost all astrophysical mass scales and study how matter falls onto them by detecting  transient X-ray flares, particularly stars being tidally-disrupted by otherwise dormant massive black holes at galactic centres;
(2) discover the X-ray photonic counterparts of gravitational-wave transients
 found with the next generation of gravitational-wave detectors and  precisely locate them;
(3) carry out systematic and sensitive surveys of high-energy transients, to
discover faint X-ray transients of various types, such as  high-redshift GRBs, supernova
shock breakout, and previously unknown transients.

The payload of EP consists of a Wide-field X-ray Telescope (WXT) with a field-of-view of
$60\times60$\,deg$^2$ in the nominal 0.5--4\,keV band,
and a Follow-up X-ray telescope (FXT) with a larger light-collecting power  than WXT,
both based on the lobster-eye MPO technology.
In addition, EP is equipped with a fast alert telemetry system,
in order to trigger multi-wavelength follow-up observations from the worldwide community.
The nominal mission lifetime is three years with a goal of five years.
For a more detailed description of the scientific goals and instrumentation of EP,
please refer to \citet{yuan14}.

%The wide-field imaging capability of WXT is achieved by using the micro-pore �lobster-eye� optics, thereby offering unprecedentedly high sensitivity and large Grasp superseding previous and existing X-ray all-sky monitors and survey missions by orders of magnitude.
%To complement this powerful ability to monitor and discover sources over a wide area,
%the Einstein Probe will also carry a smaller field-of-view ($\sim1$\,deg)
%Follow-up X-ray telescope (FXT) �
%capable of much larger light-collecting power and better energy resolution
%than the main survey telescope � with which to perform follow-up observations of newly-discovered transients.

\begin{figure}
\begin{center}
     \includegraphics[width=0.58\textwidth]{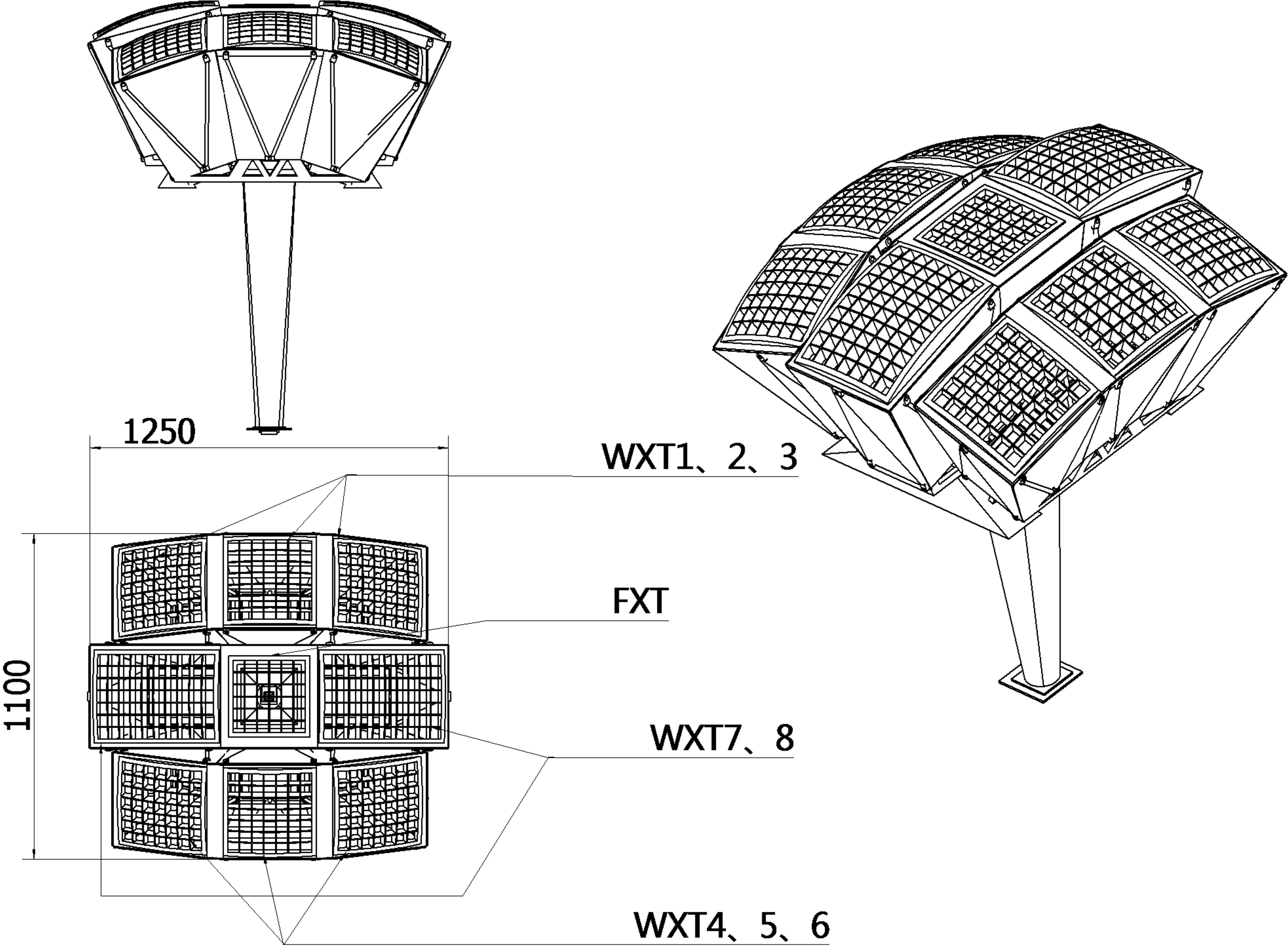}
     \includegraphics[width=0.38\textwidth]{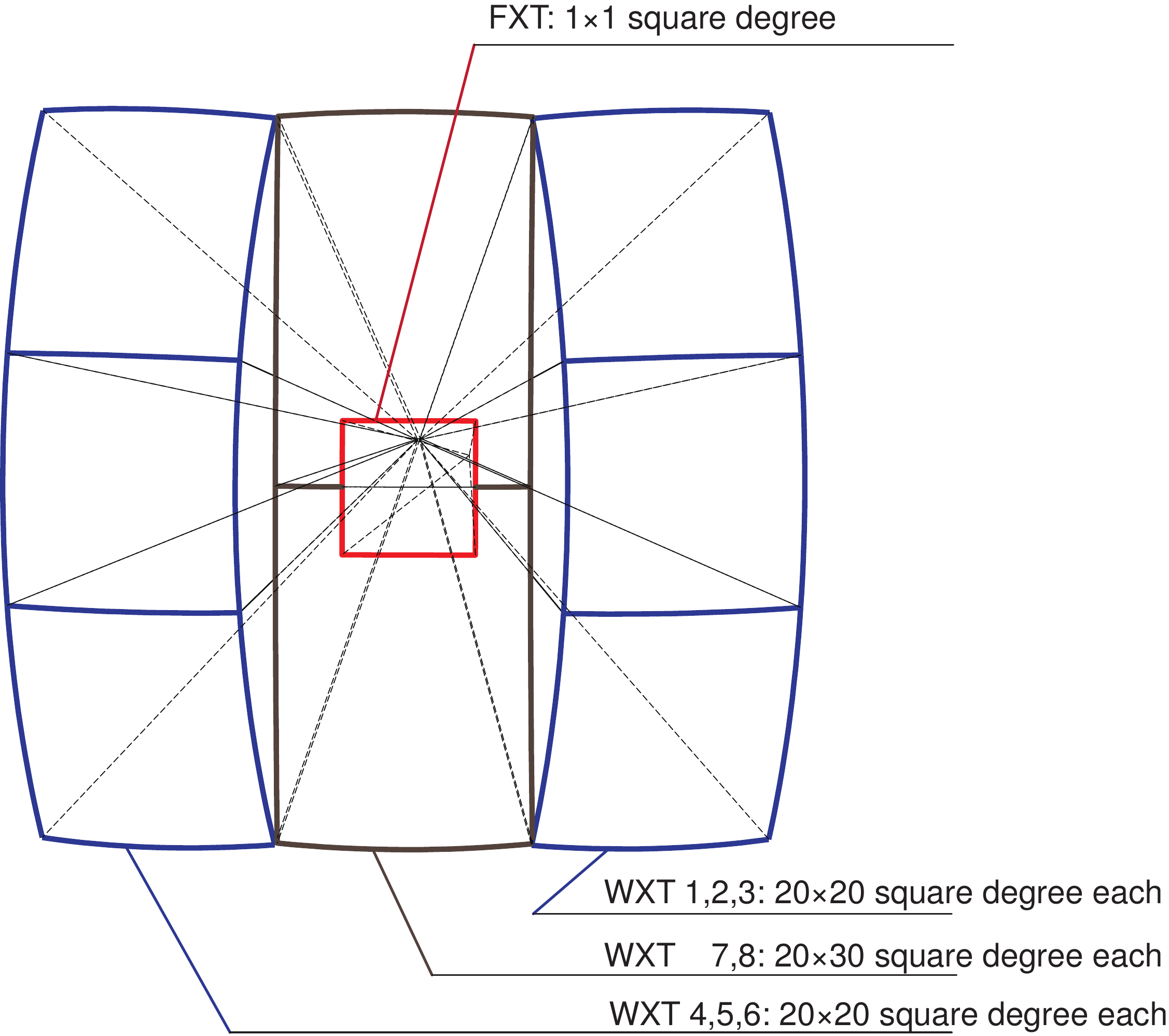}
     \caption{(left) Layout of WXT modules and FXT. (right) Sketch of the field of view of WXT (not to scale).
     [Figures adopted from \citet{yuan14}].}
     \label{fig:layout}
     \end{center}
\end{figure}

\begin{figure}
\begin{center}
     \includegraphics[width=0.44\textwidth]{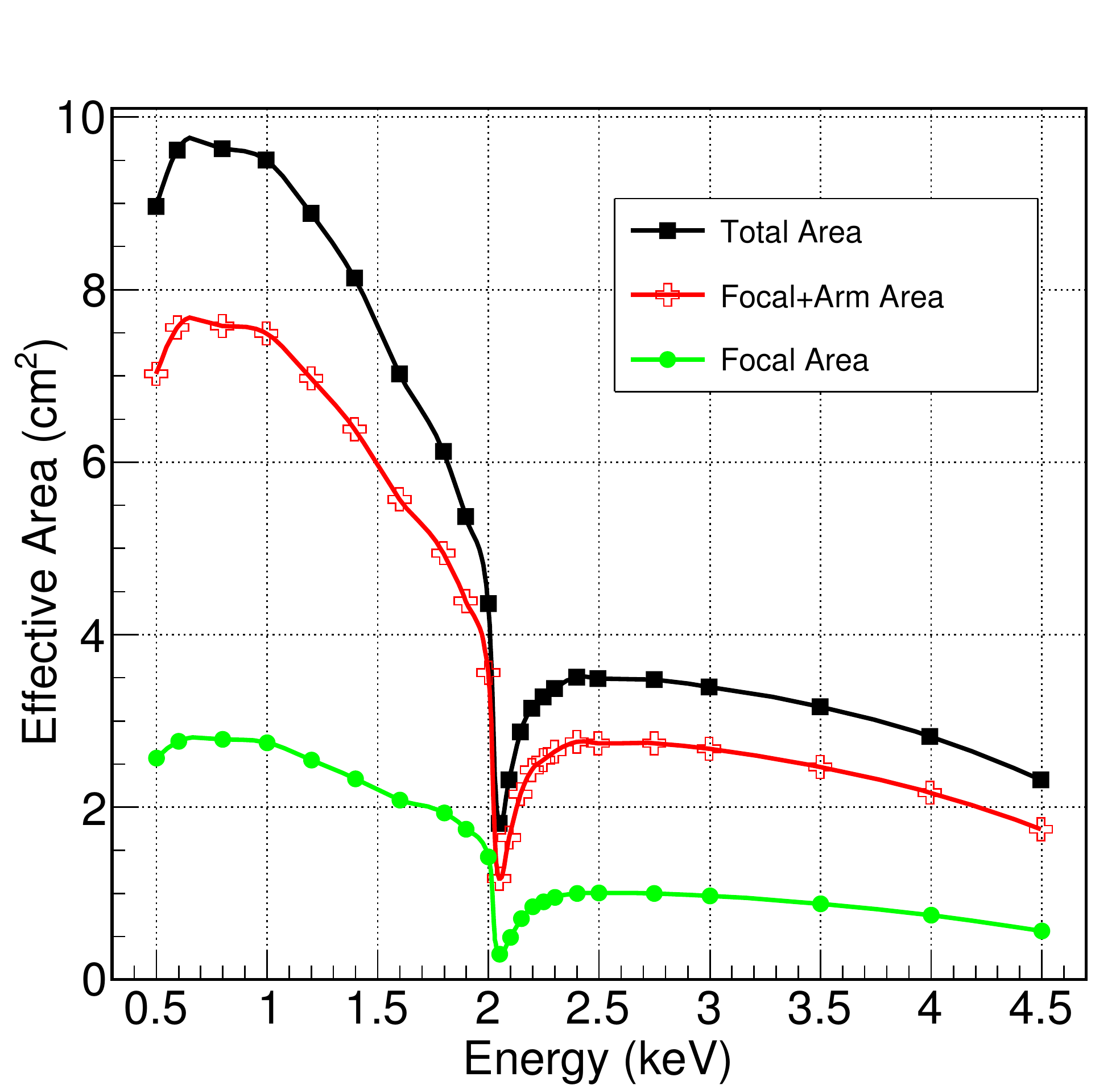}
     \includegraphics[width=0.54\textwidth]{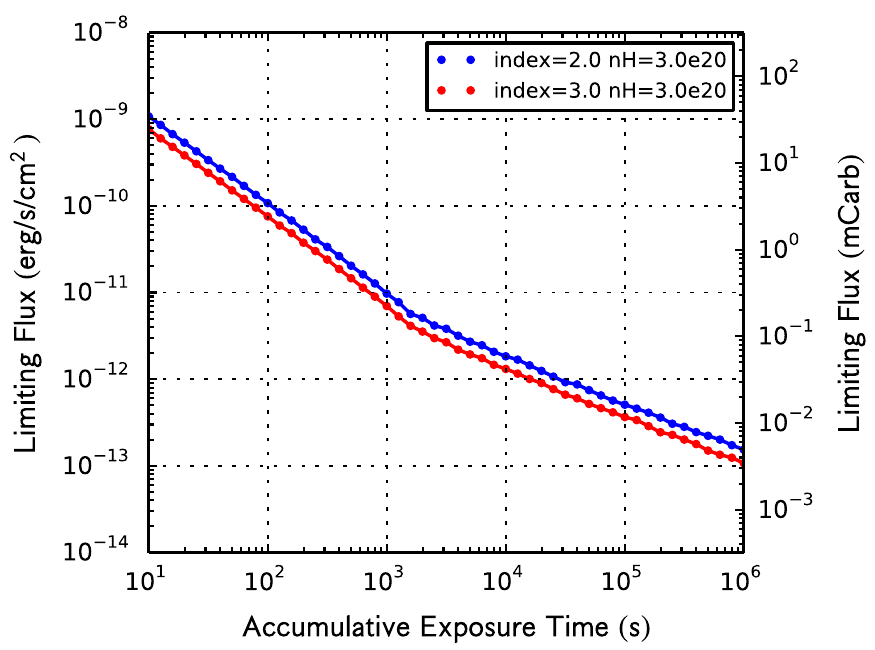}
     \caption{(left) Simulated effective area of EP/WXT with a GEM detector,
for the central focal spot (green),  central plus the cruciform arms (red), and
total (black; plus unfocused X-rays as diffuse background).
The MPO arrays are coated with Iridium, and have surface roughness of
$\sim$0.55\,nm and the tilts of pores following a Gaussian distribution with $\sigma$=0.85\,arcmin.
Xenon gas and a window of a 40\,nm-thick Si$_{3}$N$_{4}$ foil coated with 30\,nm-thick Aluminum
are used for the detector.
[Figure adopted from \citet{zhao14}].
 (right) Detecting sensitivity as a function of accumulative exposure time
 (assuming a source spectrum of an absorbed power-law with a photon index
 of $\Gamma=2$ and 3, respectively).}
     \label{fig:eff_area}
     \end{center}
     \end{figure}

The WXT telescope consists of eight modules, with FXT mounted at the center (Figure\,\ref{fig:layout}).
%WXT employs the MPO Lobster-eye optics, with a focal length of 375 mm.
Table\,\ref{tab:specification} lists the main specifications of both WXT and FXT.
The eight WXT modules make up a spherical array mosaicked by 441 concentric MPO pieces,
each of 40mmx40mm in size.
The total FoV subtends a solid angle of $60\times60$\,deg$^2$ ($\sim$1.1 steradian),
which is about 1/11 of the whole sky.
WXT has a large focal plane size, $420 \times 420$\,mm$^2$ in total.
As a baseline detector, gas detectors based on GEM (Gas electron multiplier) are currently being developed.
The effective area and sensitivity curve of WXT are shown in Figure\,\ref{fig:eff_area}.
FXT is a narrow field-of-view ($\sim1$\,deg) telescope of the same lobster-eye MPO type
with a focal length of 1.4\,m, yielding a much larger effective area ($\sim80$\,cm$^2$ at 1\,keV).
FXT has an aperture size of about 240 mm, which is mosaicked by $6\times6$ MPO pieces.
FXT will employ CCD as the focal plane detector to gain better spectral performance.

\begin{table}[]
\begin{center}
{\small \footnotesize
\caption{\label{tab:specification} Specifications of WXT and FXT}
\begin{tabular}{lcc}\hline\hline
\noalign{\smallskip}
Parameters & WXT & FXT  \\ \hline
\noalign{\smallskip}
Number of modules & 8 & 1\\
Field-of-view & $60^\circ \times60^\circ$ & $1^\circ \times1^\circ$ \\
Focal length (mm) & 375 & 1,400 \\
Angular resolution FWHM (arcmin) & $<$5 & $<$5 \\
Nominal pandpass (keV) & 0.5--4.0 & 0.5--4.0 \\
Energy resolution @1\,keV & 40\% & 100\,eV \\
Effective area (central focus + arms) (cm$^2$)&  8 @0.7\,keV & 80  @1\,keV \\
Sensitivity (erg\,s$^{-1}$\,cm$^{-2}$ @1,000\,s) & $\sim1\times10^{-11}$& $\sim3\times10^{-12}$ \\
\noalign{\smallskip}\hline\\
\end{tabular}\\
}
%a) \\
%c)  0.3--10\,keV
\end{center}
\end{table}

%\subsection{Scientific capability and mission profile}

\begin{figure}
\begin{center}
\includegraphics[width=0.55\textwidth]{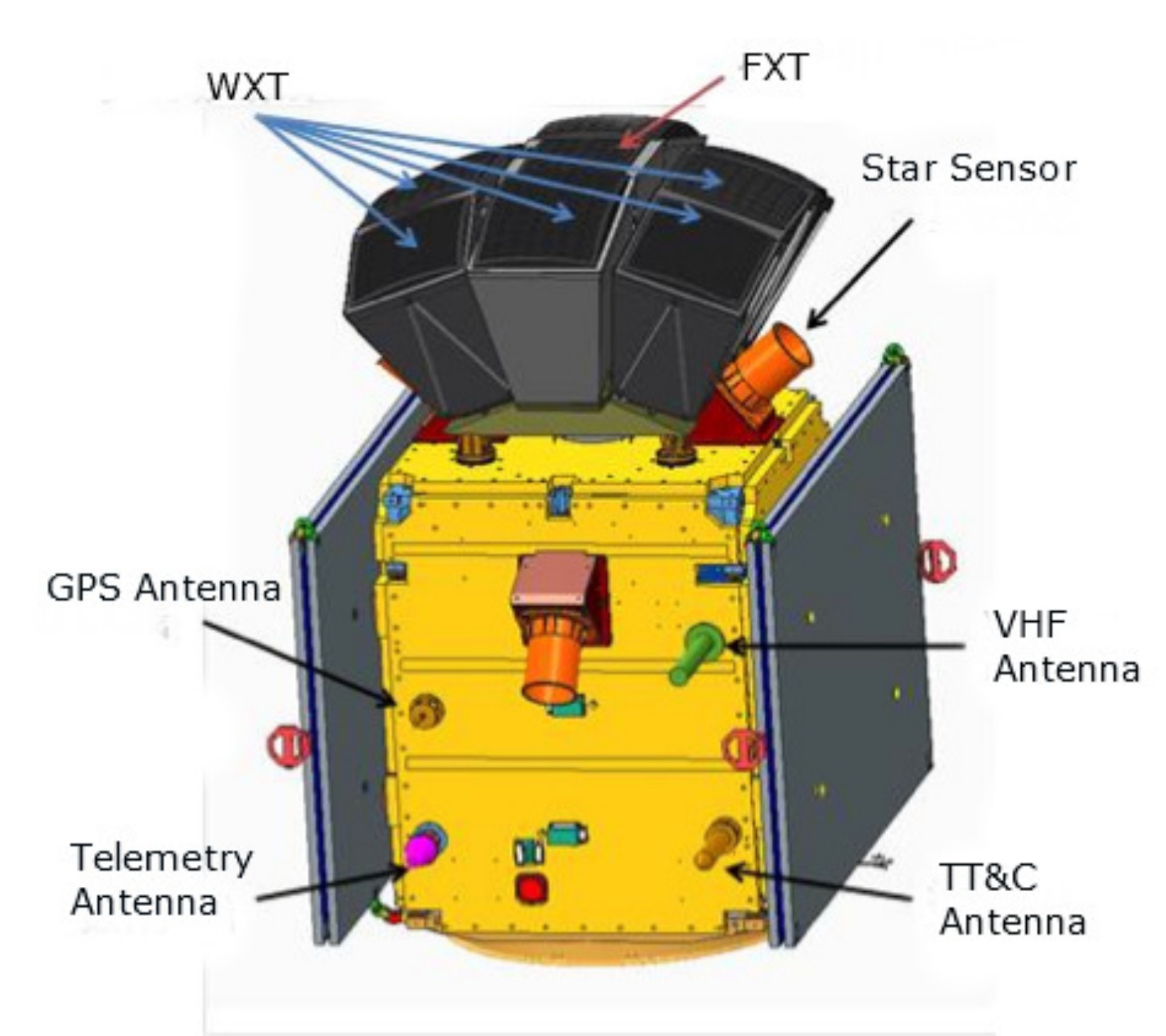}
\includegraphics[width=0.44\textwidth]{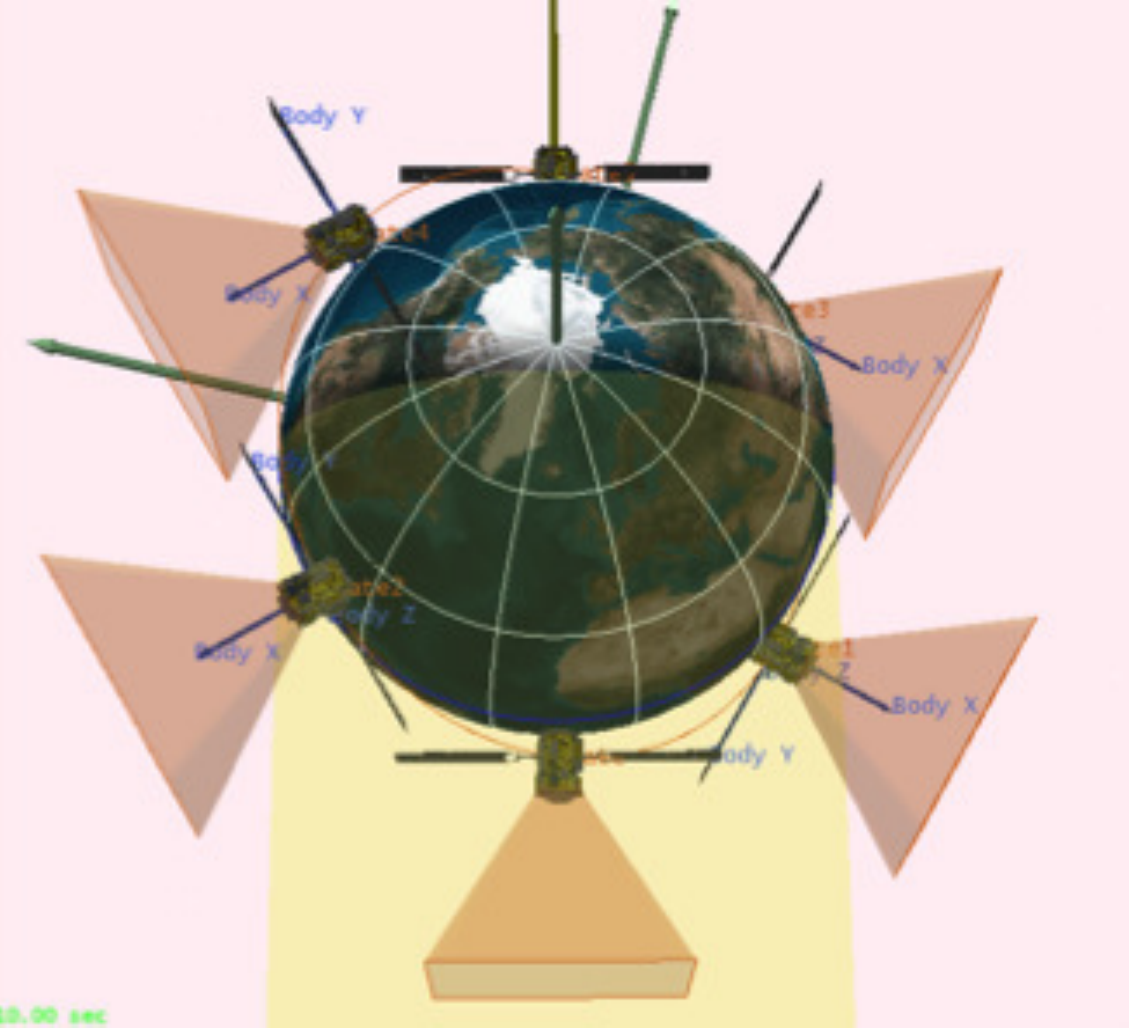}
\caption{(left) Layout of the Einstein Probe satellite. (right) Illustration of the field-of-view and pointed observations in one orbit. [Figures adopted from \citet{yuan14}].}
\label{fig:satellite}
\end{center}
\end{figure}

The layout of the EP satellite and its FoV as well as the observing mode are 
illustrated in Figure\,\ref{fig:satellite}.
The satellite will be in a circular orbit with an altitude of about 600\,km and an inclination angle less than 30 degrees.
The survey strategy of EP will be composed of a series of pointings (each of  about 11-minute exposure)
to mosaic the night sky in the directions avoiding the Sun.
Over three orbits ($\sim$97\,minutes each) almost the entire night sky will be sampled.
The pointing directions are shifted by about 1 degree per day to compensate the daily movement of the Sun on the sky. In this way, the entire sky will be covered within half-a-year's operation.
Once a transient source is detected with WXT
and is classified and triggered by the processing and alerting system onboard,
the satellite will slew to a new position to enable  pointed follow-up observations of the new source with FXT.
Meanwhile, WXT continues to monitor  the new sky region centering the position of the transient.
The alert data of transients are expected to be downlinked within one minute or so via
the fast telemetry system, for which the French VHF system is considered.

\subsection{Summary}
\label{sec:summary}

The emerging lobster-eye MPO optics provides a promising technology to realise X-ray focusing,
thus the improvement of detecting sensitivity,
while retaining large field-of-view of over thousands of square degrees.
Its application to X-ray all-sky monitors  is expected to revolutionise
the high-energy time-domain astrophysics.
The Einstein Probe mission, which is based on this technology,
will attempt to address some of the key questions in astrophysics and cosmology
by capturing faint flashes of X-ray radiation produced by energetic events
within a cosmic horizon far beyond the reach of any current and previous missions.
Its scientific impact will span a wide range of research fields
in astrophysics  and  fundamental physics,
from  stars, compact objects in our and nearby galaxies, black holes,
supernovae, GRBs, galaxies to cosmology.
In particularly, EP has the capability of detecting high-redshift GRBs at $z \sim$10 and beyond.
It is expected that EP will detect $10-30$ GRBs at $z>6$ ($3-9$ GRBs at $z>8$) per year (Wu et al., in preparation),
and will thus deliver a considerable sample of  high-redshift GRBs
with which to study the early Universe within its nominal lifetime of three years.
      % EP
\section{THESEUS}
\label{sect:theseus}

\subsection{Gamma--Ray Bursts as probes of the Early Universe}

The study of the Universe before and during the epoch of reionization represents one of the
major themes for the next generation of space and ground–based observational facilities.
Many questions about the first phases of structure formation in the early Universe will still be open in the late 2020s:
when and how did first stars/galaxies form?
What are their properties? When and how fast was the Universe enriched with metals?
How did reionization proceed?

Because of their huge luminosities, mostly emitted in the X and gamma-rays, their redshift distribution
extending at least to z $\sim$10 and their association with explosive death of massive stars and star forming regions,
Gamma--Ray Bursts (GRBs) are unique and powerful tools for investigating the early Universe: SFR evolution,
physics of re-ionization, galaxies metallicity evolution and luminosity function, first generation (pop III) stars.

The European community played a fundamental role in the enormous progress in the field of GRBs in the last 15--20 years
(BeppoSAX, HETE--2, Swift, AGILE, Fermi, plus enormous efforts in optical IR and radio follow-up)
In 2012, two European proposals for ESA Call for Small mission dedicated to GRBs and all-sky monitoring: GAME (led by Italy, SDD-based cameras + CZT-based camera + scintillator based detectors) and A--STAR (led by UK, lobster-eye telescopes + CdTe detectors).
Following all this unique esperience and efforts, a White Paper on GRBs as probes of the early Universe submitted in response to ESA Call for science theme for next L2/L3 missions \citep{Amati+al+2013} was very well considered by ESA.

The THESEUS proposal described below is a result and follow--on of all this unique esperience and efforts by
the European GRB community, reinforcing also the collaboration with extra--European (e.g., USA) GRB communities and
opening itself to tight colaboration with the cosmology and transients astrophysical communities.

\begin{figure}
  \centerline{\includegraphics[width=0.6\textwidth]{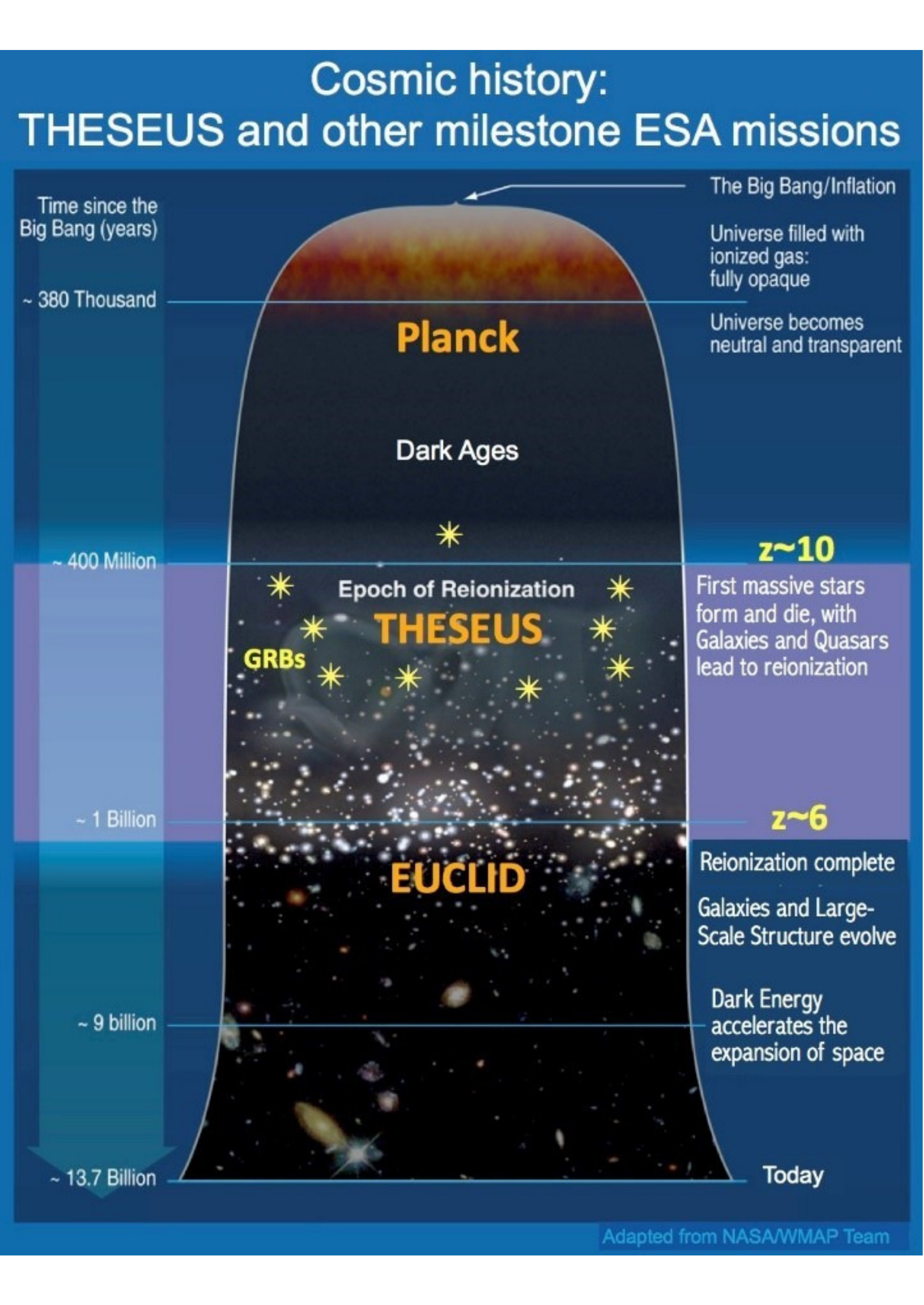}}
\caption{Example of the complementarity and uniqueness of the THESEUS mission w/r to other cosmological
measurements. (Credits: Caltech/NASA and the THESEUS Collaboration)}
\label{fig:theseus1}       % Give a unique label
\end{figure}

\subsection{The THESEUS mission concept}

The Transient High Energy Sky and Early
Universe Surveyor (THESEUS) is a mission concept  a mission concept that will be submitted to ESA in 2016 by a large international collaboration in response to the Call for next M5 mission within the Cosmic Vision Programme.
The primary scientific goals of the mission are linked to the following
Cosmic Vision themes:
4.1 Early Universe,
4.2 The Universe taking shape, and
4.3 The evolving violent Universe.

As detailed below, the main goal of THESEUS is fully exploiting GRBs as cosmological probes, thus providing
a fundamental and unique step forward in our understanding of the early Universe.
More in general, THESEUS would vastly increase the discovery space of several classes of high energy
transient phenomena over the entire cosmic history.

This is achieved via a unique payload  providing an unprecedented combination of:
(i) wide and deep sky monitoring in a broad energy band    (0.3 keV--20 MeV);
(ii) focusing capabilities in the soft X-ray band granting large grasp and high angular resolution;
and 3) on board  near-IR capabilities for immediate transient identification and first redshift estimate.

\subsection{Scientific goals}

The main scientific goals of THESUES can be summarized as follows.

{\bf  (a) Exploring the Early Universe (cosmic dawn and
reionization era) by unveiling the Gamma--Ray Burst (GRBs)
population in the first billion years}, namely
to perform unprecedented studies of the star formation history  up to z $\sim$ 8--9 and
more;
to detect and study the primordial (pop III) star population: when did the first
stars form and how did the earliest pop III and pop II stars influence their
environments?;
to investigate the re-ionization epoch, the interstellar medium (ISM) and
intergalactic medium (IGM) up to z ~ 8--9 and more: how did re-ionization proceed
as a function of environment, and was radiation from massive stars its primary
driver? How did cosmic chemical evolution proceed as a function of time and
environment?;
to investigate the properties of the early galaxies and what was the galaxies
global star formation in the re--ionization era, and
to investigate the dark energy properties and evolution.
The complementarity of THESEUS under this respect with other ``cosmology" mission
investigating the CMB or the large scale structure of the Universe is illustrated
in Figure\,\ref{fig:theseus1}.

{\bf (b) Performing an unprecedented deep survey of the soft X-
ray transient Universe} in order to:
Fill the present gap in the discovery space of new classes of transients
events, thus providing unexpected phenomena and discoveries;
Provide a fundamental step forward in the comprehension of the physics
of various classes of  Galactic and extra--Galactic transients, like, e.g.: tidal
disruption events TDE, magnetars /SGRs, SN shock break--out, Soft X--ray
Transients SFXTS, thermonuclear bursts from accreting neutron stars,
Novae, dwarf novae, stellar flares, AGNs / Blazars);
Provide real time trigger and accurate ($\sim$ 1 arcmin within a few seconds; $\sim$1 arcsec
within a few minutes) location of  (long/short) GRBs and high--energy
transients for follow-up with next-generation optical (EELT), IR (JWT),
radio (SKA), X--rays (ATHENA), TeV (CTA) or neutrino telescopes and
identify electromagnetic counterpart of  detections by next generation
gravitational wave detectors.

{\bf Additional science}.
By satisfying the requirements coming from the above main science
drivers, the THESEUS payload will also automatically be capable to
perform excellent secondary and observatory science, e.g.:
unprecedented  insights in the physics and progenitors of GRBs and their
connection with peculiar core-collapse SNe;
substantially increased detection rate and characterization of subenergetic
GRBs and X--Ray
Flashes;
IR survey and guest observer possibilities, thus allowing a strong
community involvement;
survey capabilities  of transient phenomena similar to the Large Synoptic
Survey Telescope (LSST) in the optical: a remarkable scientific synergy can
be anticipated.

\begin{table}
\caption{Main characteristics of the THESEUS/SXI instrument (Credits: P. O'Brien, J. Osborne, D. Willingale and the THESEUS Collaboration)}
\begin{center}
\label{tab:theseus1}       % Give a unique label
\includegraphics[width=0.9\textwidth]{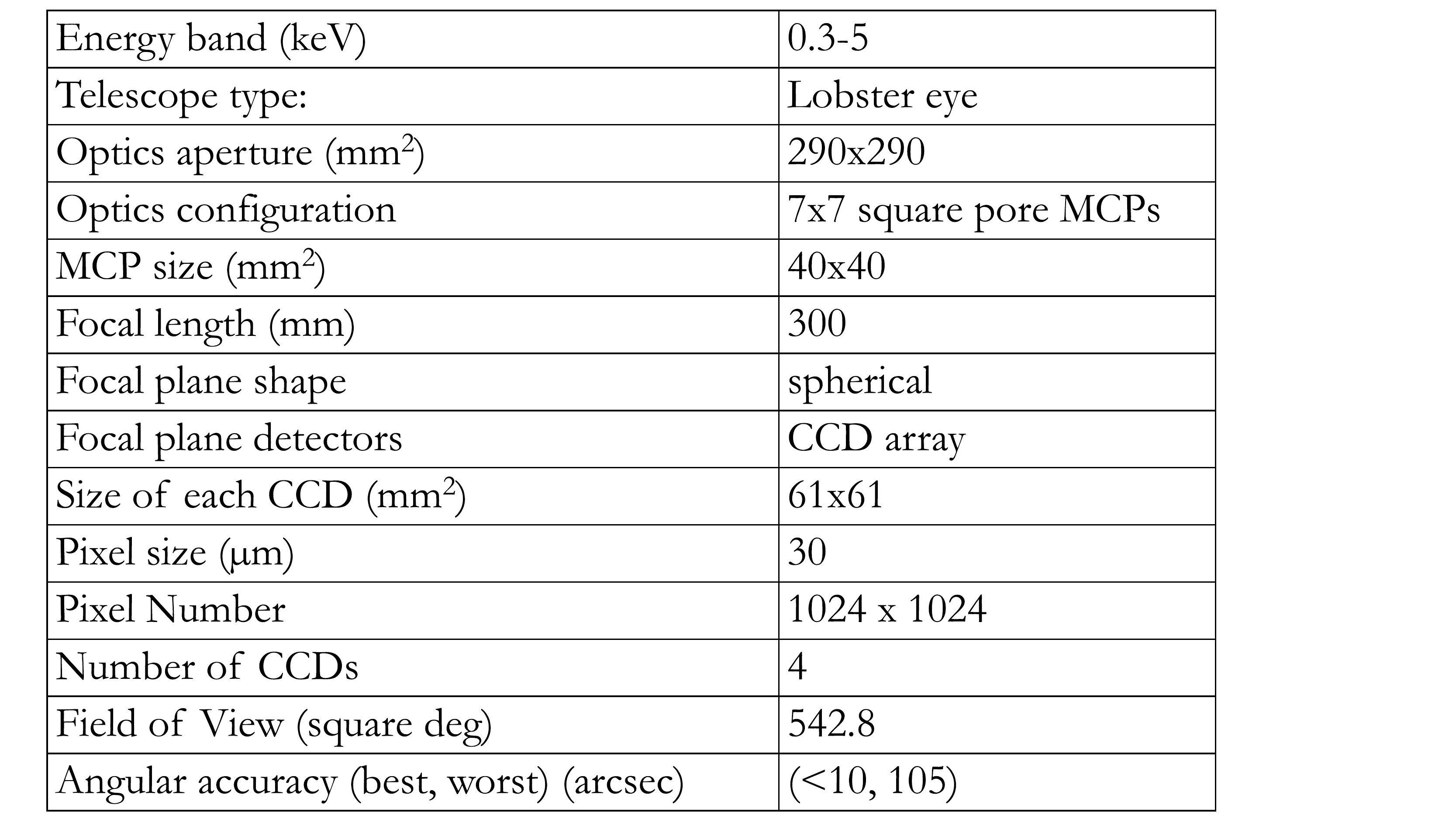}
\end{center}
\end{table}

\begin{table}
\caption{Main characteristics of the THESEUS/XGS instrument (Credits: F. Fuschino, C. Labanti, M. Marisaldi and the THESEUS Collaboration)}
\begin{center}
\label{tab:theseus2}       % Give a unique label
\includegraphics[width=0.9\textwidth]{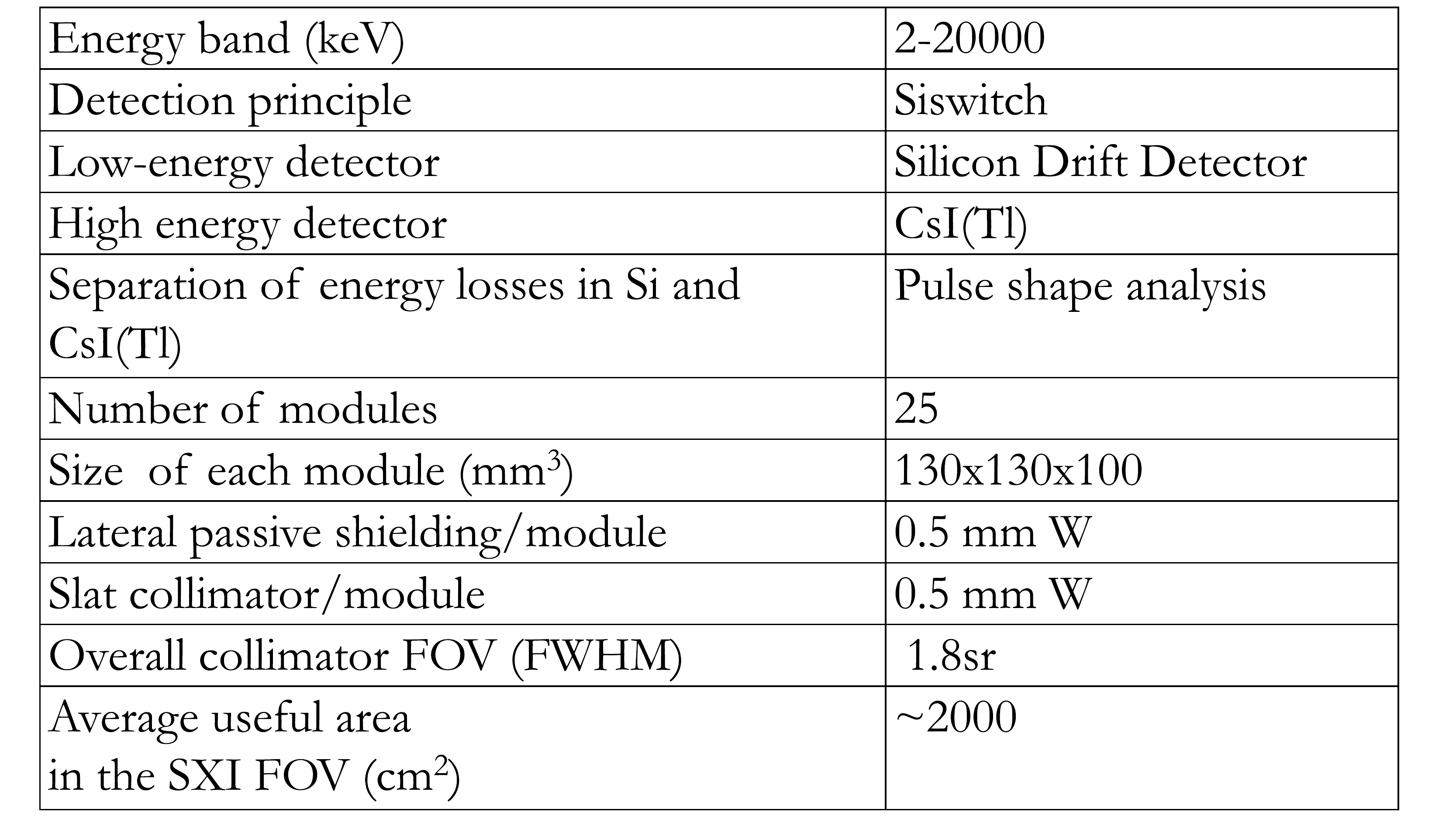}
\end{center}
\end{table}

\begin{table}
\caption{Main characteristics of the THESEUS/IRT instrument (Credits: A.J. Castro-Tirado, V. Reglero, D. Gotz and the THESEUS Collaboration)}
\begin{center}
\label{tab:theseus3}       % Give a unique label
\includegraphics[width=0.9\textwidth]{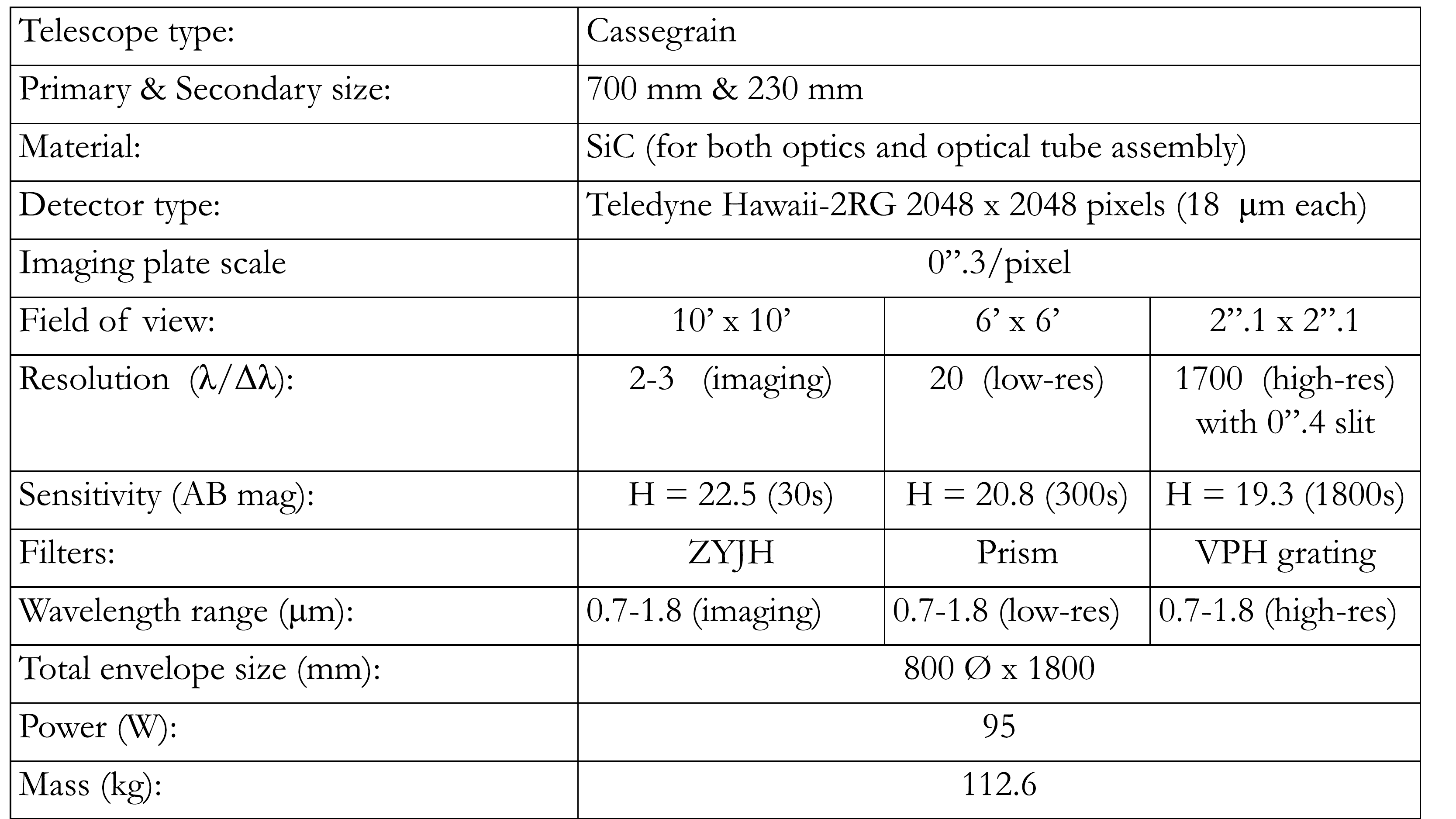}
\end{center}
\end{table}

\subsection{Payload and mission profile}

The scientific goals which come from a full exploration of the early Universe requires the detection of a factor ten more GRBs
(about 100) in the first billion years of the Universe (z $>$ 6), in the 3 years prime mission life time of THESEUS.
Such a requirement is well beyond the capabilities of current and near future GRB detectors
(Swift/BAT, the most sensitive one, has detected only very few GRBs above z = 6 in 10 years).
As supported by
intensive simulations performed by us and other works in the literature, the needed substantial increase of high--z GRBs
requires both an increase of $\sim$1 order of magnitude in sensitivity and an extension of the detector passband down to the
soft X-rays (0.5--1 keV). Such capabilities must be provided over a broad field of view ($\sim$1 sr) with a source location
accuracy $<$ 2’, in order to allow efficient counterpart detection, on-board spectroscopy and redshift measurement and optical and
IR follow-up observations.
Such performances can best be obtained by including in the payload a monitor based on the lobster--eye telescope technique, capable
of focusing soft X--rays in the 0.3--6 keV energy band over a large FOV. Such instrumentation has been under development for several
years at the University of Leicester, has an high TRL level (e.g., BepiColombo) and can perform an all--sky survey in the soft
X--rays with an unprecedented combination of FOV, source location accuracy ($<$1’) and sensitivity thus addressing both main science
goals of the mission.
An onboard infrared telescope of the 0.5--1m class is also needed, together with spacecraft fast slewing capability (e.g., 30°/min),
in order to provide prompt identification of the GRB optical/IR counterpart, refinement of the position down to a few arcs
(thus enabling follow-up with the largest ground and space observatories), on--board redshift determination and spectroscopy of the
counterpart and of the host galaxy. The telescope may also be used for multiple observatory and survey science goals.
Finally, the inclusion in the payload of a broad field of view hard X--ray detection system covering the same survey FOV as the
lobster--eye telescopes and extending the energy band from few keV up to several MeV will increase significantly the capabilities of
the mission. As the lobster-eye telescopes can be triggered by several classes of transient phenomena (e.g., flare stars, X--ray
bursts, etc), the hard X--ray detection system provides an efficient means to identify true GRBs and detect other transient sources
(e.g., short GRBs). The joint data from the three instruments will characterize transients in terms of luminosity, spectra and timing
properties over a broad energy band, thus getting fundamental insights into their physics.

Based on the above, the THESEUS payload consists of the instruments shortly described below.

{\bf Soft X--ray Imager (SXI)}: a set of Lobster Eye (0.3--6 keV) telescopes covering a total FOV of 1 sr field with 0.5--1
arcmin source location accuracy. Each module is a focusing wide field lobster eye telescope based on the optical principles described
in previous sections. The optics aperture is 290$\times$290 mm$^2$ formed by an array of 7$\times$7 square pore
Micro Channel Plates (MCPs). The MCPs are 40$\times$40 mm$^2$ and are mounted on a spherical frame with radius of curvature 600 mm (2 times the focal length of 300 mm). The open aperture provided by each plate is 38$\times$38 mm$^2$; the outer dimension of the
optics frame is 320$\times$320 mm$^2$. The focal plane of each SXI module is a spherical surface of radius of curvature 600 mm situatedat a distance of 300 mm (the focal length) from the optics aperture. The detectors for each module comprise a 2$\times$2 array of
large format CCDs baselined to be supplied by e2v technologies (UK) Ltd. Each CCD has an active area of 61$\times$61 mm$^2$;
the detectors are tilted to approximate to the spherical focal surface.

{\bf InfraRed Telescope (IRT)}: a 70 cm class near-infrared (up to 2 microns) telescope (IRT) with imaging and moderate spectral
capabilities. The telescope (optics and tube assembly) will be made of SiC, a material that has been used in other space missions
(such as Gaia, Herschel, Sentinel 2 and SPICA). Simulations using a 0.7 m aperture Cassegrain space borne NIR telescope (with a 0.23 m secondary mirror), using a Teledyne Hawaii-2RG 2048x2048 pixels detector (18 μm/pixels, resulting in 0.3 arcsec/pix plate scale)
show that, for a 22.5 (H) point like source in a single 300 s exposure one could expect a SNR of ~6.
In order to achieve such performances the telescope needs to be cooled at 240 (+/- 3) K by passive means,
conductive and radiative insulations. Regarding the instrument, the optics box needs to be cooled to 190 (+/- 5) K to by a two stage cooler for which the first stage will cool the optics and the second stage (the cold end) will cool the IR detector itself to 95 (+/-10) K: this allows the detector dark current to be kept at an acceptable level.
The mechanical envelope of IRT is a cylinder with 80 cm diameter and 180 cm height. A sun-shield is placed on top of the telescope baffle for IRT straylight protection.  The thermal hardware is compossed by a pulse tube  cooling the Detector and FEE electronics and a set of thermal straps extracting the heat from the electronic boxes and camera optics coupled to a  radiator located on the spacecraft structure. The overall telescope mass is 112.6 kg and the total power supply is 95W.

{\bf X--Gamma--rays Spectrometer (XGS)}: non-imaging spectrometer (XGS) based on SDD+CsI, covering the same FOV than the Lobster
telescope extending its energy band up to 20 MeV. The XGS consiosts of 25 modules made of scintillator bars optically coupled to an
array of Silicon Drift Detectors (SDD) PhotoDiodes (PD) tightly packaged to each other. Both SDD-PDs and scintillator detect X- and gamma-rays. The top SDD-PD, facing  the X-/gamma-ray entrance window,  is operated both as X-ray detector for low energy X-ray photons interacting in Silicon and as a read-out system of the scintillation light resulting from X-/gamma-ray interactions in the scintillator. The bottom SDD-PD at the other extreme of the crystal bar operates only as a read-out system for the scintillations.  The discrimination between energy losses in Si and CsI is based on the different  shape of charge pulses  resulting from X-ray interactions in Si or from the collection of the scintillation light  thanks to their different timing properties (Marisaldi et al. 2005).
Each bar is made of scintillating crystal 5$\times$5$\times$30 mm$^3$ in size. Each extreme of the bar is covered with a
PD for the read-out of the scintillation light, while the other sides of the bar are wrapped with a light reflecting material convoying the scintillation light towards the PDs. The scintillator material is CsI(Tl) peaking its light emission at about 560 nm. The PD is realized with the technique of Silicon Drift Detectors with an active area of 5$\times$5 mm$^2$ so matching the scintillator
cross section.

\begin{figure}
  \centerline{\includegraphics[width=0.7\textwidth]{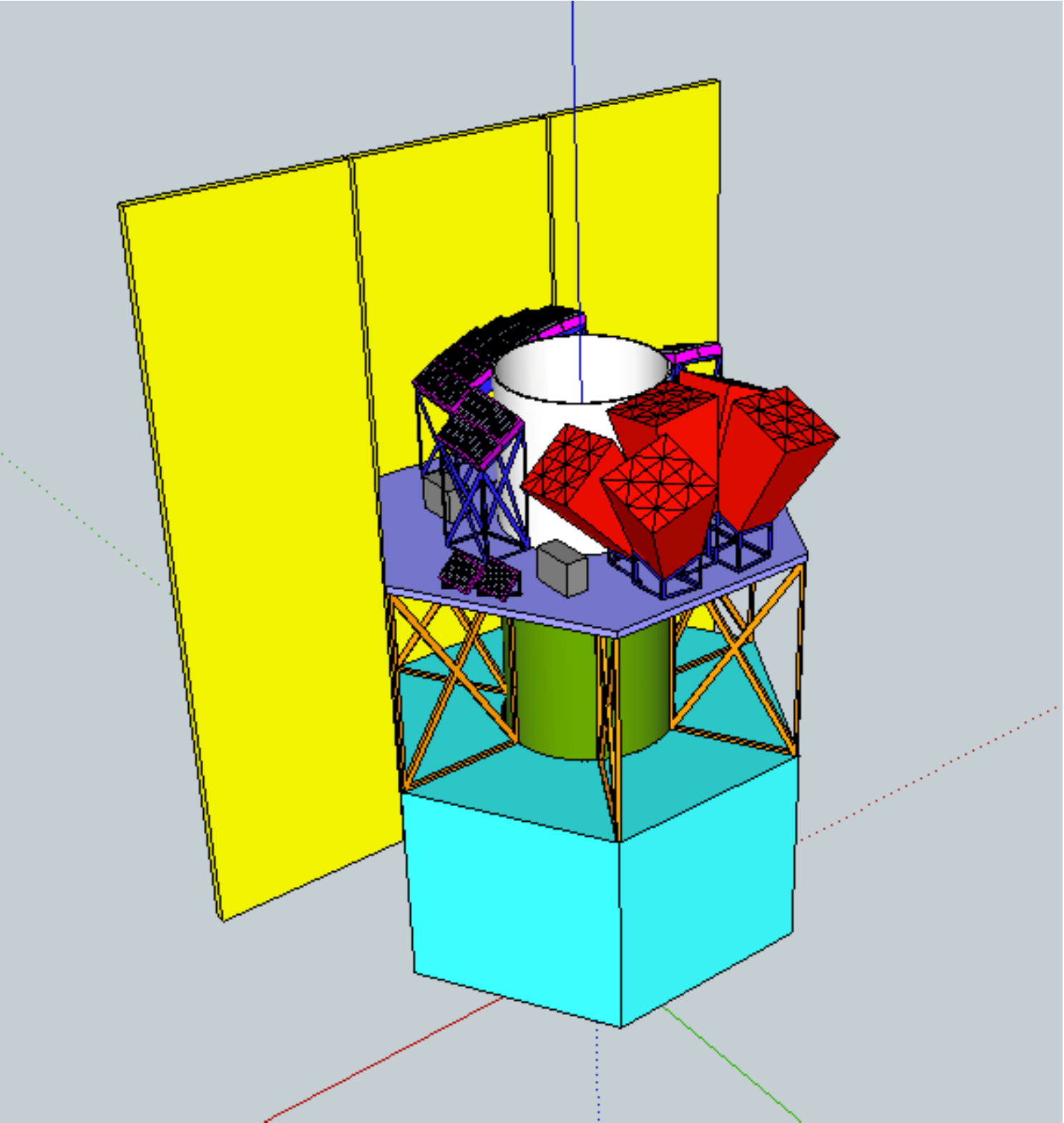}}
\caption{Schematic view of the payload and satellite in THESEUS: the central green and white tube contains the IRT
detectors, optics and buffle; the red modules are the SXI; the black--pink modules are the XGS. (Credits: F. Fuschino, C. Labanti and the THESEUS Collaboration)}
\label{fig:theseus2}       % Give a unique label
\end{figure}

\begin{figure}
  \centerline{\includegraphics[width=0.9\textwidth]{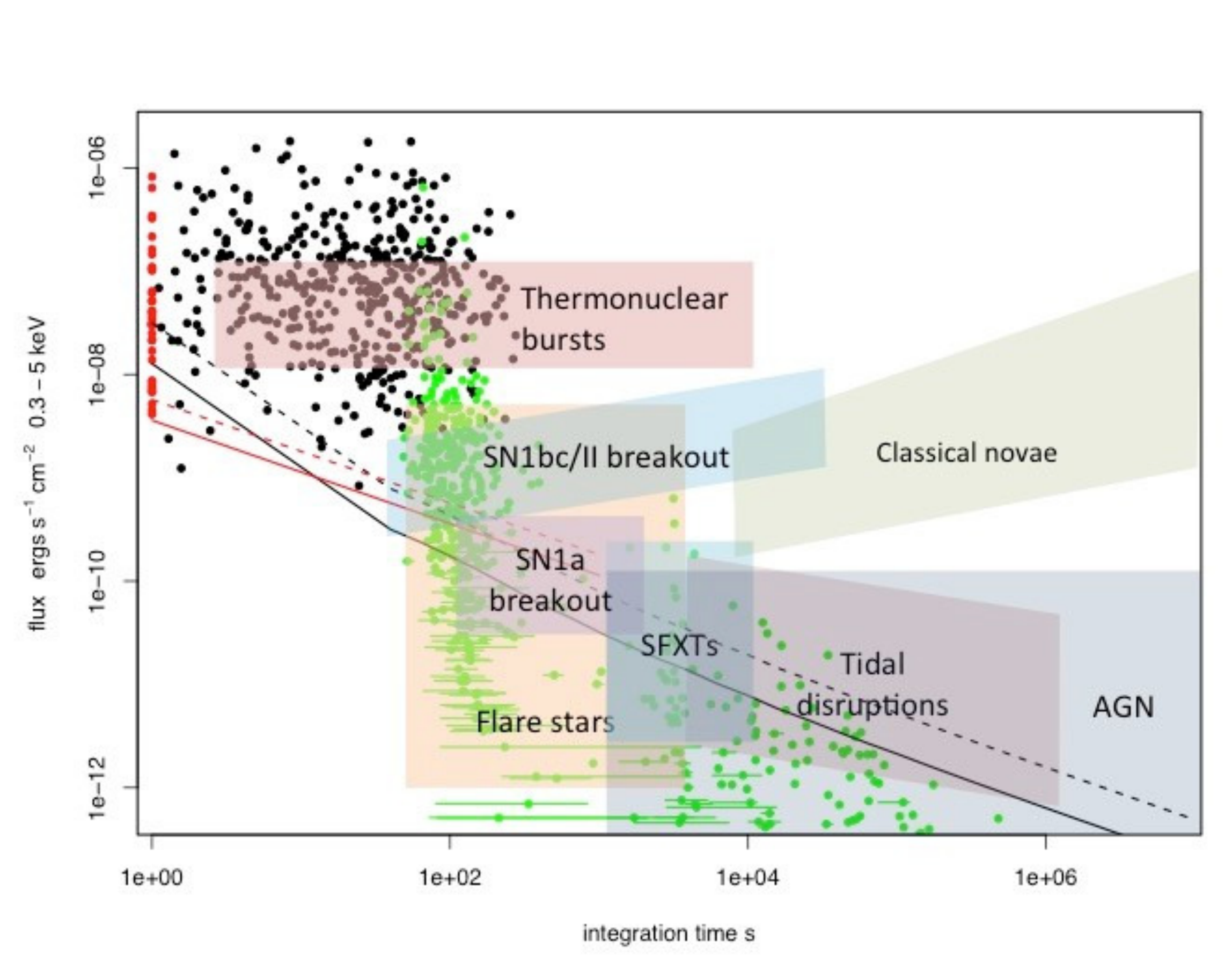}}
\caption{Sensitivity of the SXI (black curves) and XGS (red) vs. integration time. The solid curves assume a source column density of
5$\times$10$^{20}$ cm$^{-2}$ (i.e. well out of the Galactic plane and very little intrinsic absorption).
The dotted curves assume a source column density of 10$^{22}$ cm$^{-2}$ (significant intrinsic absorption).
The black dots are the peak fluxes for Swift BAT GRBs plotted against T90/2. The flux in the soft band 0.3-10 keV was estimated using the T90 BAT spectral fit including the absorption from the XRT spectral fit. The red dots are those GRBs for which T90/2 is less than 1 second. The green dots are the initial fluxes and times since trigger at the start of the Swift XRT GRB light-curves. The horizontal lines indicate the duration of the first time bin in the XRT light-curve. The various shaded regions illustrate variability and flux regions for different types of transients and variable sources. (Credits: D. Willingale, P. O'Brien, J. Osborne and the THESEUS Collaboration)}
\label{fig:theseus3}       % Give a unique label
\end{figure}

\begin{figure}
  \centerline{\includegraphics[width=0.9\textwidth]{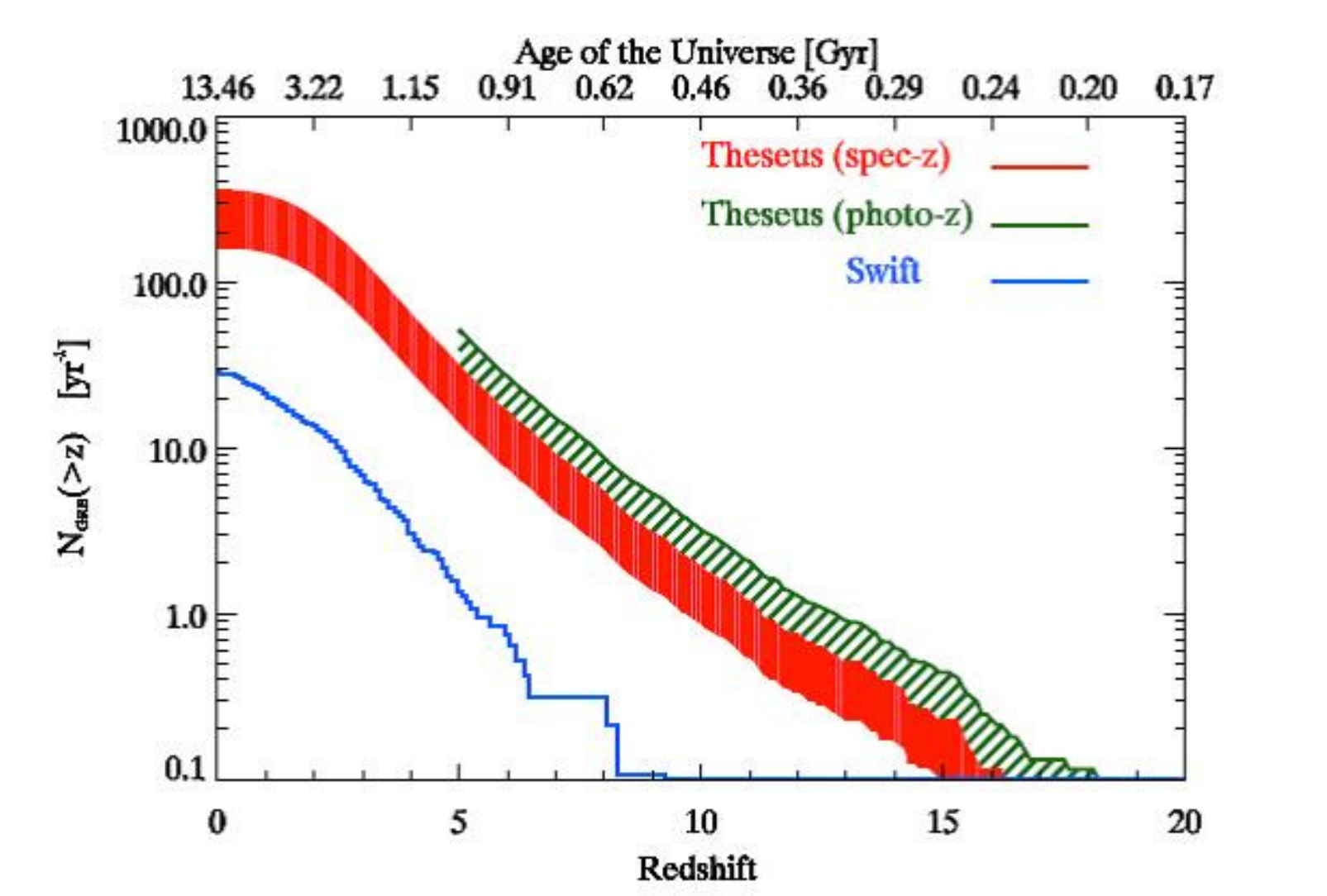}}
\caption{The annual rate of GRBs predicted for THESEUS SXI (red) compared to Swift (blue). The upper scale shows the age of the Universe. For Swift the actual number of known redshifts is approximately one third that plotted and none were determined on board (the blue curve has been linearly scaled upwards to match the total Swift trigger rate). For THESEUS the red region uses the simulations from Ghirlanda et al. (2015) and adopts the instrument sensitivity for the SXI. (Credits: G. Ghirlanda, R. Salvaterra and the THESEUS Collaboration)}
\label{fig:theseus4}       % Give a unique label
\end{figure}

The main instruments characteristics are summerized in Tables\,\ref{tab:theseus1}, \ref{tab:theseus2} and \ref{tab:theseus3}, and a possible payload accomodation sketch is
shown in Figure\,\ref{fig:theseus2}. Sensitivity curves of the SXI and XGS are jointly shown in Figure\,\ref{fig:theseus3}.
Examples of expected performances in terms of GRB detection rate as a function of redshift and rate of different classes of
transients are shown in Figure\,\ref{fig:theseus4} and Table\,\ref{tab:theseus4}, respectively.

\begin{table}
\caption{Theseus detection rates for different astrophysical transients and variables (Credits: J. Osborne, P. O'Brien, D. Willingale and the THESEUS Collaboration)}
\label{tab:theseus4}       % Give a unique label
\centerline{\includegraphics[width=\textwidth]{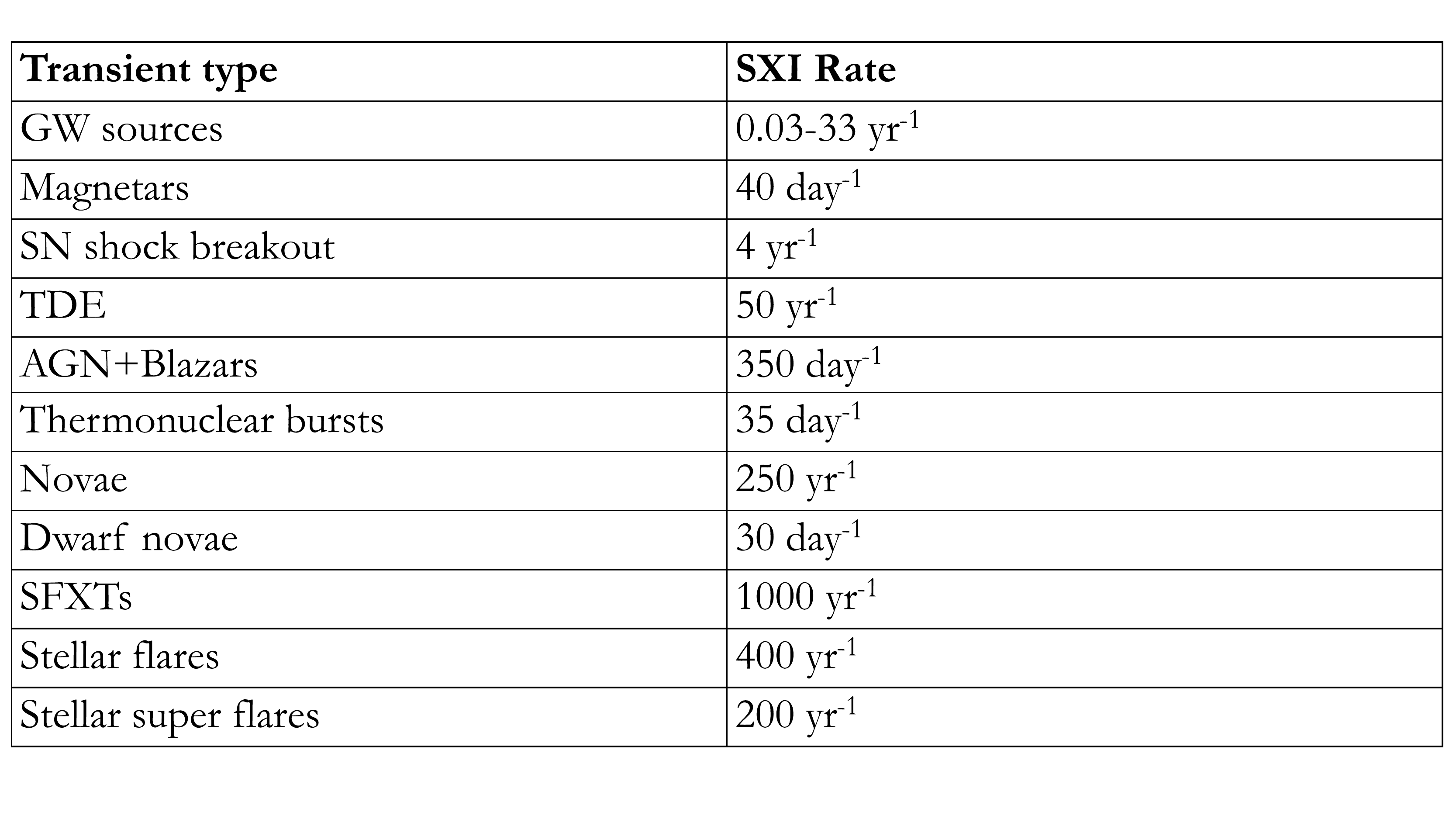}}
\end{table}

\subsection{Mission profile and consortium}

The proposed mission profile includes an
an onboard data handling (OBDH) system capable of detecting, identifying and localizing likely transients in the SXI and XGS FOV;
the capability of promptly (within a few tens of seconds at most) transmitting to ground the trigger time and position of GRBs
(and other transients of interest); and a spacecraft slewing capability of ~30°/min).
The baseline launcher / orbit configuration is a launch with Vega to a low inclination low Earth orbit (LEO, $\sim$600 km, $<5$°),
which has the unique advantages of granting a low and stable background level in the high-energy instruments, allowing the
exploitation of the Earth’s magnetic field for spacecraft fast slewing and facilitating the prompt transmission of transients
trigger and positions to ground. The basic observing strategyASI antenna in Malindi and, as an option, the brazilian antenna in Alcantara were
proposed as ground stations. The basic observing strategy is based on alternating anti--sun and $\sim$ polar
pointing, a compromise between maximum sky coverage, optimization of follow--up observations from the ground,
instruments requirements (thermal, etc.).
Considered prompt downlink options include: NASA/TDRSS, ESA/EDRS, WHF network, IRIDIUM network,
ORBCOMM. MOC and SOC were proposed to be managed by ESA, while the SDC was proposed to be managed by ASI (ASDC).

The total payload mass of THESEUS, including  all contingencies, was estimated to be $\sim$350 kg, and the total spacecraft
dry mass about 1000 kg (power about 230 W and 800 W, respectively). The foreseen telemetry budget is fully compatible with
the capabilities of the X--band, which will be the standard for next ESA M missions.

The THESEUS payload consortium for the ESA/M4 proposal is organized as follow.
The SXI will be responsibility of the UK (led by the University of Leicester), the XGS will be responsibility of Italy
(led by INAF, the University of Ferrara and INFN), and the IRT will be the responsibility of Spain
(a consortium led by IAA-CSIC including UV, INTA, UGR and UMA) for the camera and ESA for the optics.
The core consortium includes also Poland (led by CBK) and Denmark (led by DTU) for payload data handling hardware and software.
Czech Repubblic (led by CTU) is also involved for contributions to the SXI, as well as Slovenia (SPACE-Sl) for communications.
France (CNES, CEA) is available to provide the SVOM network of VHF antennae and associated control center.
A junior international contribution is foreseen by USA (NASA) for the TDRSS system and contributions to the XGS and IRT camera.
Finally, Hungary and Ireland have declared its interest in investigating their contributions to different payload components during assessment phase.
  % Theseus

\begin{acknowledgements}
The authors are grateful to the International Space Science Institute of Beijing (ISSI-Beijing), 
its executive director Prof. M. Falanga and all the staff for hosting and funding the workshop ``Gamma-Ray Bursts: a tool to Explore the Young Universe" held in Beijing from April 10 to 15 2015.
W. Yuan and C. Zhang  thank R. Willingale, J.P. Osborne and P. O'Brien for their contribution to the EP project
and acknowledge support of the ``Strategic Priority Research Program on Space Science" 
(Grant number No. XDA04061100) of the Chinese Academy of Sciences. 
B. Cordier and D. G\"otz acknowledge  financial support of the UnivEarthS Labex program at
Sorbonne Paris Cit\'e (ANR-10-LABX-0023 and ANR-11-IDEX-0005-02). 
Sun acknowledges support from the 973 program 2014CB845802 and NSFC 11503028. 
\end{acknowledgements}

\bibliographystyle{aps-nameyear}
\bibliography{reference}
\nocite{*}

\end{document}